\documentclass{bmcart}
\usepackage[utf8]{inputenc} 
\usepackage{amsfonts}
\usepackage{amsmath}
\usepackage{geometry}
\usepackage{hyperref}
\usepackage{amssymb}
\usepackage{adjustbox}
\usepackage{graphicx}
\usepackage{caption}
\usepackage{subcaption}
\usepackage{courier}
\usepackage[mathscr]{eucal}
\usepackage{xcolor}
\usepackage{array}
\usepackage{url}
\usepackage{stackengine}
\usepackage{placeins}
\usepackage{lscape} 
\usepackage{ntheorem}
\theoremstyle{break}
\newtheorem{theorem}{Theorem}
\newcommand{\XX}{\mathbb{X}}

\begin{document}
\begin{frontmatter}

\begin{fmbox}
\dochead{Research}

\title{Cost Effective Reproduction Number Based Strategies for Reducing Deaths from COVID-19}

\author[
   addressref={aff1},          
   email={thron@tamuct.edu}  
]{\inits{}\fnm{Christopher} \snm{THRON}}
\author[
   addressref={aff2},
   corref={aff2},                      
   noteref={n1},
   email={vianney.mbazumutima@imsp-uac.org}
]{\inits{}\fnm{Vianney} \snm{MBAZUMUTIMA}}
\author[
   addressref={aff3},
   email={vargas\_tamayo98@yahoo.com}
]{\inits{LVT}\fnm{Luis V.} \snm{TAMAYO}}
\author[
   addressref={aff4},
   email={leonardt@imsp-uac.org}
]{\inits{}\fnm{L\'eonard} \snm{TODJIHOUNDE}}


\address[id=aff1]{%
  \orgname{Department of Sciences and Mathematics, Texas A $\&$ M University-Central Texas}, 
  \street{ Killeen TX 76549 USA},                     
  \city{Austin},                               
  \cny{USA}                                    
}
\address[id=aff2]{%
  \orgname{Institute of Mathematics and Physical Sciences, IMSP-B\'enin, Abomey Calavi University},
  \city{Porto-Novo},
  \cny{B\'enin}
}
\address[id=aff3]{%
  \orgname{ Department of Sciences and Mathematics, Texas  A\&M University-Central Texas},
  \street{Killeen TX 76549 USA},
  \city{Austin},
  \cny{USA}
}
\address[id=aff4]{%
  \orgname{Institute of Mathematics and Physical Sciences, IMSP-B\'enin, Abomey Calavi University},
  \city{Porto-Novo},
  \cny{B\'enin}
}

\begin{artnotes}
\note[id=n1]{Equal contributor} 
\end{artnotes}

\end{fmbox}


\begin{abstractbox}

\begin{abstract} 
In epidemiology, the effective reproduction number $R_e$ is used to characterize the growth rate of an epidemic outbreak. If $R_e >1$, the epidemic worsens, and if $R_e< 1$, then it subsides and eventually  dies out.  In this paper, we investigate properties of $R_e$ for a modified  SEIR model of COVID-19 in the city of Houston, TX USA, in which the population is divided into low-risk and high-risk subpopulations. The response of $R_e$ to two types of control measures (testing and distancing) applied to the two different subpopulations is characterized. A nonlinear cost model is used for control measures, to include the effects of diminishing returns.
 Lowest-cost control combinations for reducing instantaneous $R_e$ to a given value are computed.  We propose three types of heuristic strategies for mitigating COVID-19 that are targeted at reducing  $R_e$, and we exhibit the tradeoffs between strategy implementation costs and number of deaths. 
We also consider two variants of each type of strategy: basic strategies, which consider only the effects of controls on $R_e$, without regard to subpopulation; and high-risk prioritizing strategies, which maximize control of the high-risk subpopulation. Results showed that of the three heuristic strategy types, the most cost-effective involved setting a target value for $R_e$ and applying sufficient controls to attain that target value. This heuristic led to strategies that begin with strict distancing of the entire population, later followed by increased testing. Strategies that maximize control on high-risk individuals were less cost-effective than basic strategies that emphasize reduction of the rate of  spreading of the disease.  
The model shows that delaying the start of control measures past a certain point greatly worsens strategy outcomes. We conclude that the effective reproduction can be a valuable real-time indicator in determining cost-effective control strategies. 
 
\end{abstract}



\end{abstractbox}
%

\end{frontmatter}



\section{Introduction}
One of the major concerns of the World Health Organization(WHO) is the prevention of  large epidemics or pandemics. Various techniques and different human, economic, and material resources  are deployed in order to eradicate epidemics as soon as possible. For example, smallpox destabilized the world for centuries \cite{owidsmallpox}, but was completely eliminated worldwide by 1977 thanks to efforts by WHO and other organizations \cite{fenner1988smallpox,henderson2013lessons}. However, in spite of man's best efforts, some diseases have evaded control, and continue to threaten entire populations both locally and internationally. A recent example of this is  Coronavirus-19, which was discovered in  Wuhan City, China, in December 2019 \cite{malik2016mid,2020updated}, spread throughout the world within a few weeks, and  was declared a pandemic by WHO\cite{sohrabi2020corrigendum} on 30th January 2020. 

COVID-19 poses special difficulties in that a significant proportion of infectious cases are asymptomatic. Infectious asymptomatic cases may spread the infection without being detected. Also, asymptomatic cases may persist even after all known cases of the disease have been eradicated.
Asymptomatic cases of COVID-19 may constitute a large proportion of the infected individuals.
 There is a wide range of estimates of the proportion of cases that are asymptomatic.     Early reports from China from testing of residents and overseas arrivals suggested that $40\%-80\%$ of infections showed no symptoms \cite{day2020covid,ferguson2020report}.
Comprehensive testing was performed in the city of Vo' before and after lockdown   showed that  about $43\%$ of  infections detected were asymptomatic \cite{lavezzo2020sup}.
The authors of \cite{he2020proportion} reviewed of $41$ studies with a total of $50,155$ confirmed COVID-19 cases, and found the pooled percentage of asymptomatic infection was $15.6\% (95\% CI: 10.1\%‐23.0\%)$.
In \cite{chen2020Epid}, infection rates due to contact with asymptomic carriers was estimated at $4.11\%$ (6 infections from 146 contacts) compared to $6.30\%$ for symptomatic cases (126 infections from 2001 contacts).
Besides asymptomatic cases, a presymptomatic infectious phase of $1-4$ days  is estimated in \cite{byrne2020inferred}, while the  estimated asymptomatic infectious period was $4-9.5$ days. 
Since the discovery of COVID-19, numerous measures and resources  were deployed for its eradication. Vaccines and curative medicines were lacking, but some measures such as social distancing and testing were effective in slowing its spread.
Health specialists recommend using both strategies.  The effectiveness and costs associated with both strategies are discussed below.

There exist many various types of  social distancing, but  the most basic involves maintaining distance in public spaces, mask-wearing, and quarantine for symptomatic individuals and their contacts. More severe measures include  banning public gatherings, restricting population movement, closing businesses, and stay-at-home orders. 
The effectiveness of distancing measures has been investigated by  researchers using simulations based on mathematical models. In \cite{kruse2020optimal}  an  SIR model that includes lockdown policies was studied analytically. The authors concluded that the optimal policy depends only on the shadow price difference between infected and susceptible individuals. They furthermore concluded that more extreme measures applied over a short time horizon are more effective than less extreme measures over a longer horizon. In \cite{matrajt2020eval}, an age-structured mathematical model was developed for investigating the effectiveness of social distancing interventions to stop the spread of COVID-19 using 4 scenarios.  It was found that the number of the new infections, hospitalisations and deaths were all decreased by distancing measures.  \cite{prem2020effect} uses the data from Wuhan city and found that when the contact patterns are  changed as a result of distancing, the epidemic peak is delayed and new cases of coronavirus disease 2019 are decreased.  Other research works that have explored the importance of using the  social distance strategy against COVID-19  are \cite{alvarez2020simple,kantner2020b,kissler2020social,kra2020eff,wu2020nowcasting}. Although social distancing measures have saved many lives, they also have incurred significant costs for society. Economic activity has decreased, producing widespread hardship and  unemployment. Several researchers have investigated these costs in order to better understand economic consequences of the COVID-19 epidemic, especially the costs associated with distancing and testing measures.   \cite{toda2020suscept} uses a SIR model to simulate  transmission rates for various countries under various   social distancing strategies, and  estimates the associated prices during an epidemic period. \cite{RRA1731} estimates the economic costs caused by social distancing strategy in fighting against COVID-19, where five  main social-distancing policies have been considered. Results show a decline in average income in the range of $4.6-18.6\% $, depending on level of distancing.
As an alternative and supplement to distancing, testing and tracing is another viable and complementary approach. The effectiveness of tests to detect the presence of SARS-CoV-2 virus and antibodies to SARS-­CoV-2 has also been studied by a number of researchers\cite{jarrom2020Eff,lopes2020eff,hometesting}.
Reference\cite{grassly2020comp} established a mathematical  model of SARS-CoV-2 that includes PCR(Polymerase Chain Reaction) testing, and estimates the reduction in the effective reproduction number  achieved by testing and isolating symptomatic individuals, regular screening of high-risk groups irrespective of symptoms, and quarantine of contacts of laboratory-confirmed cases identified through test-and-trace protocols.

Although testing avoids the economic slowdown and social costs associated with distancing, it is still not without costs.
In \cite{nisha}, it is estimated the  cost of   \$51 or \$100  per diagnostic test depending on the type of test, while \cite{cent2020medic} gives an approximate cost of \$100 paid by medicare for each laboratory tests for detecting SARS-CoV-2, and \cite{ducomparative} quotes a price of \$5 per test, but without mentioning the false positive and false negative rates.

\subsection{Basic reproduction number and effective reproduction number}
In theoretical epidemiology, the basic parameter used to characterize the rate of spread of a disease is called the \emph{reproduction number}. The basic reproduction number  $R_0$ is defined as the average number of secondary infections
which one typical infected individual would generate if the population were completely
susceptible\cite{far2001esti}. In multicompartment models of disease dynamics, $R_0$  is computed as the dominant eigenvalue
of a positive linear operator.  The concept of basic reproduction number ($R_0$) was first introduced in 1886 \cite{bockh1886stat} and  has been used in multitudes of studies of infectious diseases. Some recent examples include 
\cite{choi2003simple,chowell2004basic,hef2005persp,holme2015basic,riley2003trans,van2002reprod,wesley2009basic,wonham2004epid}. For the recent pandemic COVID-19, many researchers have estimated $R_0$ using different approaches for various countries and regions \cite{al2020basic,katul2020global,li2020basic,linka2020reproduction,liu2020reproductive,rahmanpreliminary,rahman2020basic}.

In addition to $R_0$, the \emph{effective reproduction number} (denoted by $R_e$) is also of interest. $R_e$ is defined as  the number of secondary infections produced by a single infectious when the population has both susceptible and non-susceptible individuals (non-susceptible may include infectious, immune, vaccinated, etc.) and/or control methods have been implemented. Several previous studies have estimated $R_e$ for various scenarios. The effective reproduction number for COVID-19 of India and its states has been determinated in \cite{tyagi2020estimation} using  Real-Time Bayesian Method. \cite{ochoa2020effective} determines the effective reproduction number for COVID-19 at the first 10 days of Latin American countries where the highest was in Ecuador($R_e =3.95$) and the smallest in Peru ($R_e =2.36$) and make a comparison with one of Spain($R_e= 2.9$) and Italy($R_e=2.83$). The effective reproduction number is evaluated in \cite{na2020probabilistic}
 by using a probabilistic methodology which consider only the daily death statistics of a given country. 

In our work, we will focus on the effects of different levels of testing and distancing measures on the effective reproduction number of COVID-19, as well as these measures' economic costs. 
The organization of the paper is as follows. Section~\ref{sec:2} describes the COVID-19 model with and without controls, computes basic and effective reproduction numbers, and  estimates model parameters from available data.  Section~\ref{sec:3} gives simulations and interpretations of daily and long-term scenarios, including sensitivity analysis and comparison of different control strategies.
Finally, Section~\ref{sec:4}  summarizes our findings and gives concluding remarks.

\section{COVID-19 epidemic model formulation and mathematical properties}\label{sec:2}

In this section, we present the multicompartment models of COVID-19 and, identify the parameters  and estimated control costs used in simulation. In addition, using  the next  generation matrix,  we  calculate the basic and effective reproduction number.

\subsection{Multicompartment model}
\begin{figure}[h!]
  \centering
    \includegraphics[width=9cm]{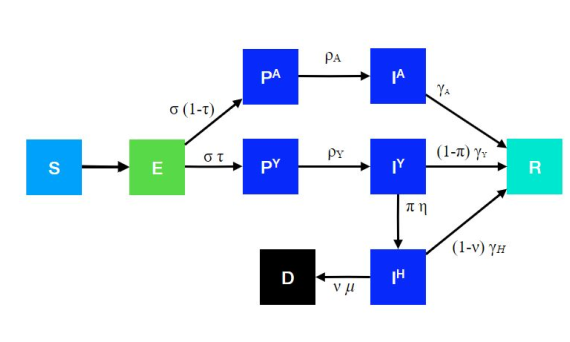}
    \caption{COVID-19 transmission schema\cite{haoxi20}\label{fig:model}}
  \end{figure}

In this section, we describe the deterministic compartmental model that was used to model the transmission dynamics of COVID-19.  The model is based on \cite{haoxi20}, and partitions the entire population into subpopulations according to age and risk group and,  where each subpopulation is further subdivided into the following compartments: susceptible$(S)$, exposed $(E)$, pre-symptomatic infectious $(P^Y)$, pre-asymptomatic infectious $(P^A)$, symptomatic infectious $(I^Y)$, asymptomatic infectious $(I^A)$, symptomatic infectious that are hospitalized $(I^H)$, recovered $(R)$, and deceased $(D)$. In our model, two subpopulations are identified, namely low risk and high risk. It is assumed that the survivors have permanent immunity, and  that dead individuals are not infectious. The subpopulation model is diagrammed in Figure~\ref{fig:model}, 
and the explicit equations are as follows:
\begin{equation}\label{eqn:1}
\begin{aligned}
\frac{dS_j}{dt}&=-\sum_{i=0}^1\frac{1}{N_i} \left(I^Y_i\omega^Y +I^A_i\omega^A+P^Y_i\omega^{PY}+P^A_i\omega^{PA}\right)\phi_{ji}\beta S_j,\\
\frac{dE_j}{dt}&=\sum_{i=0}^1\frac{1}{N_i} \left(I^Y_i\omega^Y +I^A_i\omega^A+P^Y_i\omega^{PY}+P^A_i\omega^{PA}\right)\phi_{ji}\beta S_j-\sigma E_j, \\
\frac{dP^A_j}{dt}&=(1-\tau)\sigma E_j-\rho^A P^A_j,\\
\frac{dP^Y_j}{dt}&=\tau\sigma E_j-\rho^Y P^Y_j, \\
\frac{dI^{A}_j}{dt}&=\rho^A P^A_j-\gamma^A I^A_j, \\
\frac{dI^{Y}_j}{dt}&=\rho^Y P^Y_j-(1-\Pi_j)\gamma^Y I^Y_j-\Pi_j \eta I^Y_j,  \\
\frac{dI^{H}_j}{dt}&=\Pi_j \eta I^Y_j-(1-\nu_j)\gamma^H I^H_j - \mu \nu_j I^H_j, \\
\frac{dR_j}{dt}&=\gamma^{A}I^{A}_j+(1-\Pi_j)\gamma^{Y}I^{Y}_j+(1-\nu_j)\gamma^H I^H_j+\\
&~~- \mu  \nu_{j}I_j^{H}   r  \left(1 - \frac{\theta / r}{ \max(  \nu_{0}I_{0}^{H} +  \nu_{1}I_{1}^{H} , \theta / r) } \right),\\
\frac{dD_j}{dt}&= \mu  \nu_{j}I_{j}^{H}  \left (1 + r  \left(1 - \frac{\theta / r}{ \max(  \nu_{0}I_{0}^{H} +  \nu_{1}I_{1}^{H} , \theta / r) } \right) \right),
\end{aligned}
\end{equation}
where  $j$ is the subpopulation index (0=low risk, 1=high risk), and 
\begin{equation}\label{eq:NiDef}
N_i=S_i+E_i+P^A_i+P^Y_i+I^A_i+I^Y_i+I^H_i+R_i.
\end{equation}
The initial conditions require all compartment populations to be nonnegative.
 The interpretation and numerical values of the parameters in \eqref{eqn:1} are listed in Table \ref{table:1}. 	
The model \eqref{eqn:1} has non-negative solutions  contained in the feasible region $\Gamma =\{ S_j,  E_j , P^A_j,  P^Y_j, I^A_j,  I^Y_j , I^H_j , R_j, D_j\}\in\mathbb{R}^{18}_+$.

The contact matrix $\Phi = [\phi_{ji}]$, $(i,j) \in \{0,1\}$ is defined by
\begin{align}\label{eqn:phi00}
\Phi&=
\left(\begin{matrix}
\phi_{00}&\phi_{01}\\
\phi_{10}&\phi_{11}\\
\end{matrix}\right)
=\left(\begin{matrix}
10.52&2.77\\
9.4&2.63\\
\end{matrix}\right),
\end{align}
where $\phi_{ji}$ represents the mean number of contacts per day experienced by individuals in group $j$  from individuals of group $i$. The  matrix values in \eqref{eqn:phi00} were obtained  by averaging the contacts between low and high risk individuals over all age groups, using Tables A.4.1-4  and Figure A3  in  \cite{haoxi20} for contact rates and age-specific high risk proportions, respectively.  
\newpage

\begin{table}[h!]
\caption{Baseline  parameters used in the model (based on  \cite{haoxi20}).}\label{table:1}
\centering 
\scalebox{0.8}{
    \begin{tabular}{|c|l|l|l|}
        \hline          
Parameters&Interpretation&Values\\
\hline           
 $\beta$  &baseline transmission rate & 0.0640\\
 \hline 
$ \gamma^A$ &recovery rate on asymptomatic compartment & Equal to $\gamma^Y$\\
\hline 
$\gamma^Y$ &recovery rate on symptomatic non-treated
compartment &$\frac{1}{\gamma^Y}= 4.0$\\
\hline 
$\tau$  &symptomatic proportion& $0.55$\\
\hline 
$\sigma$&exposed compartment exit rate&$\frac{1}{\sigma}\sim 2.9$\\
\hline 
$\rho^A$ &pre-asymptomatic compartment exit rate &Equal to $\rho^Y$\\
\hline 
$\rho^Y$ &pre-symptomatic compartment exit rate & $\frac{1}{\rho^Y}=2.3$\\
\hline 
$P$ &proportion of pre-symptomatic transmission&$0.44$\\
\hline 
$\omega^Y$&relative infectiousness of symptomatic individuals&$1.0$\\
\hline 
$\omega^A$ &relative infectiousness of infectious individuals in
compartment $I^A$&$0.66 $\\
\hline 
$\omega^P$&relative infectiousness of pre-symptomatic individuals&$\omega^P=\frac{P}{1-P}\frac{\tau \omega^Y[YHR/\eta+(1-YHR)/\gamma^Y]+(1-\tau)w^A/\gamma^A}{\tau\omega^Y/\rho^Y+(1-\tau)\omega^A/\rho^A}$\\
\hline
$IFR$ &infected fatality ratio, age specific (\%)&$[0.6440,6.440]$\\
\hline
$YFR$ & symptomatic fatality ratio, age specific (\%)&$[1.130, 11.30]$\\
\hline 
$\gamma^H$ &recovery rate in hospitalized compartment &$\frac{1}{\gamma^H}\sim 10.7$\\
\hline 
$YHR$ &Symptomatic case hospitalization rate \% &$[4.879,48.79]$\\
\hline 
$\Pi$&rate of symptomatic individuals go to hospital, age-specific &$\Pi=\frac{\gamma^Y*YHR}{\eta+(\gamma^Y-\eta)YHR} $\\
\hline 
$\eta$&rate from symptom onset to hospitalized&$0.1695$\\
\hline 
$\mu$&rate at which terminal patients die&$\frac{1}{\mu}= 8.1$\\
\hline 
$HFR$&hospitalized fatality ratio, age specific (\%) & $[4, 23.158]$\\
\hline 
$\nu$&death rate on hospitalized individuals, age specific &$\nu=\frac{\gamma^H HFR}{\mu+(\gamma^H-\mu)HFR}$ \\
\hline
$\theta$& total ventilator capacity in all hospitals & 3000 \cite{barker2020}\\
\hline 
$1/r$& number of deaths from people who are put on respirators & 1/3\\
\hline 
\end{tabular}}
\end{table}       	

There are a few differences between our model and the model in \cite{haoxi20}. In our model the definition of $N_i$ in \eqref{eq:NiDef} does not include $D_i$, since the individuals who have died are no longer in the active population.
In addition, the last two equations in our model  include additional terms that reflect the additional mortality that occurs when the ventilator capacity (represented by the parameter $\theta$) is exceeded.  Note that this change does not affect the reproduction number of the system.
The modifications are derived based on the following assumptions:
\smallskip

\begin{itemize}
\item
All patients that are at risk of dying are put on respirators, if respirators are available;
\item
A fixed fraction of patients that need respiration and are put on respirators nonetheless die. According to the literature, this fraction is about 1/3;  In the model we introduce the parameter $r$, which is the inverse of this fraction, thus $r \approx 3$;
\item
All patients that need respiration but are not put on respirators will die;
\item
Respirators are allocated proportionately to the low and high risk patients who need them.
\end{itemize}
\smallskip

Let $I_{0}^{H}, ~ I_{1}^{H}$ be the number of each group that is hospitalized. We already have that $ \nu_{0}, \nu_{1}$ are the death rates for hospitalized low and high risk, respectively. It follows that there are $r  \nu_{0}I_{0}^{H}$ and  $r   \nu_{1}I_{1}^{H}$  low and high risk patients respectively that need respirators. Letting $n_0,~n_1$ be the number of patients in each group who are on respirators, it follows that the number of terminal patients that die without respiration is $ (r  \nu_{0}I_{0}^{H} + r  \nu_{1}I_{1}^{H} - n_0 - n_1)$.

It remains to solve for $n_0$ and $n_1$. The constraint on total respirators gives
$n_0 + n_1 \leq  \theta. $ 
According to our assumption of proportionate respirator allocation,
$n_0 /~ n_1 =   \nu_{0}I_{0}^{H} /  \nu_{1}I_{1}^{H} $. We may distinguish two cases.
 First, if  $r ( \nu_{0}I_{0}^{H} +  \nu_{1}I_{1}^{H}) \leq \theta$, then we have $n_j = r  \nu_{j}I_{j}^{H}$  for $j=0,1$.
Otherwise, $n_0 + n_1 = \theta$ which implies
\begin{equation}
n_j = \frac{ \theta   \nu_{j}I_{j}^{H} }{( \nu_{0}I_{0}^{H} +  \nu_{1}I_{1}^{H})}, ~~ j = 0,1.
\end{equation}
We may combine these two cases into the single equation:
\begin{equation}
n_j = \frac{ r\theta   \nu_{j}I_{j}^{H}}{\max(r   \nu_{0}I_{0}^{H} + r   \nu_{1}I_{1}^{H} , \theta)},~~ j = 0,1,
\end{equation}
so that
the number of terminal patients in group $j$ that are denied respirators is $\omega_j$, where
\begin{equation}
 \omega_j = r  \nu_{j}I_{j}^{H} -  \frac{ r\theta  \nu_{j}I_{j}^{H}}{\max(r   \nu_{0}I_{0}^{H} + r   \nu_{1}I_{1}^{H} , \theta)}.
\end{equation}  
Low- and high-risk patients that not denied respirators are terminal at rates $\nu_0$ and $\nu_1$, respectively. Therefore we have:
\begin{equation}
    \begin{aligned}
 \frac{dD_j}{dt} &= \mu \nu_j I_{j}^{H}  + \mu \omega_{j}(1-\nu_j) = \mu  \nu_{j}I_{j}^{H}  \left (1 + r  \left(1 - \frac{\theta / r}{ \max(  \nu_{0}I_{0}^{H} +  \nu_{1}I_{1}^{H} , \theta / r) } \right) \right),
    \end{aligned}
\end{equation}
which is identical to the equation for $\frac{dD_j}{dt}$ in \eqref{eqn:1}. The equation for $\frac{dR_j}{dt}$ must be similarly adjusted   by an amount $-\mu\omega_j(1-\nu_j)$ to offset the increased number of deaths due to insufficient respirators.

\subsection{COVID-19 epidemic model formulation under controls}\label{sec:3}
The use of the control measures has an important effect at a certain level on the spread of the COVID-19 epidemic. In order to study disease mitigation, we introduce the effects of two controls: social distancing and COVID testing. Social distancing (denoted by $v_j$) will reduce the overall infectivity, while COVID testing (denoted by $u_j$) will reduce the infectivity of the asymptomatic  and presymptomatic  infectious compartments. The model with controls is identical to \eqref{eqn:1}, except the first two equations are modified as follows:
\begin{equation}\label{eqn:1control}
\begin{aligned}
\frac{dS_j}{dt}&=-\sum\limits_{i=0}^1\frac{1}{N_i} \left(I^Y_i\omega^Y +(1-u_i) [I^A_i\omega^A + \omega^P(P^Y_i\omega^{Y}+P^A_i\omega^{A})]\right)(1-v_j)\beta \phi_{ji}S_j,\\
\frac{dE_j}{dt}&=\sum\limits_{i=0}^1\frac{1}{N_i} \left(I^Y_i\omega^Y +(1-u_i) [I^A_i\omega^A + \omega^P(P^Y_i\omega^{Y}+P^A_i\omega^{A})]\right)(1-v_j)\beta \phi_{ji} S_j-\sigma E_j.
\end{aligned}
\end{equation}
We shall use $X$ to denote the vector
$X = [X_0,X_1]$   of all infected classes, where we define $X_j = [E_j,P^A_j,P^Y_j,I^A_j, I^Y_j,I^H_j]$ and $X'=[X'_0,X'_1]$ the vector of all uninfected classes with $X'_j=[S_j, R_j]$ where $j=0,1$ corresponds to low and high risk subpopulations respectively, with susceptible$(S_j)$, exposed $(E_j)$, pre-symptomatic infectious $(P_j^Y)$,
pre-asymptomatic infectious $(P_j^A)$, symptomatic infectious $(I_j^Y)$, asymptomatic infectious $(I_j^A)$, symptomatic infectious that are hospitalized $(I_j^H)$, recovered $(R_j)$, and deceased $(D_j)$.

In order to take into account the cost associated to the model \eqref{eqn:1control}, we define 
\begin{equation}\label{eqn:Costs}
\begin{aligned}
 \alpha_j(u_j,N^A_j)&=
\begin{cases}
0&~~\text{if}~~u_j=0,\\
a_{j0}+ a_{j1}N^A_j u_j + a_{j2}N^A_j( u_j)^2 &~~\text{if}~~0<u_j\leq u_{j}^{(max)},
	\end{cases}\\
\beta_j(v_j)&= 
 b_{j1}N_jv_j + b_{j2}N_j(v_j)^n \qquad \qquad \qquad \text{if}~~0<v_j\leq v_{j}^{(max)},
\end{aligned}
\end{equation}
where $N^A_j =S_j+ E_j +P^A_j+ P^Y_j+I^A_j$ is the number of asymptomatic individuals in subpopulation. 
The functions $\alpha_j$ and $\beta_j$ are intended to model the costs associated with COVID testing and  social distancing respectively for subpopulations $j=0,1$. The coefficient $ a_{j0}$ represents the fixed cost when the testing program  is implemented; $a_{j1}$ is the  testing cost per person,  $u_j$ is the fraction of asymptomatic individuals in population j that are subject to testing,  and $a_{j2}$ represents the increasing marginal  cost per person incurred as the testing program becomes more intensive (reflecting the law of diminishing returns). The cost function $\beta_j$ in  \eqref{eqn:Costs} reflects the economic cost of social distancing measures, and $b_{j1}$, $b_{j2}$ reflect per-capita costs. $\beta_j$ is modeled as a  nonlinear function, since marginal costs will increase as the severity of distancing measures increases (for example, low-level distancing measures such as wearing masks incurs much less expense than serious measures such as closing stores and stay-at-home orders). The exponent $n >1$ is chosen to reflect these nonlinear effects.  The parameter $v_j$  expresses the proportionate reduction in contacts that result from the implemented measures. 

\subsection{Estimation of  basic and effective reproduction numbers }\label{eq:reprod}
In this section, we will present the calculation of basic and effective reproduction numbers ($R_0$ and $R_e$, respectively) using the next generation matrix technique of Van Den Driessche. 
\subsubsection{Computation of basic reproduction number}
The basic reproduction number $(R_0)$ is the average number of secondary infections produced by a typical case of an infection in a population where everyone is susceptible. It  is affected by the following  factors: the rate of contacts in the host population, the probability of infection being transmitted during contact, the duration of infectiousness.
Using the next generation matrix defined in \cite{van2002reprod} (see also \cite{van2017reproduction}), we establish $R_0$ for the system in \eqref{eqn:1}.  
Recall that $X$ and $X'$ are vectors representing infected and uninfected compartments respectively. 
We have
\begin{eqnarray}
\frac{d X}{dt}&=&\mathcal{F}(\XX)-\mathcal{V}(\XX),\label{eqn:28Aug}\\
\frac{d X'}{dt}&=&\mathcal{W}(\XX),
\end{eqnarray}
where $\XX=(X,X')$, $\mathcal{F}(\XX)$ represents the vector of in-flows into infected compartments (including new infections) and $\mathcal{V}(\XX)$ is the vector of out-flows.  The functions $\mathcal{F}$ and $\mathcal{V}$ are chosen so that $ \mathcal{F}(\XX) \geq 0$ and $\mathcal{V}(\XX) \geq 0$.
 We denote the disease free equilibrium by $(0,\bar{X'})$. Replacing in Equation \eqref{eqn:28Aug}, we have $\mathcal{F}(0,\bar{X'})=0$ and $\mathcal{V}(0,\bar{X'})=0$.
The next generation matrix is given by $FV^{-1}$, where
$$F=\left.\left(\frac{\partial \mathcal{F}}{\partial X}\right)\right|_{(0,\bar{X'})}~\text{and}~V=\left.\left(\frac{\partial \mathcal{V}}{\partial X}\right)\right|_{(0,\bar{X'})}.$$
 The basic reproduction number $R_0$ is computed as the spectral radius of the matrix  $FV^{-1}$.
From System \eqref{eqn:1}, $F$ and $V$ may be evaluated as (note in the disease-free case, $S_j=N_j$):
\begin{align}
F&=
\left(\begin{matrix}\label{eqn:Fmx}
F_{00}&F_{01}\\
F_{10}&F_{11}\\
\end{matrix}\right),
\end{align}
with
\begin{align}\label{eqn:Fjj}
F_{jj}&=
\left(\begin{matrix}
0&\beta \omega^{P}\omega^A\Phi_{jj}&\beta \omega^P\omega^{Y}\phi_{jj} &\beta \omega^{A}\phi_{jj}&\beta \omega^{Y}\phi_{jj}\\
(1-\tau)\sigma &0&0& 0&0\\
\tau\sigma&0 &0 & 0&0\\
0 & \rho^A & 0&0&0\\ 
0&0&\rho^Y&0&0\\
\end{matrix}\right),
\end{align}

\begin{align}\label{eqn:F1-jj}
F_{1-j,j}&=
\left(\begin{matrix}
0&\frac{\beta \omega^P\omega^{A}N_{1-j}}{N_j}\phi_{1-j,j}&\frac{\beta \omega^P\omega^{Y}N_{1-j}}{N_j}\phi_{1-j,j} & \frac{\beta \omega^{A}N_{1-j}}{N_j}\phi_{1-j,j}&\frac{\beta \omega^{Y}N_{1-j}}{N_j}\Phi_{1-j,j}\\
0 &0&0& 0&0\\
0&0 &0 & 0&0\\
0 & 0 & 0&0&0\\ 
0&0&0&0&0\\
\end{matrix}\right),
\end{align}

 and 

\begin{align}\label{eqn:Vmx}
V&=
\left(\begin{matrix}
V_{00}&V_{01}\\
V_{10}&V_{11}\\
\end{matrix}\right),
\end{align}
with

\begin{align}\label{eqn:Vjj}
V_{jj}&=\left(\begin{matrix}
\sigma&0 &0 & 0&0\\
0 &\rho^A&0& 0&0\\
0&0 &\rho^Y & 0&0\\
0 & 0& 0&\gamma^A&0\\ 
0&0&0&0&(1-\Pi_j)\gamma^Y+\Pi_j\eta\\
\end{matrix}\right); \qquad
V_{1-j,j}=\textbf{0}_{5 \times 5}.
\end{align}

The following theorem guarantees that  the spectral radius is equal to the dominating eigenvalue $ FV^{-1}$, which is real, positive, and has an associated eigenvector that is also real and positive: 

\begin{theorem}(Perron-Frobenius theorem)\cite{stephen2009boyd}\label{thm:PF}: 
Suppose  $A \in \mathbb{R}^{n\times n}$ is nonnegative and regular, $i.e. A^k >0$ for some $k$, then
\begin{enumerate}
\item[$\bullet$]there is an eigenvalue $\lambda_{pf}$ of $A$ with geometric multiplicity 1 that is real and positive, with positive left and right eigenvectors
\item[$\bullet$]for any other eigenvalue $\lambda$, we have $|\lambda|<\lambda_{pf}$.
\end{enumerate}
\end{theorem}
\subsubsection{Effective reproduction number}
As an epidemic progresses, there will be an increasing proportion of the population which has recovered from the disease and hence has some degree of immunity. When this happens, the basic reproduction does not accurately reflect the number of secondary cases produced by an infection. Our calculation of $R_0$ also does not include the introduction of controls. In order to obtain an estimate of the effective reproduction number $R_e$, we modify \eqref{eqn:Fjj}-\eqref{eqn:F1-jj} based on \eqref{eqn:1control} as follows:

\begin{align}\label{eqn:Fjj-1}
F_{jj}&=
\scriptsize
\left(\begin{matrix}
0&\frac{(1-u_j)(1-v_j)\beta \omega^P\omega^{A}S_j}{N_j}\Phi_{jj}&\frac{(1-u_j)(1-v_j)\beta \omega^P\omega^{Y}S_j}{N_j}\Phi_{jj} &\frac{ (1-u_j)(1-v_j)\beta \omega^{A}S_j}{N_j}\Phi_{jj}&\frac{(1-v_j)\beta \omega^{Y}S_j}{N_j}\Phi_{jj}\\
(1-\tau)\sigma &0&0& 0&0\\
\tau\sigma&0 &0 & 0&0\\
0 & \rho^A & 0&0&0\\ 
0&0&\rho^Y&0&0\\
\end{matrix}\right),
\end{align}

\begin{align}\label{eqn:F1-jj-1}
F_{1-j,j}&=
\frac{(1-v_{1-j})\beta S_{1-j}}{N_j}\Phi_{1-j,j}\left(\begin{matrix}
0&(1-u_j) \omega^P\omega^{A}&(1-u_j)\omega^P\omega^{Y} & (1-u_j)\omega^{A}&\omega^{Y}\\
0 &0&0& 0&0\\
0&0 &0 & 0&0\\
0 & 0 & 0&0&0\\ 
0&0&0&0&0\\
\end{matrix}\right).
\end{align}

The effective reproduction number $R_e$ is computed as the spectral radius of the matrix  $FV^{-1}$, where $F$ and $V$ are specified using \eqref{eqn:Fmx} and \eqref{eqn:Vmx}, where $F_{jj}$ and $F_{1-j,j}$ are given by Eqs. \eqref{eqn:Fjj-1} and \eqref{eqn:F1-jj-1}, respectively.

\subsection{Control cost estimates}\label{sec:5}
In order to optimize control costs according to the model described above, it is necessary to find realistic estimates of the coefficients $a_{jk}$ and $b_{jk}$ in \eqref{eqn:Costs}. In this section, we present 
results from the literature on which our coefficient estimates are based.

In \cite{thunstrom2020ben}, the reduction in U.S. Gross Domestic Product (GDP) due to distancing was estimated as \$7.2 trillion over a 30-year period, assuming a 3\% discount rate. This 
corresponds to a \$3 per day cost per person and  the distancing was assumed to reduce contact rate by 40\%.  The article also presents an estimate by Goldman-Sachs of 6\% reduction in U.S. GDP due to mortality, morbidity, and productivity losses due to distancing measures. Since the mean yearly income in Houston is  \$31,576 per person \cite{quickfac2019},  
this would translate to a cost of roughly \$5/day/person which includes mortality and morbidity impacts in addition to distancing.  
Reference \cite{RRA1731} estimates between 4.6--18.6\% decline in income in Texas for distancing measures, depending on severity of measures. 
If we suppose that these declines correspond to contact rate reductions of 40\% and 80\% respectively, this would indicate that doubling the severity of distancing measures roughly quadruples the cost. This would correspond to a purely quadratic cost function $\beta_j$ in \eqref{eqn:Costs}, which implies $b_{j1}=0$ and $n=2$. This quadratic cost function reflects the economic principle of diminishing returns: increasingly strict methods incur disproportionately higher costs. 

Economic cost is only part of the total cost incurred by distancing measures. There are social and political costs as well. People feel oppressed by distancing measures, and many people view such measures as infringements on their freedom\cite{aljazeera2020}. Supportive connections with family and friends are disrupted \cite{katerine2020ell}. In the U.S., many anti-distancing demonstrations have taken place \cite{everytown2020} as people are not convinced that such severe measures are necessary.  The costs of distancing are also not distributed evenly among the population, and low-income individuals are often the hardest hit when retail sales diminish and  restaurants and shops are closed down as part of distancing measures.

Testing does not have the same social costs as distancing, but has its own problems. The gold standard test for COVID is the molecular Polymerase Chain Reaction (PCR) test, which requires lab analysis. Results of the test are not immediately available, and often take up to five days to obtain. The lag means that any infection acquired between the test administration and the results will go undetected. Therefore, testing also involves some quarantining, which has its own costs. Additionally, a testing program is most effective if it is coupled with contact tracing, so that others exposed to possible infection are identified and tested as well.  Some contacts may missed, which diminishes the effectiveness of the control strategy. In addition, PCR tests can yield both false negative and false positive results: estimates of error rates range from 2-33\% and 0.8-4\% for false negative and false positive rates per test, respectively \cite{surkova2020false}. PCR tests are also relatively expensive, at about \$50 per test. A cheaper test with 15-minute turnaround is available (the BinaxNOW Covid-19 Ag Card, produced by Abbott Laboratories), but has relatively high false negative and false positive rates  of 2.9\% and 1.5\% respectively \cite{abbott2020d}, and is thus most useful in locations where prevalence is high \cite{menachem2020fast}.

 From a study by Campbell et al \cite{campbell2020active} on testing costs in Canada, we may infer  a per-test total cost estimate (including test + personnel) of \$57 for testing only, and 
\$69 for testing plus tracing \cite{campbell2020active}.  The article also mentions the possibility of reducing costs by 40\% through introduction of pooled sampling and other efficiency measures. One must also take into consideration that the economic cost of testing is of a somewhat different nature from the economic costs of distancing.  Distancing costs correspond to lost productivity and reduced consumption, while testing costs are paid to companies which will produce higher wages and profits, thus returning benefits to the economy. 

Two questions remain vis-a-vis testing, namely the frequency of testing and the effectiveness. A study by Jiang et al. \cite{jiang2020cost} recommends performing three PCR tests before discharging patients into the general population (their conclusion is based on presumed false negative rate of 29\%, which is high compared to other studies). Many studies appear to recommend weekly or monthly testing of key subpopulations (\cite{campbell2020active}, \cite{neilan2020clin}). In \cite{neilan2020clin}, the most intensive testing regimen was estimated to produce a $63\%$ reduction in infections, but at a cost that is 4 times as high as a less strict regimen. The cost of the most intensive regimen was between 2--2.5 billion dollars per million people per 180 days, which comes out to between \$11--\$14 per day per person.
Reference \cite{RRA1731} estimates a decline in income of 4.6--18.6\% due to distancing measures, depending on severity. Using a figure of \$31,576 for mean income per person in Houston registers, this gives a daily cost \$4--\$16.  Based on those simple cost examples, we arrived at baseline cost coefficients for testing and distancing controls as specified in Table \ref{table:2} that will be used on the rest of the work. 

 \begin{table}[h!]
\caption{Testing and social distancing control cost and level parameters}\label{table:2}
\centering 
\scalebox{1}{
    \begin{tabular}{|c|l|c|}
        \hline 
Parameters&Interpretation&Values\\
\hline     
$a_{00},a_{10}$&minimum testing cost per person & \$0\\  
\hline
$a_{01},a_{11}$ &linear testing cost coefficient &\$2.3/person/day\\
\hline
$a_{02},a_{12}$ &quadratic rate of increase of per capita testing cost &\$27/person/day$^2$\\
\hline    
  $b_{01},b_{11}$&constant per capita social distancing costs &\$0\\ 
 \hline
 $b_{02},b_{12}$&quadratic rate of increase of per capita social distancing cost&\$40/person/day$^2$\\ 
 \hline      
$u_0^{max},u_1^{max}$ & maximum testing control level&$0.66 $\\
\hline
$v_0^{max},v_1^{max}$& maximum social distancing control level& $0.8 $\\
 \hline
\end{tabular}}
\end{table}   

\section{Simulations results and discussion}\label{sec:3}

A number of simulations were performed to analyze the relations $R_e$, control levels, and control costs, and to explore the use of $R_e$ in determining cost-effective strategies.  All simulations used the model described in Section \ref{sec:2}, with the parameters given in Tables~\ref{table:1} and \ref{table:2}.  The simulations performed can be classified into three groups.   In the first group of simulations, we first characterize the response of $R_e$ and control cost  to individual control levels.  In the second group we investigate the sensitivity of these relationships to important parameters. In the third groups, we simulate long-term strategies for epidemic mitigation that are based on the findings from previous simulations, and determine their cost-effectiveness.

\subsection{Dependence of $R_e$ on control level and control cost for  individual control strategies}\label{eq:depend}
In this subsection we investigate the behavior of $R_e$ depending on the level of testing and distancing controls applied to  low and high risk population groups. Our computations  were based on \eqref{eqn:Fmx}--\eqref{eqn:F1-jj-1}, with high-risk and low-risk populations of 1.34 million and 423,000 corresponding to demographics of the city of Houston, TX.

\begin{figure}[h!]
  \centering
    \includegraphics[width=4in]{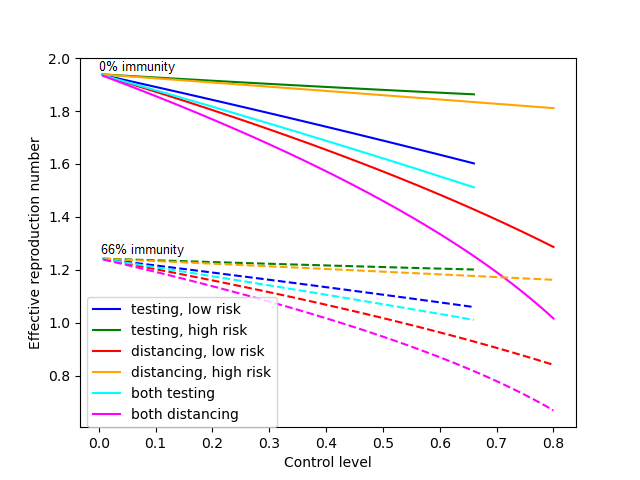}
   \caption{Dependence of $R_e$ on control level for six control strategies at $0\%$ and $66.6\%$ immunity\label{fig:RevsCtrl}}
  \end{figure}

Figure~\ref{fig:RevsCtrl} shows  $ R_e$ as a function of control level for six different control strategies, at two different population immunity levels. The first four strategies employ a single nonzero control (either testing or distancing) on a single group (either low- or high-risk).    In the last two strategies, both population groups are subject to the same control (testing and distancing, respectively). Solid curves correspond to these six strategies when the entire population has $0\%$ immunity (i.e. $100\%$ of the population is susceptible), and the dotted curves are for a population with $67\%$ immunity  respectively ($33\%$ susceptible).   The curves show that when no herd immunity is present,  none of the six strategies is sufficient to bring  $R_e$ below 1. When herd immunity reaches $67\%$  of the population, only high levels of distancing for either the low-risk population or the entire population can bring     $R_e$ below $1$. The figure shows that controls that are applied only to the high-risk group do not  significantly reduce $R_e$ . Note however that these graphs do not take into account the reduced deaths in the high-risk population, because they only show the effect on overall $R_e$ and not the number of deaths incurred by infection.  It is also clear that applying the same strategy to the entire population rather than just the low-risk group greatly increases the effectiveness of the strategy. 

\begin{figure}[h!]
 \centering
 \captionsetup[subfigure]{width=1.0\textwidth}
 	\begin{subfigure}{.43\textwidth} 
 	\centering
 	\includegraphics[width=1.95in]{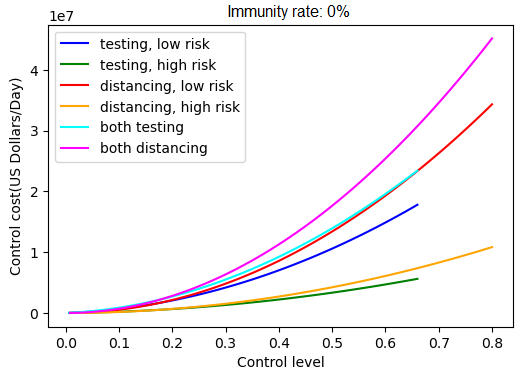}
 	\vspace*{2mm}
 	\caption{ Control costs at 0\% immunity}\label{fig:FigLabelForRefInDoc}
 	\end{subfigure}
 	\begin{subfigure}{0.43\textwidth}
 	\centering
 	\includegraphics[width=1.95in]{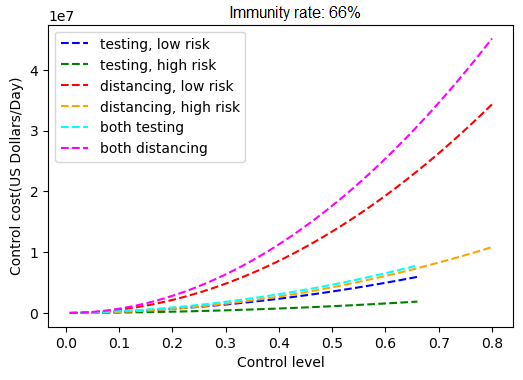}
 	\vspace*{2mm}
 	\caption{Control costs at 67\% immunity}\label{fig:Fig2Label}
 	\end{subfigure}

 \caption{ Control cost as a function of control level for six strategies, for two levels of immunity.\label{fig:CostFigure}}
 \end{figure}

Figures~\ref{fig:CostFigure}(a) and (b)  show the daily cost associated with each level of  the six control strategies defined above, applied at $0\%$ and 66.6\% immunity  respectively. We notice that costs for distancing are consistently higher than corresponding costs for testing: for example, distancing  costs for the entire population (which is the most expensive strategy, for a given control level)  is higher than the cost for testing the entire population at the same level of control. Comparing Figures~\ref{fig:CostFigure}(a) and (b), we see that distancing costs do not depend on immunity level, but testing costs are reduced by more than 50\% when immunity is increased from $0\%$ to $66.6\%$.

\begin{figure}[h!]
  \centering
   \includegraphics[width=3.2in]{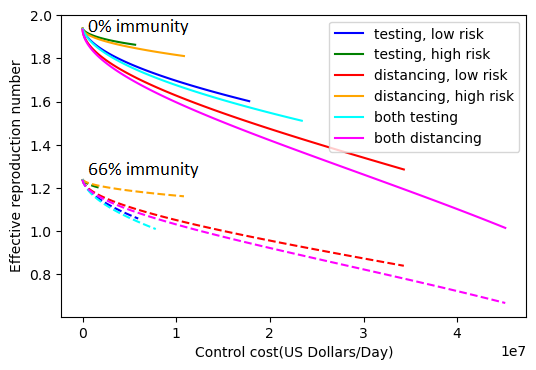}
 \caption{Dependence of $R_e$ on daily implementation cost for six strategies at $0\%$ and $67\%$ immunity \label{fig:RevsCost}}
  \end{figure}

In Figure~\ref{fig:RevsCost}, $R_e$ is plotted as  function of  daily implementation cost for the different control strategies, where the control costs depend on the control levels through the relations 
\eqref{eqn:1}-\eqref{eqn:1control}. At $0\%$ immunity, the most cost-effective strategy is distancing applied to the entire population: this is true regardless of expenditure level. However, the situation changes at  $67\%$ immunity: in this case, testing of the entire population is most cost-effective. However, the effects of universal testing are limited, and cannot reduce $R_e$ below 1 even at the highest possible testing level. As with Figure~\ref{fig:RevsCtrl}, these results do not account for the greater percentage of deaths among the high risk population, because they only include implementation costs and not the costs associated with hospitalization or death.

  \newpage
Figures~\ref{fig:RevsCtrl}~-~\ref{fig:RevsCost} above show that strategies applied to the entire population have a much greater effect and are more cost-effective than strategies applied to a single population group.  Therefore, in the following analysis, we consider only testing and distancing strategies that are applied to the entire population, and not to individual subpopulations.
 
\begin{figure}[hbt]
 \centering
 \captionsetup[subfigure]{width=1.0\textwidth}
 	\begin{subfigure}{.49\textwidth} 
 	\centering
 	\includegraphics[width=2.3in]{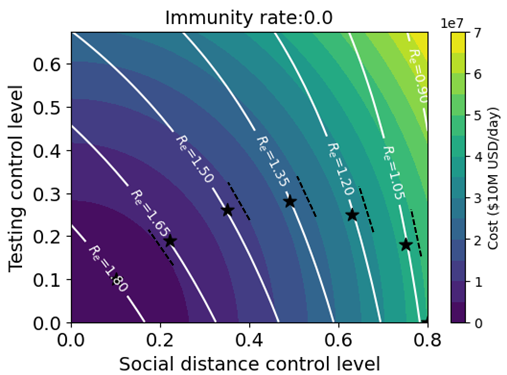}
 	\vspace*{2mm}
 	\caption{ Control costs and $R_e$ values at 0\% immunity}\label{fig:Fig6Label}
 	\end{subfigure}
 	\begin{subfigure}{0.45\textwidth}
 	\centering
 	\includegraphics[width=2.3in]{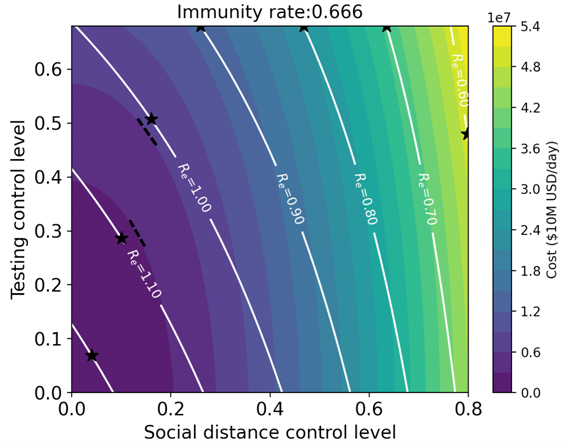}
 	\vspace*{2mm}
 	\caption{Control costs and $R_e$ values at 67\% immunity}\label{fig:Fig7Label}
 	\end{subfigure}
 \caption{$R_e$ and cost dependence on testing and social distancing control levels at two different population immunity levels:   Stars locate points on $R_e$ contours corresponding to minimum cost for the given value of $R_e$.\label{fig:DistCostTestLevelCost}}
 \end{figure}

Figures~\ref{fig:DistCostTestLevelCost}(a),(b) show the dependence of $R_e$ and daily implementation cost on social distance and testing control levels, for two different levels of population immunity.  The white contour lines represent different values of $ R_e$ which are achieved by the different distancing and testing control levels indicated on the $x$ and $y$ axes respectively. The color scale indicates the cost for that combination.  As control levels increase, $R_e$ decreases but daily costs increase.  The shapes of the constant-$R_e$  contours in  Figure~\ref{fig:DistCostTestLevelCost}(a) (0\% immunity)  resemble those in Figure~\ref{fig:DistCostTestLevelCost}(b) (67\% immunity), but the $R_e$ values for corresponding contours differ by a ratio of roughly $2/3$. Similarly, cost contours in the two figures resemble each other, but the cost values differ by a similar factor.  According to Figure~\ref{fig:DistCostTestLevelCost}(a), a daily control cost of over 60 million dollars is required to reduce  $R_e$ below $1$, while Figure~\ref{fig:DistCostTestLevelCost} (b) shows that only about $10$ million dollars per day is required to achieve the same $R_e$ level at $67\%$ immunity. 
The black stars on each $R_e$ contour line  mark the (distancing, testing) control combinations that achieve minimum cost for the corresponding $R_e$ value. Note that these optimum points may be visually identified as the points on the contour lines where the contour lines are parallel to the nearest constant-cost contour, which is represented as a boundary between two colors.  

\newpage
\subsection{Sensitivity of effective reproduction number and control costs to model parameters}\label{eq:sensit}
In this section we analyse the sensitivity of the effective reproduction number $R_e$ and control costs to important parameters and cost coefficients. Three key parameters that we will analyze are the infectivity  $\beta$; symptomatic proportion $\tau$; and  relative infectiousness of asymptomatic individuals $\omega^A$. These parameters are difficult to estimate exactly, so it is important to determine their effect on model outcomes. In the following analysis, perturbations of $\pm 25\%$ are applied to each parameter, under two different levels of herd immunity.

\begin{figure}[h!]
  \centering
\includegraphics[width=3.4in]{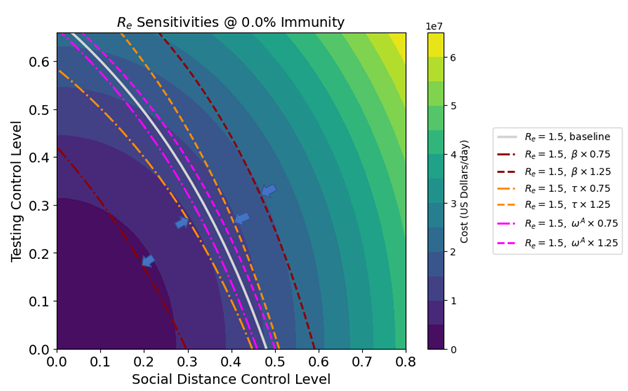}
\caption{Sensitivity of $R_e$ at $0\%$ immunity  under different values of $\beta, \tau$ and $ \omega^A$. Blue arrows indicate minimum-cost control solutions for different contours.\label{fig:Levelpara00}}
  \end{figure}

\begin{figure}[h!]
  \centering
\includegraphics[width=3.4in]{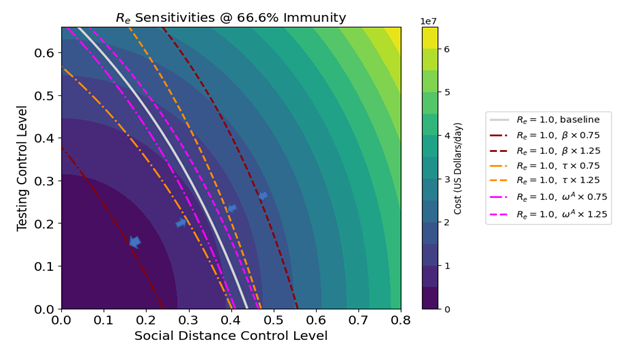}
 \caption{Sensitivity of $R_e$ at $66.6\%$ immunity  under different values of $\beta, \tau$ and $ \omega^A$. Blue arrows indicate minimum-cost control solutions for different contours.\label{fig:Levelpara66}}
  \end{figure}

 Figure \ref{fig:Levelpara00} shows the sensitivity of $ R_e$ at $0\% $ immunity under different  values of $\beta, \tau$ and $\omega^{A}$. The six colored curves represent the control levels that produce an effective $R_e = 1.5$ when the three parameters are individually varied by $\pm 25\%$.  The two curves for $\beta = 1 \pm .25$ times the baseline value are widely separated, showing that the predicted effect of controls on costs depends strongly on the value of $\beta$ used in the model. Indeed, $25\%$   shifts in the value of $\beta$ produce changes in control levels that exceed $25\%$. In contrast, the model parameter $\omega^A$ has little effect on control level estimates, while $\tau$ only has a large effect when the testing control level is high.  These same observations apply  to  Figure~\ref{fig:Levelpara66}, which shows the effect of $\beta, \tau$ and $\omega^{A}$ on the $R_e=1$ curve in a population with $67\%$  immunity.

From the positions of the blue arrows in Figures~\ref{fig:Levelpara00} and~\ref{fig:Levelpara66}, we may conclude that the values of $\beta$ and $\tau$ used in the model have a much greater effect on the optimum distancing controls than on testing. For example, at 0\% immunity a variation of $\pm 25\%$ in $\beta$ gives an optimal distancing control range of $0.34 \pm 40\%$, and a variation in optimal testing control of $0.25 \pm 30\%$.  These results indicate that it may be difficult in practice to accurately determine optimal control levels that can produce a given $R_e$ value for the system.

\begin{figure}[h!]
  \centering
\includegraphics[width=3.5in]{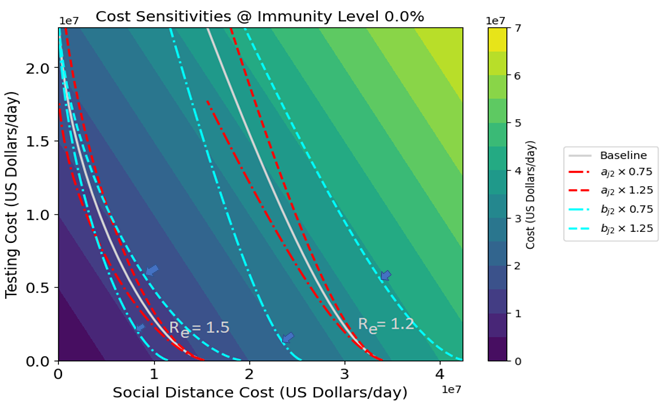}
\caption{Cost sensitivity from quadratic testing cost $a_{j2}$ and quadratic distancing cost $b_{j2}$  at $0\%$ herd immunity, for two different levels of $R_e$.\label{fig:Levelimm00}}
 \end{figure}

\begin{figure}[h!]
  \centering
\includegraphics[width=3.5in]{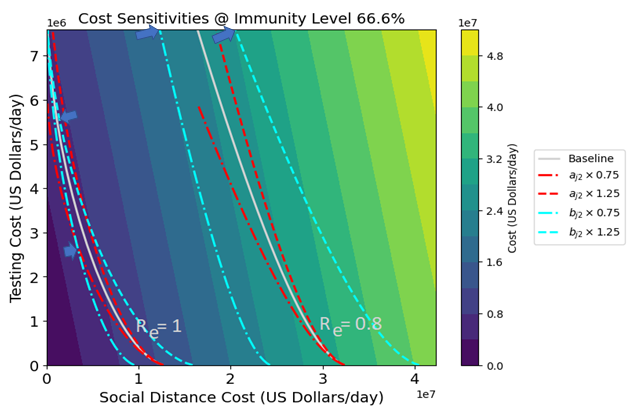}
\caption{Cost sensitivity from quadratic testing cost $a_{j2}$ and quadratic distancing cost $b_{j2}$  at $0\%$ herd immunity, for two different levels of $R_e$.\label{fig:Levelimm66}}
  \end{figure}

Figures~\ref{fig:Levelimm00}--\ref{fig:Levelimm66} represent the sensitivity of costs to changes in the quadratic cost coefficients for testing and distancing ($a_{j2}$ and $b_{j2}$ respectively). The horizontal and vertical axis scales represent control costs for social distancing and testing respectively (as in the previous figures, both distancing controls have the same values, as do both testing controls). Each figure shows  two white contours showing constant values of  $R_e$ with the baseline cost parameter values. Both $a_{j2}$ and $b_{j2}$ are varied by $\pm 25\%$, corresponding to the red and blue contours respectively. The shades of color indicate total cost for each mix of testing and distancing strategies:  in this case, the lines of constant cost are straight lines.  Optimum (cost-minimizing) operating points for the different  values of $R_e$ are indicated by arrows, as in previous figures. Regardless of immunity level, the  shifts in costs and optimum strategy point are much greater when $b_{j2}$ is varied than when $a_{j2}$, indicating a greater sensitivity of the system to distancing quadratic costs than testing quadratic costs. For example, for $R_e =1.2$ in Figure~\ref{fig:Levelimm00}, with baseline parameters the control costs along the contour vary from $35-40$ million dollars per day. When the quadratic testing cost $a_{j2}$ is increased by $25\%$, then control costs still lie in the same range. However, when the quadratic distancing cost $b_{j2}$ is increased by the same percentage, the cost range shifts upwards to $40-45$ million dollars per day.   The two figures closely resemble each other: however, it should be noted that  Figure~\ref{fig:Levelimm66} which portrays $67\%$ immunity is showing $R_e$ values that are only about $67\%$ as big as the $R_e$ values shown in Figure~\ref{fig:Levelimm00} which shows $0\%$ immunity. Also, the testing cost scale for Figure~\ref{fig:Levelimm66} has been reduced by roughly $67\%$, although the distancing cost scale remains the same.

 \newpage

\subsection{Instantaneous control strategies for optimizing effective reproduction number}\label{eqn:instantCost}
\begin{figure}[h!]
 \centering
 \captionsetup[subfigure]{width=1.0\textwidth}
  	\begin{subfigure}{0.45\textwidth}
 	\centering
 	\includegraphics[width=2.5in]{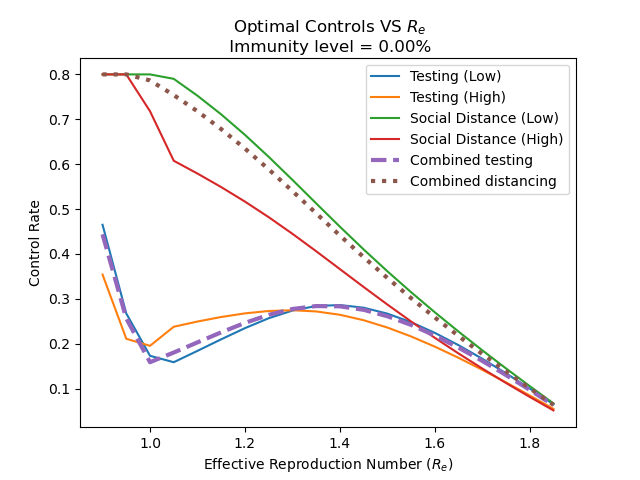}
 	\caption{Optimal control levels vs. $R_e$}\label{fig:Fig11Label}
 	\end{subfigure}
	\begin{subfigure}{.45\textwidth} 
 	\centering
 	\includegraphics[width=2.5in]{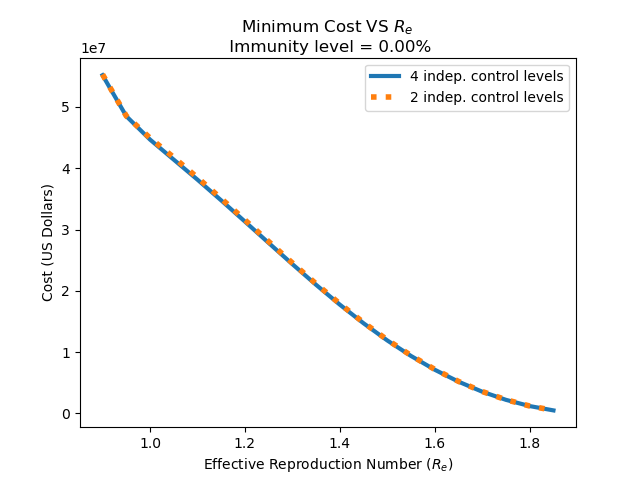}
 	\caption{Control costs vs. $R_e$ for instantaneous optimal strategies }\label{fig:Fig10Label}
 	\end{subfigure}
 \caption{Optimal control levels (a) an associated control costs (b)  as a function of population's current $R_e$ value  for populations with $0\%$  immunity. Solid lines correspond to strategies in which both controls 
can be applied at different levels on the low-risk and high-risk population subgroups, while dashed lines are for strategies in which the same control levels are applied to both subgroups. \label{fig:Optmix0}}
 \end{figure}

\begin{figure}[h!]
 \centering
 \captionsetup[subfigure]{width=1.0\textwidth}
  	\begin{subfigure}{0.45\textwidth}
 	\centering
 	\includegraphics[width=2.5in]{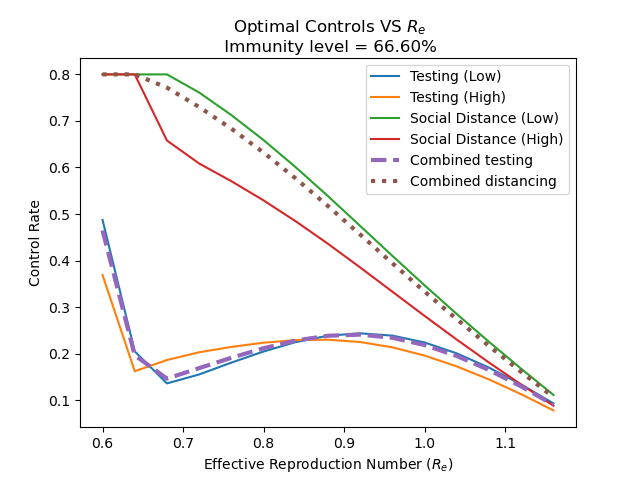}
 	\caption{Optimal control levels vs. $R_e$ }\label{fig:Fig11Label}
 	\end{subfigure}
	\begin{subfigure}{.45\textwidth}
 	\centering
 	\includegraphics[width=2.5in]{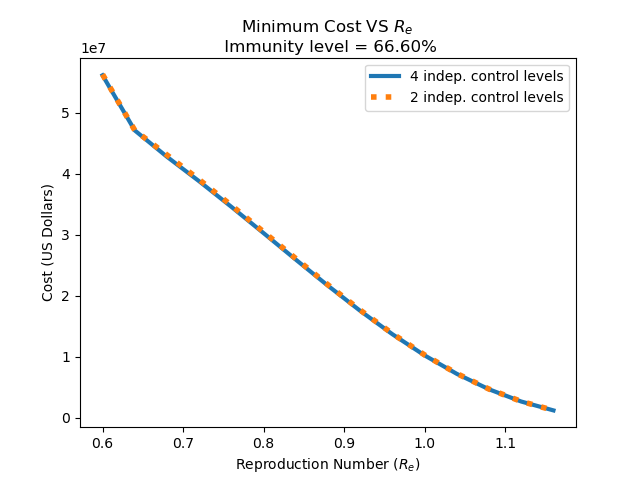}
 	\caption{Control costs vs. $R_e$ for instantaneous optimal strategies }\label{fig:Fig10Label}
 	\end{subfigure}

\caption{Optimal control levels (a) an associated control costs (b)  as a function of population's current $R_e$ value  for populations with $0\%$  immunity. Solid lines correspond to strategies in which both controls 
can be applied at different levels on the low-risk and high-risk population subgroups, while dashed lines are for strategies in which the same control levels are applied to both subgroups.\label{fig:Optmix66}}
 \end{figure}

Equations \eqref{eqn:Fmx}-\eqref{eqn:F1-jj-1} can be used to find optimum control levels associated with a given value of $R_e$.  This optimization was implemented in Python using  the \texttt{minimize} function from the  \texttt{scipy.optimize} package in Python for constrained minimization.  Figures \ref{fig:Optmix0}(a) and \ref{fig:Optmix66}(a) show the optimal levels of four controls (low and high risk testing, low and high risk distancing) associated with different instantaneous $R_e$ values for 0\% and 67\% population immunity, respectively.  In the figures, solid lines indicate optimal control levels when all four controls are allowed to vary independently; while dashed and dotted  lines show control levels when the controls on low and high risk groups are constrained to be the same. The costs associated with these optimal control strategies (both strategies where all four controls vary independently and those for which high and low risk controls are the same) as a function of $R_e$ are shown in  Figures \ref{fig:Optmix0}(b) and \ref{fig:Optmix66}(b) for 0\% and 67\% immunity, respectively. As expected, lower values of $R_e$ require higher levels of control, and incur greater costs. At all levels, distancing controls are applied at a higher level than testing, especially for values of $R_e$ near 0.7 where the optimal testing levels show a dip. When four controls are allowed to vary independently, social distancing for the low-risk subpopulation is applied at a higher level than for the high-risk subpopulation. This is due to the fact that according to \eqref{eqn:phi00} low-risk individuals have greater contact rates, and thus are more influential in spreading the disease. However, when the same level of control is applied to both low and high risk subpopulations, the costs are nearly the same as shown in Figures  \ref{fig:Optmix0}(b) and \ref{fig:Optmix66}(b). This indicates that costs are not highly sensitive to the specific strategies used.

\newpage

\subsection{Optimizing long-term strategies that target effective reproduction number reduction}\label{eqn:optinstantCost}

We may define three different classes of long-term strategies that target $R_e$ reduction. For all strategies, control measures are begun at a certain time, and  continue until the total infective population is reduced below a given level, to prevent resurgence of the disease. All strategies set control measure levels on a daily basis, so that the intensity of measures varies from day to day.   

In the first class of control strategies, a maximum daily budget is fixed to spend on control measures. Daily expenditure is constant, except in cases where the maximum possible controls cost is less than the allocated budget. The strategy for each day is determined as the set of control measures that does not exceed the budget, but which reduces $R_e$ as much as possible. The user-defined parameters for these strategies are  the daily maximum budget and the date at which control starts.

In the second class of control strategies, during the active control period  a combination of distancing and testing measures are used to reduce the $R_e$ level to a constant fraction of the level that  would be achieved without control. For example if at day 40 the computed $R_e$ value without control is 1.4 and the target fraction is $0.8$, then sufficient testing and distancing controls are applied to reduce $R_e$ to a value of $1.4 \cdot 0.8 =  1.12$.  The combination of testing and distancing controls used to achieve this value is computed using the same algorithm as was used to compute Figures~\ref{fig:DistCostTestLevelCost}. The user-defined  parameters for these strategies are the $R_e$ ratio and the date at which control starts.

The third class of strategy resembles the second, except that instead of targeting a given fraction of $R_e$, the daily control measures are chosen so as to achieve a fixed  $R_e$ value between 0 and 1.  If it is not possible to achieve the target $R_e$ even with the maximum control limits, then maximum controls are applied. The user-defined  parameters for these strategies are the $R_e$ target value and the date at which control starts.

We used simulations to evaluate and compare the effectiveness of these three classes of strategies. Simulations used the parameters in Table~\ref{table:1}. In addition, the simulation assumed an exposed population of 150 low-risk and 50 high-risk individuals at time $t=0$, out of a total population of 1.34 million low-risk and 423,000 high-risk individuals.  Treatment is continued until the number of exposed and infectious individuals reaches 10, at which point it is assumed that the disease can be contained by targeted measures without the need for population-wide control.  The simulation was continued for 180 days. Control strategies were updated on a daily basis.

We also simulated three parallel strategies in which high-risk individuals were given maximum protection.  In these strategies, applying controls to the high-risk population was prioritized.  Specifically, controls were only applied to the low-risk population  if the target budget or $R_e$ value could not be met through control measures applied to the high-risk population. For example, suppose that  we use the third strategy and the target $R_e$ value on day 20 is 0.9. In a case where distancing and testing applied only to the high-risk population is sufficient to achieve the target, then on day 20 no controls are applied to the low-risk population.  On the other hand, in a case where maximum distancing and testing on the high-risk population still fails to reach  $R_e=0.9$, then on day 20 maximum control would be applied to the high-risk population, and additional controls would also be placed on the low-risk population so that $R_e=0.9$ can be achieved.

\begin{figure}[htb]
 \centering
 \captionsetup[subfigure]{width=1.0\textwidth}
 	\begin{subfigure}{.48\textwidth}
 	\centering
 	\includegraphics[width=2.5in]{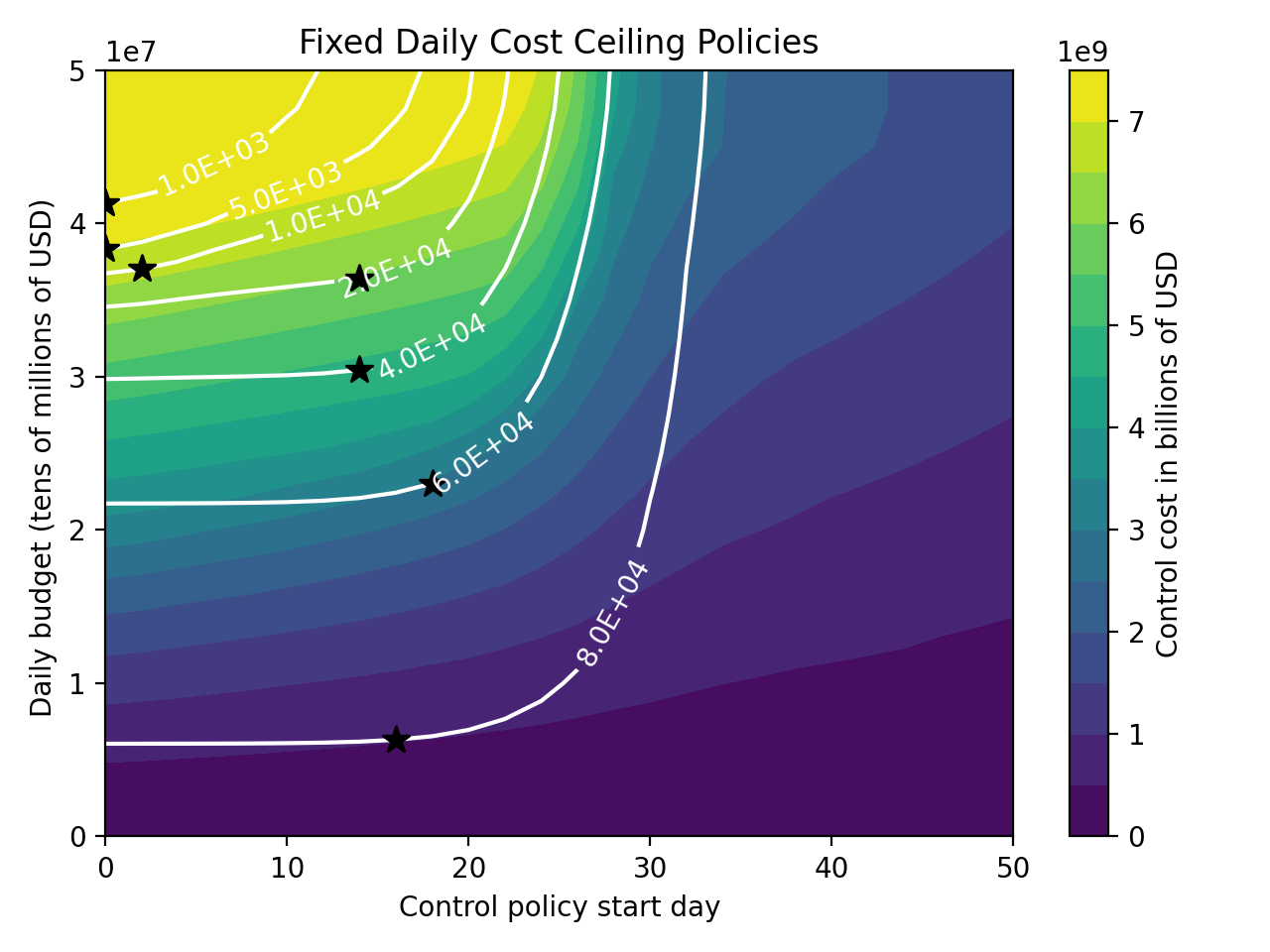}
 	\caption{Strategies that minimize $R_e$ }\label{fig:Fig10Label}
 	\end{subfigure}
 	\begin{subfigure}{0.48\textwidth}
 	\centering
 	\includegraphics[width=2.5in]{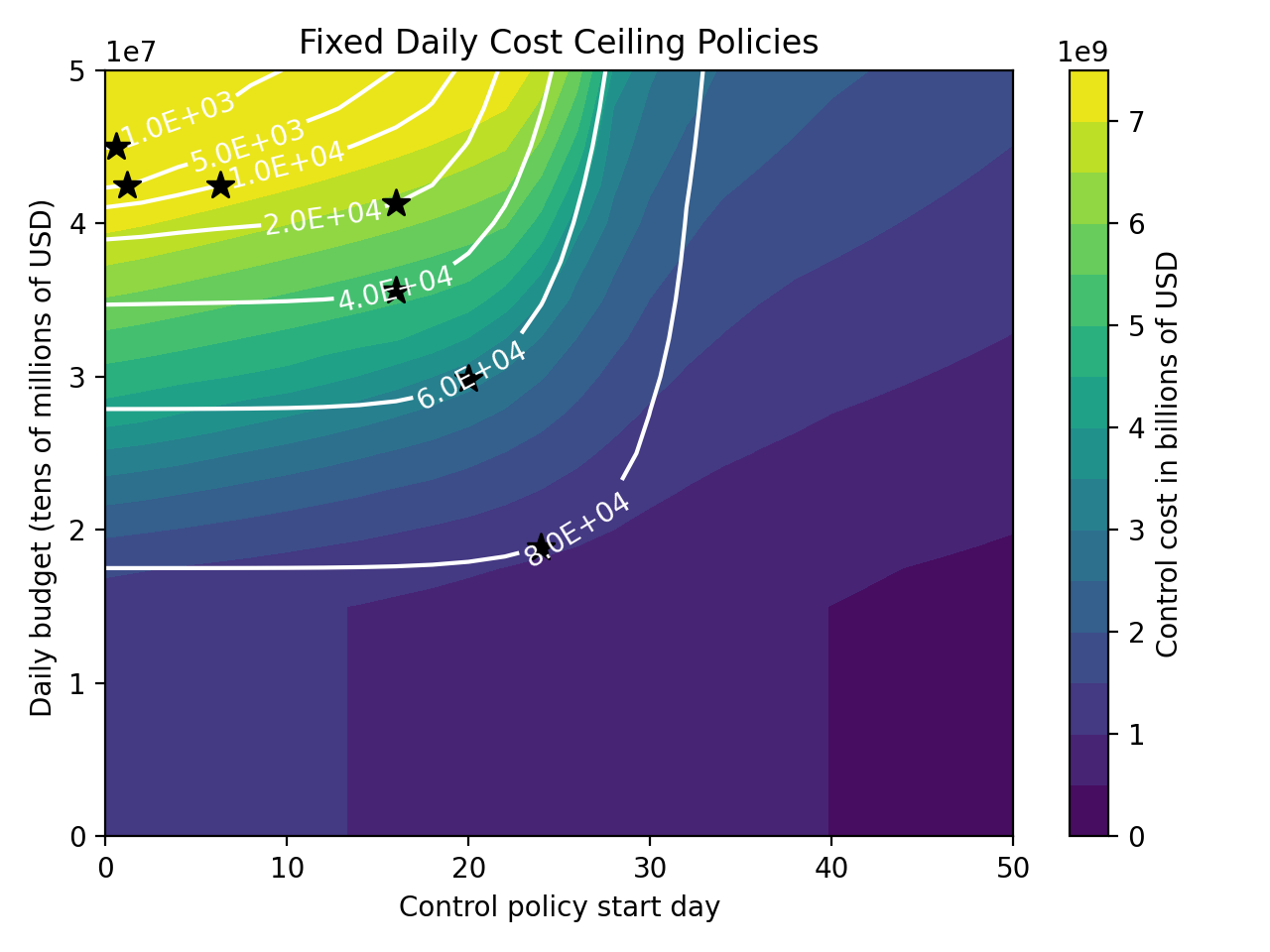}
 	\caption{Strategies that prioritize high-risk}\label{fig:Fig11Label}
 	\end{subfigure}
 \caption{Daily budget and starting day for adjusting the number of deaths.\label{fig:budget}}
 \end{figure}

\begin{figure}[htb]
 \centering
 \captionsetup[subfigure]{width=1.0\textwidth}
 	\begin{subfigure}{.48\textwidth}
 	\centering
 	\includegraphics[width=2.5in]{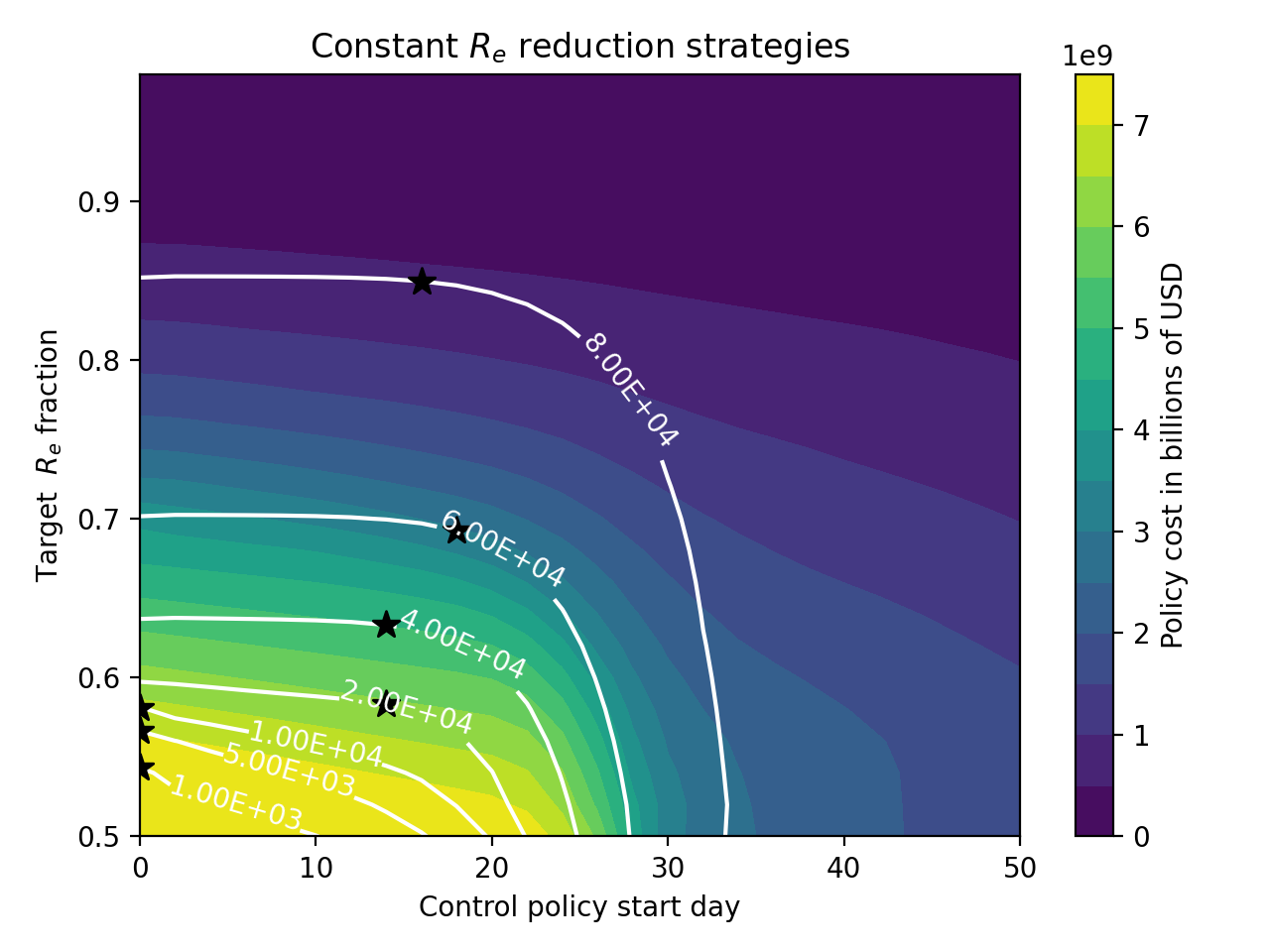}
 	\caption{Strategies that minimize $R_e$ }\label{fig:Fig10Label1}
 	\end{subfigure}
 	\begin{subfigure}{0.48\textwidth}
 	\centering
 	\includegraphics[width=2.5in]{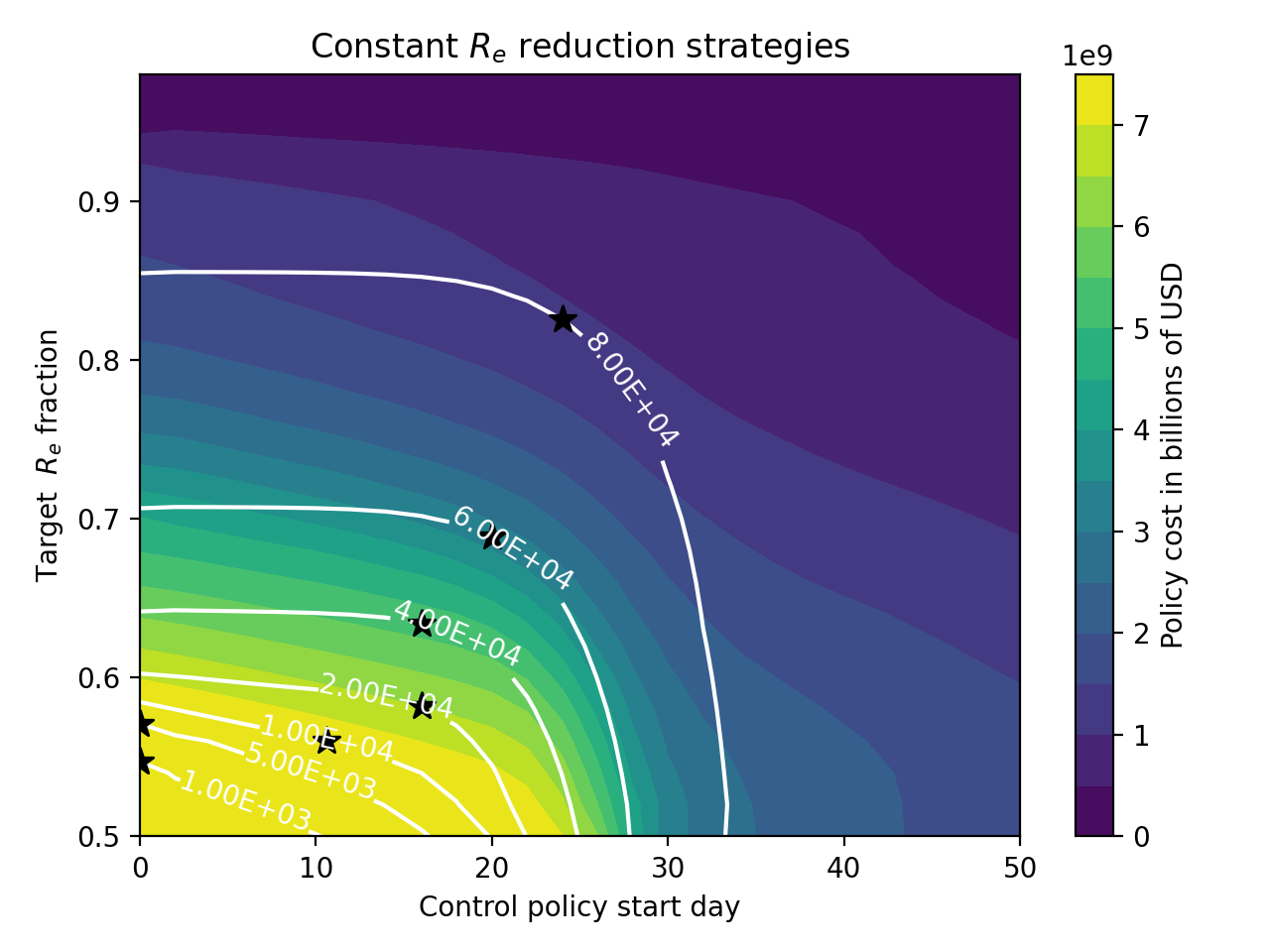}
 	\caption{Strategies that prioritize high-risk}\label{fig:Fig11Label2}
 	\end{subfigure}
\caption{Deaths number and costs depending on target $R_e$ fraction and control starting day strategies.\label{fig:Re_frac}}
 \end{figure}

\begin{figure}[h!]
 \centering
 \captionsetup[subfigure]{width=1.0\textwidth}
 	\begin{subfigure}{.48\textwidth}
 	\centering	\includegraphics[width=2.5in]{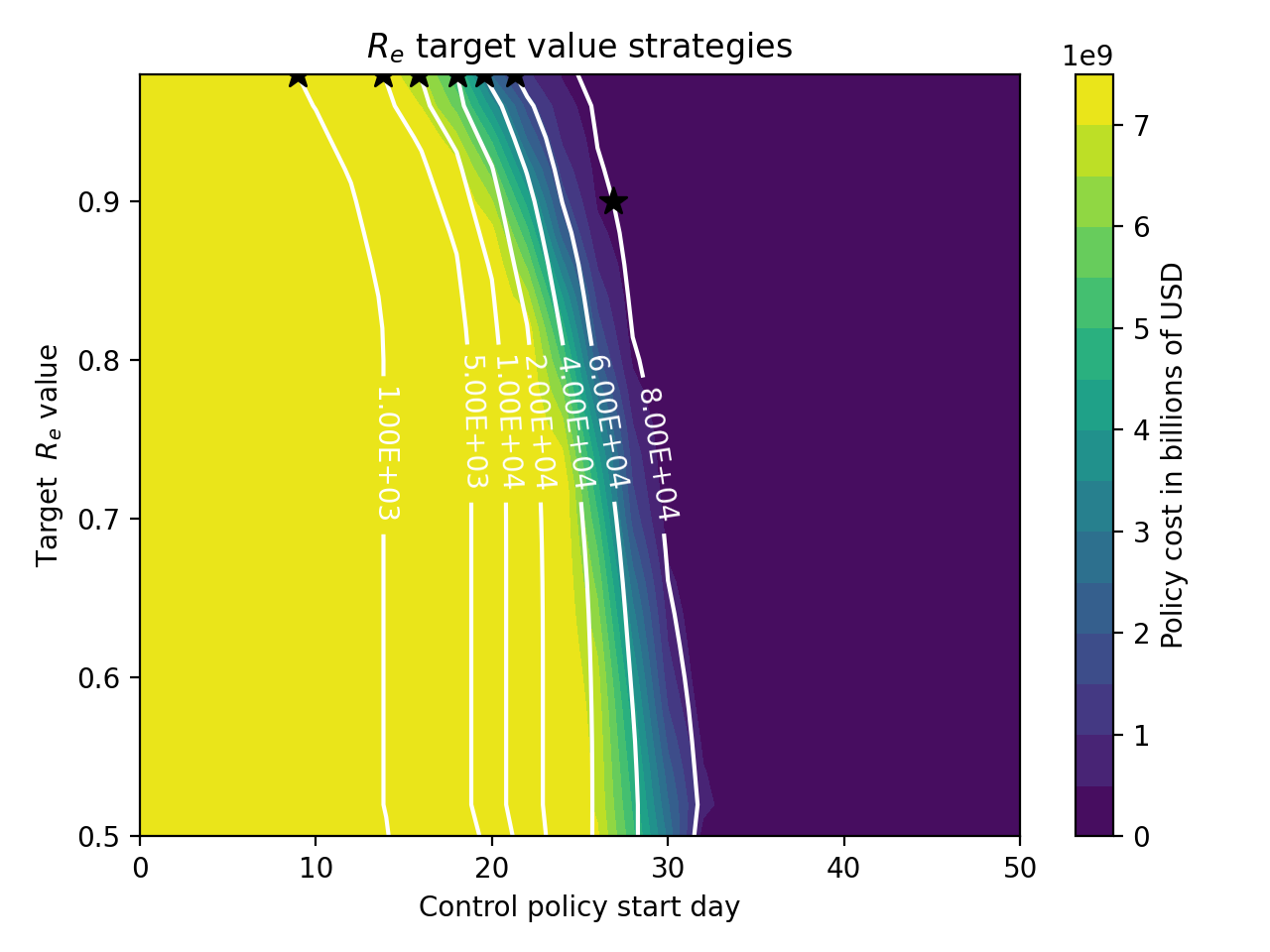}
 	\caption{Strategies that minimize $R_e$}\label{fig:Fig10Label3}
 	\end{subfigure}
 	\begin{subfigure}{0.48\textwidth}
 	\centering
 	\includegraphics[width=2.5in]{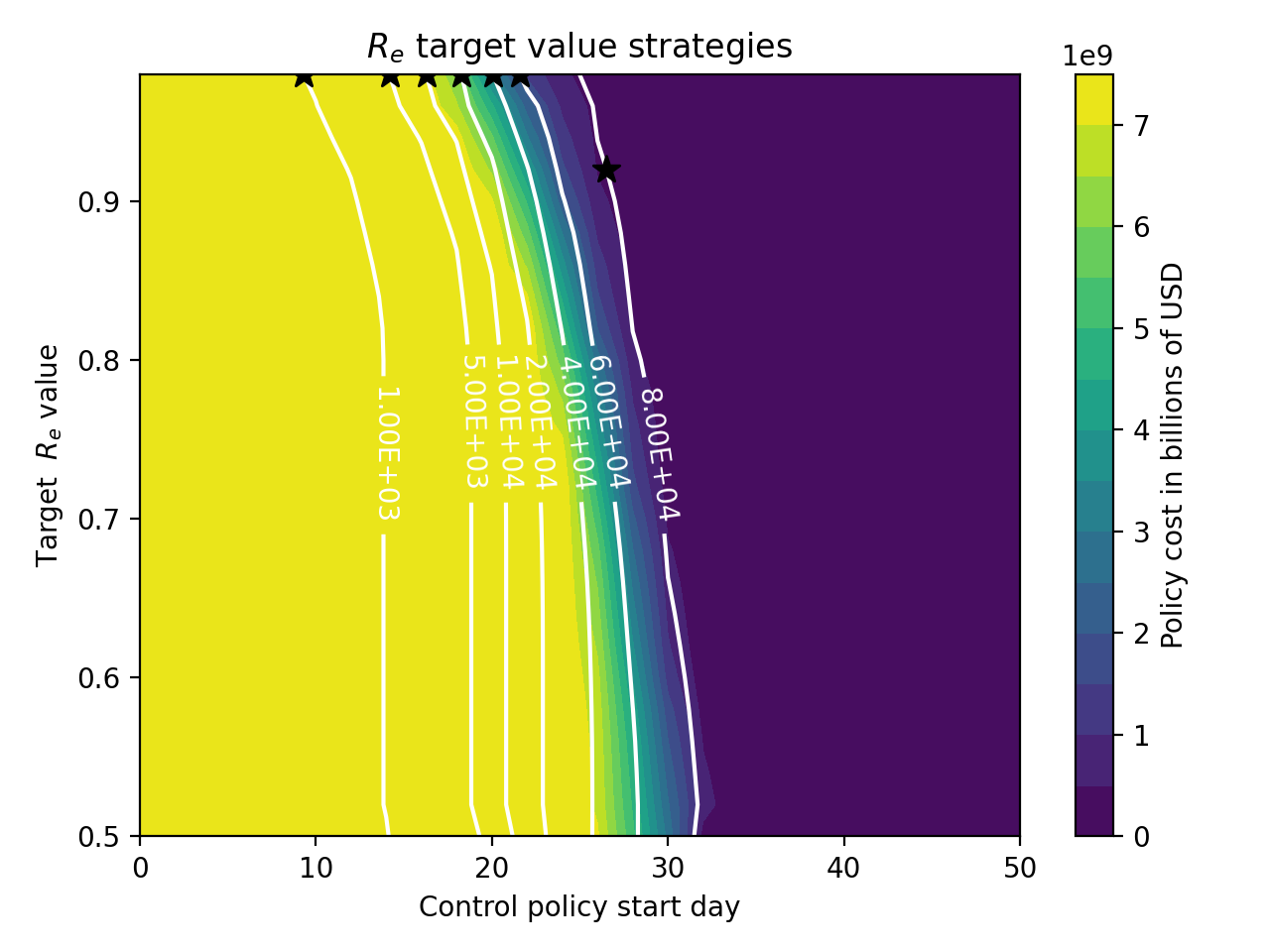}
 	\caption{Strategies that prioritize high-risk}\label{fig:Fig11Label4}
 	\end{subfigure}
\caption{Deaths number and costs depending on target $R_e$ value and control starting day strategies.\label{fig:Re_tgt}}
 \end{figure}

Figures~\ref{fig:budget}--\ref{fig:Re_tgt} show the cost and timing characteristics of the two variants of the three types of strategies considered. Each plot shows the costs (color level) and deaths (white contour lines) for each combination of policy start day ($x$ axis) and policy severity level ($y$ axis). For each figure, Subfigure (a) shows the regular case where controls are chosen to reduce $R_e$  at the lowest cost; while Subfigure (b) shows the results of policies that first prioritize controls on high-risk individuals. 

Figure \ref{fig:budget} shows results for daily budget-based strategies. The black stars on each white contour indicate  the control policy start date and daily budget that minimize the total control cost for the number of deaths specified by the contour. From Figure \ref{fig:budget}(a) we see that regular strategies give a death range of 1000--80,000  and a cost range of about 7.5--0.6 billion USD. Optimum start dates range from day 0 for 1000 deaths to day 19 for 60,000 deaths. For the high-risk prioritizing strategies showed in Figure \ref{fig:budget}(b), costs are about 0.5 billion USD higher, while the optimum policy start dates are slightly later, ranging up to day 23 for 80,000 deaths. However, regardless of start date and death level the regular strategy  will cost less than a high-risk prioritizing strategy with the same start date and deaths.  Both figures show the critical role of start date.  For example, if a regular strategy with a daily budget \$30 million  is started at day 15,  then deaths are limited to 40,000 and the total cost is \$5.5 billion. However, if control is delayed for one week, then to acheive the same number of deaths the daily budget must be raised to \$35 million, and the total cost is about \$6.5 billion. If the start date is after day 30, it is not possible to achieve less than 80,000 deaths if the daily budget is limited to \$50 million. On the other hand, if the control policy is started too early then the total cost is also increased: a control policy starting at day 0 that obtains 40,000 deaths has a daily budget limit of \$30 million, and total cost of about \$5.5 billion which is \$0.5 more than the policy starting on day 20. In general, a lower death target will require an earlier start date for the cost-minimizing strategy.

Figure \ref{fig:Re_frac} shows results for strategies that produce a constant proportional reduction in $R_e$ on a daily basis. Most of the observations for budget-based strategies also apply to these strategies as well.  From \ref{fig:Re_frac}(a), we see that 
an  early intervention is critical in reducing the number of deaths for aggressive strategies that produce large decreases in $R_e$; for example, if control starts after day 18 it is not possible to obtain less than 5,000 deaths. Start date is much less important for milder strategies: for basic strategies that attain 60,000 or more deaths, day 18 is optimal.
High-risk prioritizing strategies are expensive: reducing to 5,000 deaths requires at least \$7.5 billion USD, regardless of start date. Compared to basic strategies, optimal start dates are delayed: for 60,000 deaths, the optimal start date for the basic strategy is day 18, with total cost about \$4 billion, while the optimal high-risk prioritizing strategy for 60,000 deaths starts on day 23, with total cost about \$4.5 billion.  For high-risk prioritizing strategies, between start dates 20 and 33 the number of deaths increases rapidly, while the cost decreases rapidly. In this start date range, both deaths and costs are nearly independent of the target $R_e$ fraction. 

 \newpage
 Figures~\ref{fig:Re_tgt} (a) and (b)  show costs and deaths for control strategies that  target fixed $R_e$ values, for basic control strategies and strategies that prioritize the high risk group respectively. These figures resemble  each other, showing that strategies targeting high-risk give nearly the same results as basic strategies. The cost and start date for optimal strategies at each death level is not strongly dependent on the target value of $R_e$, although for most death levels setting the target $R_e=1$ gives the lowest cost control.
Optimal strategies producing lower deaths must be initiated earlier (e.g. to attain 1000 deaths, it is best to start on day 10 using target $R_e=1$, while the optimal strategy corresponding to 80,000 deaths begins around day 30, with $R_e \approx 0.9$.  

\begin{figure}[h!] 
 \centering
 \captionsetup[subfigure]{width=1.0\textwidth}
 	\begin{subfigure}{1.\textwidth} 
 	\centering
 	\includegraphics[width=4.5in]{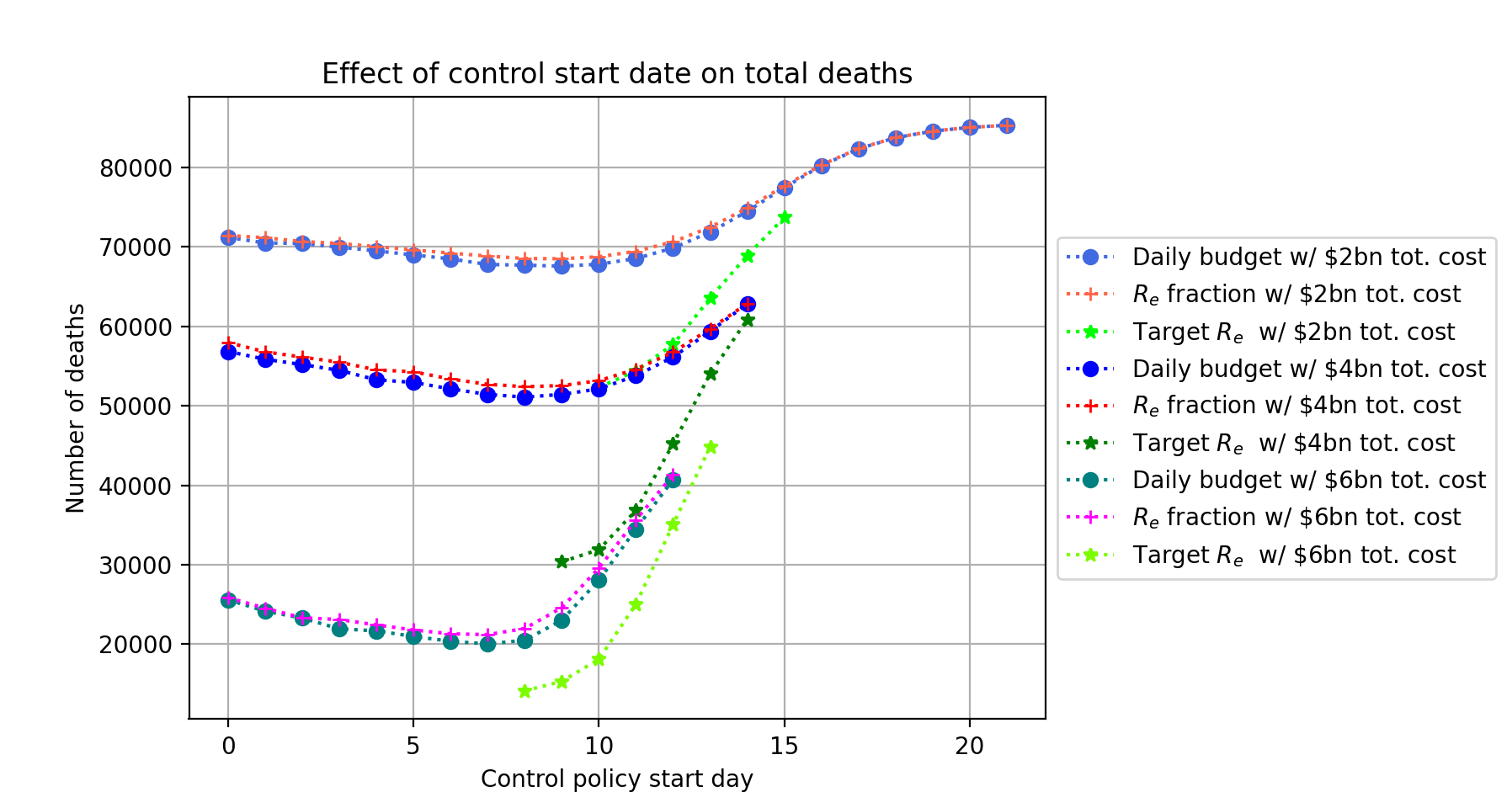}
 	\vspace*{2mm}
 	\caption{Strategies that minimize $R_e$ \label{fig:Fig12aLabel}}
 	\end{subfigure}
 	\begin{subfigure}{1.\textwidth}
 	\centering
 	\includegraphics[width=4.5in]{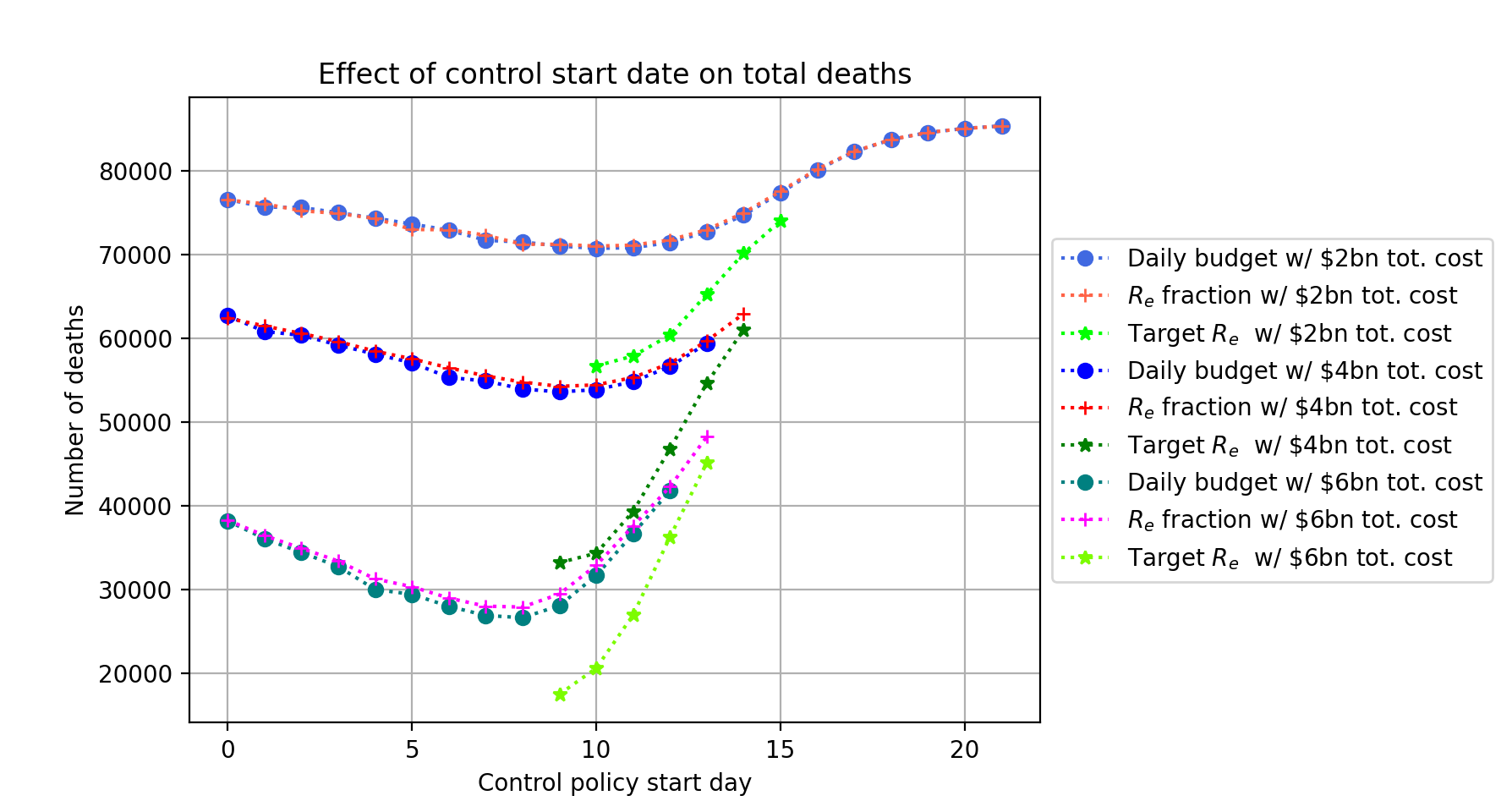}
 	\vspace*{2mm}
 	\caption{Strategies that prioritize high-risk}\label{fig:Fig12bLabel}
 	\end{subfigure}
 \caption{Number of the deaths obtained from different control strategy types depending on the control start day, for different budget levels. \label{fig:StartDateImpactOnDeaths}}
 \end{figure}

Figures~\ref{fig:StartDateImpactOnDeaths}(a) and (b) show the impact of starting control day on the deaths and costs that can be obtained from the different types of optimal strategies. 
In each figure, the three strategy types (constant budget, constant $R_e$ fraction, and constant $ R_e$ target) are compared at 3 different levels of total control cost: \$2, \$4  and \$6 billion USD. The  vertical axis shows deaths resulting from each of the strategies at the given control start date with the given total control cost. As before, Subfigures (a) and (b) correspond to basic and high-risk prioritizing strategies, respectively.

Figure~\ref{fig:StartDateImpactOnDeaths}(a)  shows that for all three cost levels, the basic $R_e$ target strategy achieves the lowest number of deaths at the latest optimal start date, indicating the superior performance of this type of strategy. However, if the $R_e$ target strategy is delayed past the optimal start date, the effectiveness is drastically reduced. For example, a target $R_e$ strategy with total cost \$4 billion starting at day 10 can reach 33,000 deaths, but if the start date is delayed by 4 additional days the numbers of deaths rises to almost 60,000.  Apart from $R_e$ target strategies, the other two strategies produce similar deaths for the same cost and start date. For all strategies, earlier optimal start times are associated with higher-cost strategies that attain fewer deaths.

Figure~\ref{fig:StartDateImpactOnDeaths}(b)  compares strategies that apply maximum control to high-risk individuals. Similar relations  between strategies hold as were noted for Figure~\ref{fig:StartDateImpactOnDeaths}(a). A comparison between Figures (a) and (b) shows that  high-risk prioritizing strategies yield higher deaths for similar costs, and are thus less effective:
for high-risk prioritizing strategies at the \$2, \$4, and \$6 billion USD level respectively, optimal deaths for $R_e$ target strategies are 57,000, 33,000, and 17,000 compared to 53,000, 30,000, and 15,000 which may be obtained by basic $R_e$ target strategies at the same cost levels.
Once again, the constant budget and $R_e$ fraction strategies yield similar outcomes.

\newpage

\begin{figure}[h!]
 \centering
 \captionsetup[subfigure]{width=1.0\textwidth}
 	\begin{subfigure}{.48\textwidth}
 	\centering	\includegraphics[width=2.5in]{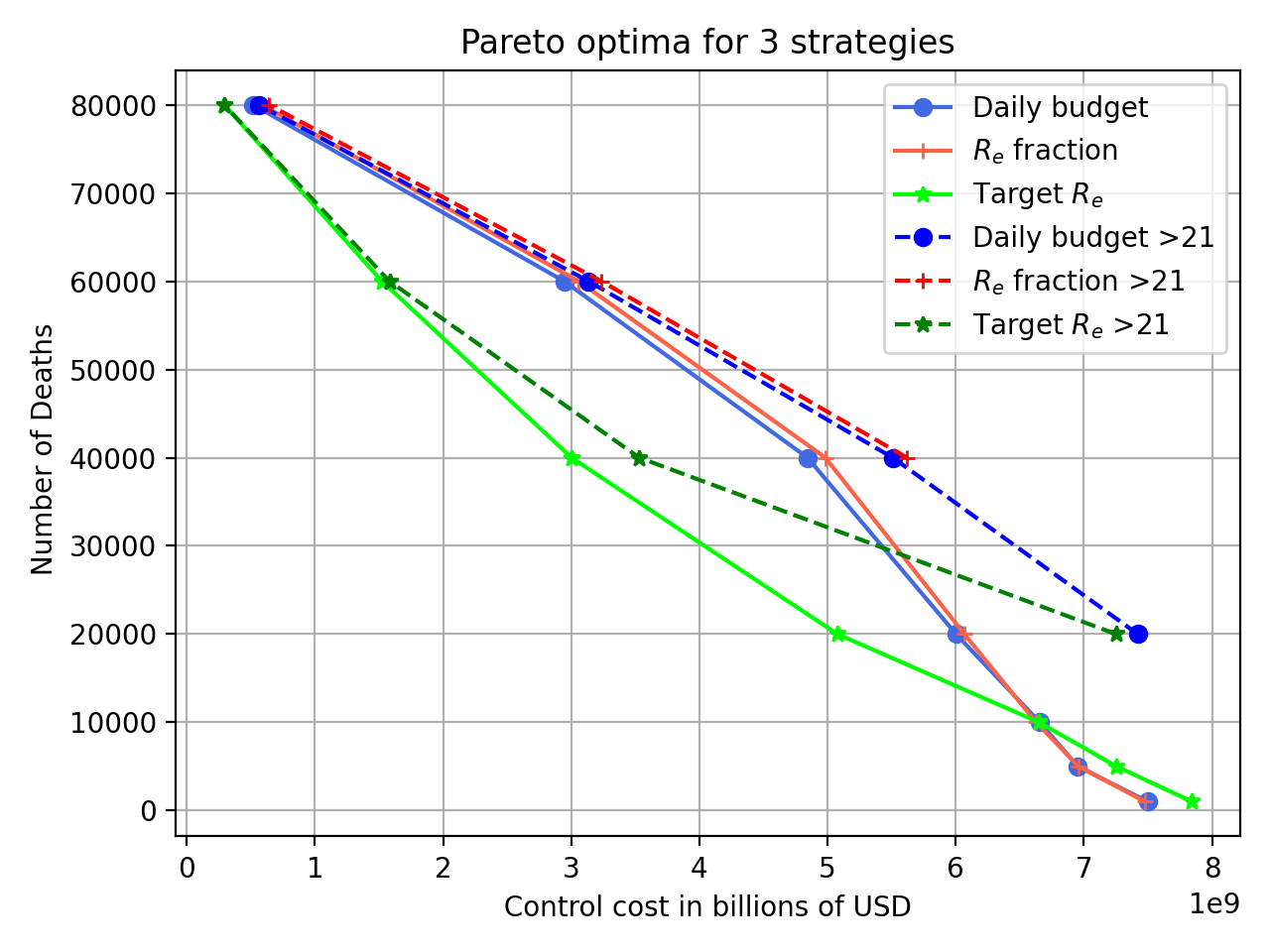}
 	\caption{Basic strategies that minimize $R_e$}\label{fig:Fig10Label3}
 	\end{subfigure}
 	\begin{subfigure}{0.48\textwidth}
 	\centering
 	\includegraphics[width=2.5in]{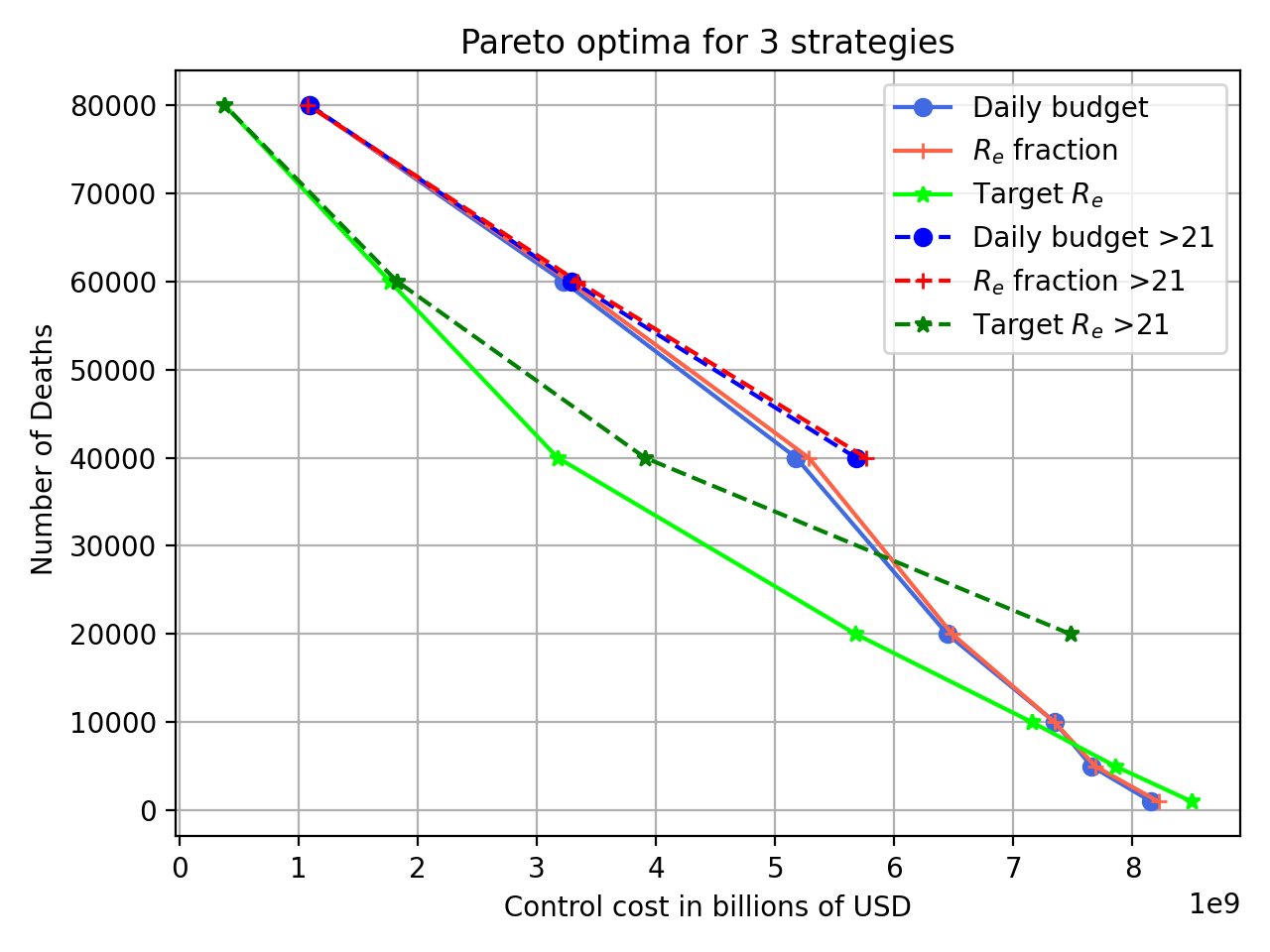}
 	\caption{Strategies that prioritize high-risk}\label{fig:Fig11Label4}
 	\end{subfigure}
\caption{Number of the deaths associated to control cost depending on the starting control date for the three different strategies. \label{fig:StartDateControlCostDeaths}}
 \end{figure}

Figures~\ref{fig:StartDateControlCostDeaths}  shows Pareto fronts for the 3 strategy types, where deaths and control costs are the two competing factors. For each strategy type, two Pareto  fronts are shown: one front is based on overall optimal strategies, while the other front restricts strategies to those which begin after  21 days.  The $x$ axis  representes  the control cost while the  vertical $y$ axis shows the number of deaths depending on the current cost.  As above, Subfigures (a) and (b) give results for basic and high risk-prioritizing strategies, respectively.

Figure~\ref{fig:StartDateControlCostDeaths}(a) 
In most cases, target $R_e$ is best except for very high control costs and low deaths. The  advantage of target $R_e$ is especially large for mid-range strategies that produce about 30,000 deaths at a cost of \$4 billion USD. In contrast, to obtain the same number of deaths with the other two strategy types costs \$5.5 billion. Alternatively, using the \$4 billion for constant-budget or target $R_e$ fraction will produce an additional 20,000 deaths compared to a strategy with target $R_e$ value. Delaying the control start date has large costs:  compared to the above-mentioned target $R_e$ strategy which reaches 30,000 deaths and costs \$4 billion, a 21-day delayed target $R_e$ strategy will either cost an additional \$1.5 billion at the same death level, or will result in an additional 9,000 deaths for the same cost. 
The observations for~\ref{fig:StartDateControlCostDeaths}(a) are still valid for ~\ref{fig:StartDateControlCostDeaths}(b) except that  the cost-to-death tradeoffs are slightly more unfavorable.  

\begin{landscape}
\begin{figure}
    \begin{subfigure}[b]{0.24\columnwidth}    
    \includegraphics[width=1.95in]{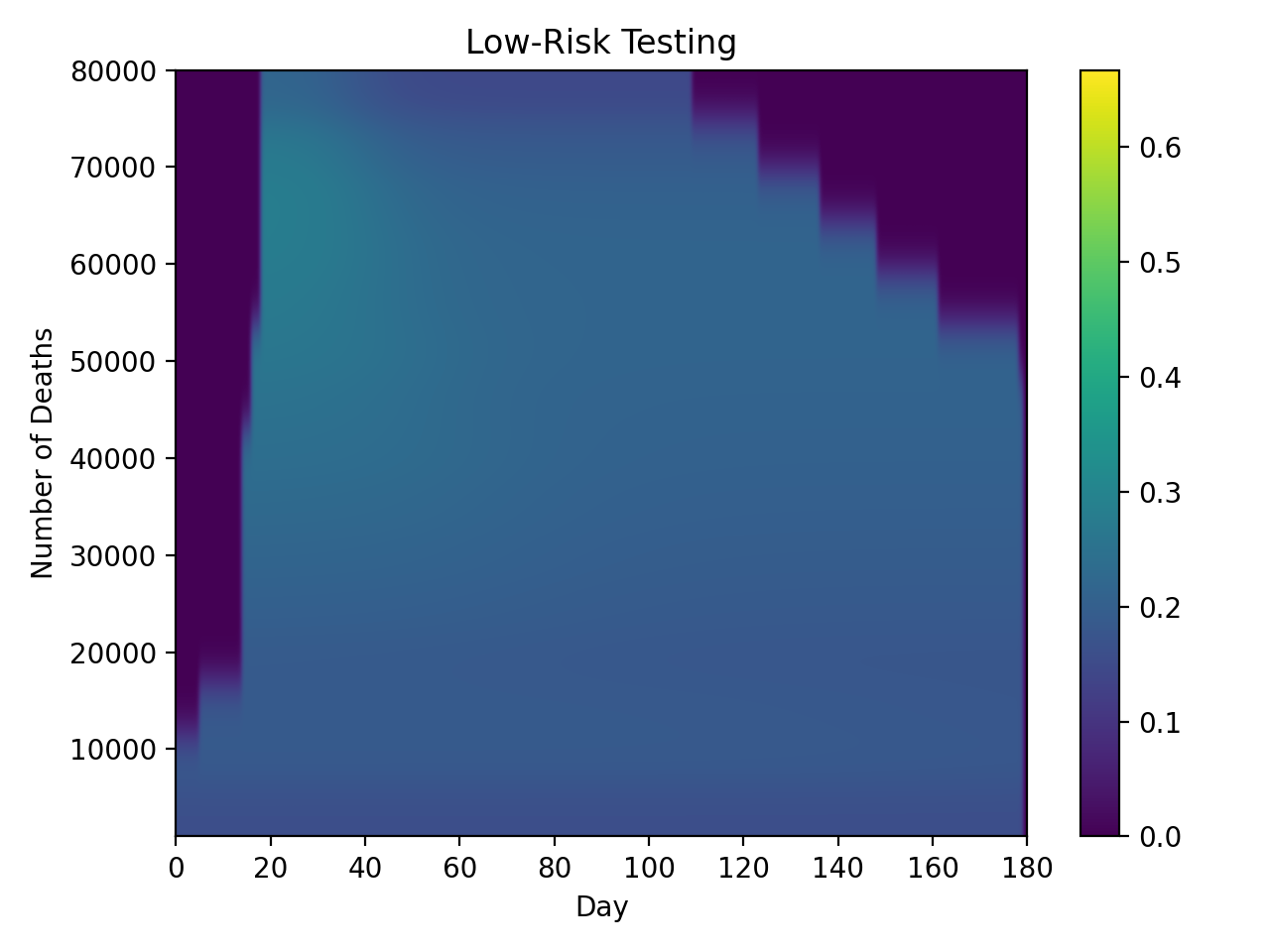}  
    \end{subfigure}
    \hfill
    \begin{subfigure}[b]{0.24\columnwidth}
        \includegraphics[width=1.95in]{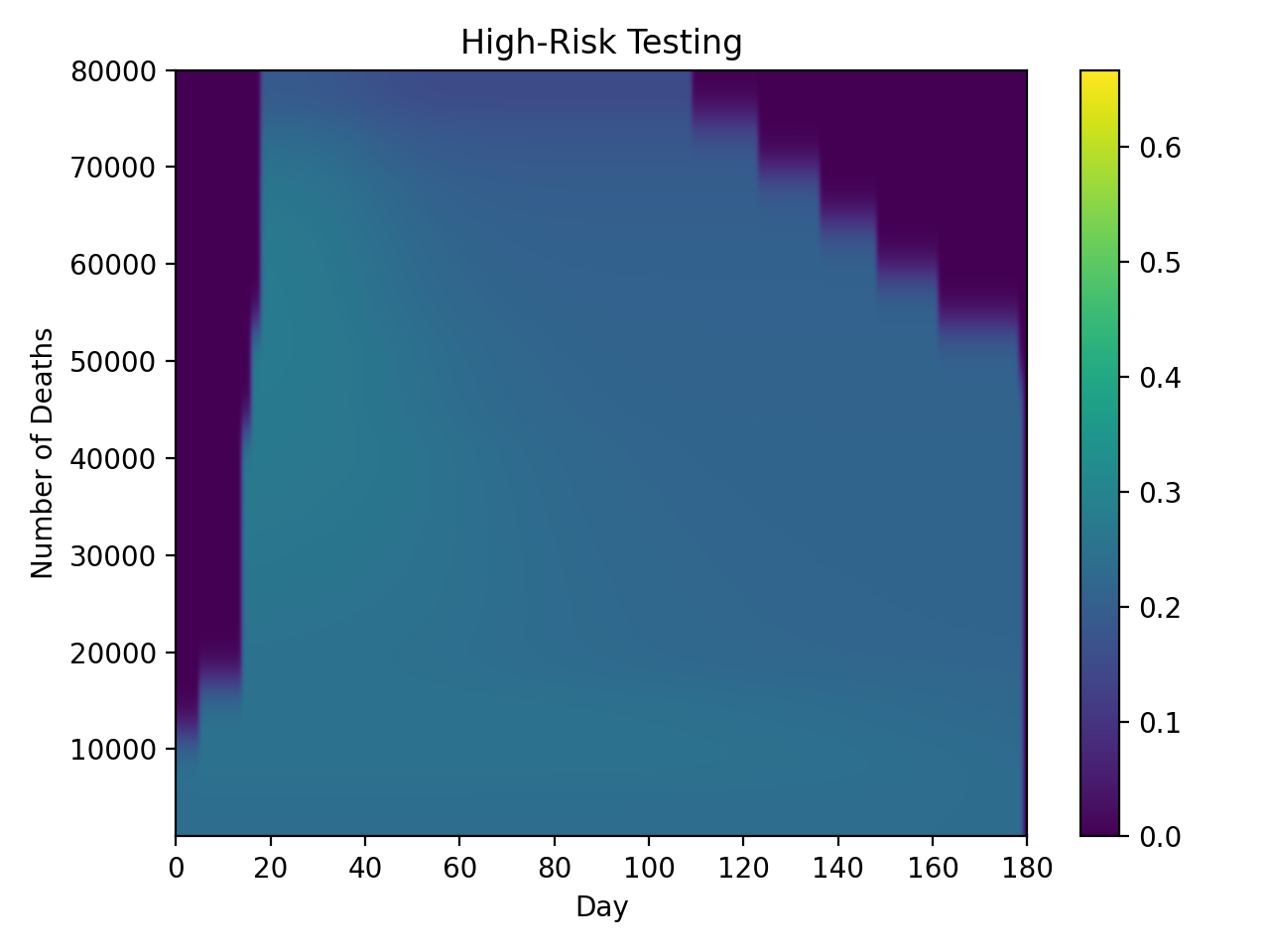}
    \end{subfigure}
 \hfill
    \begin{subfigure}[b]{0.24\columnwidth}
        \includegraphics[width=1.95in]{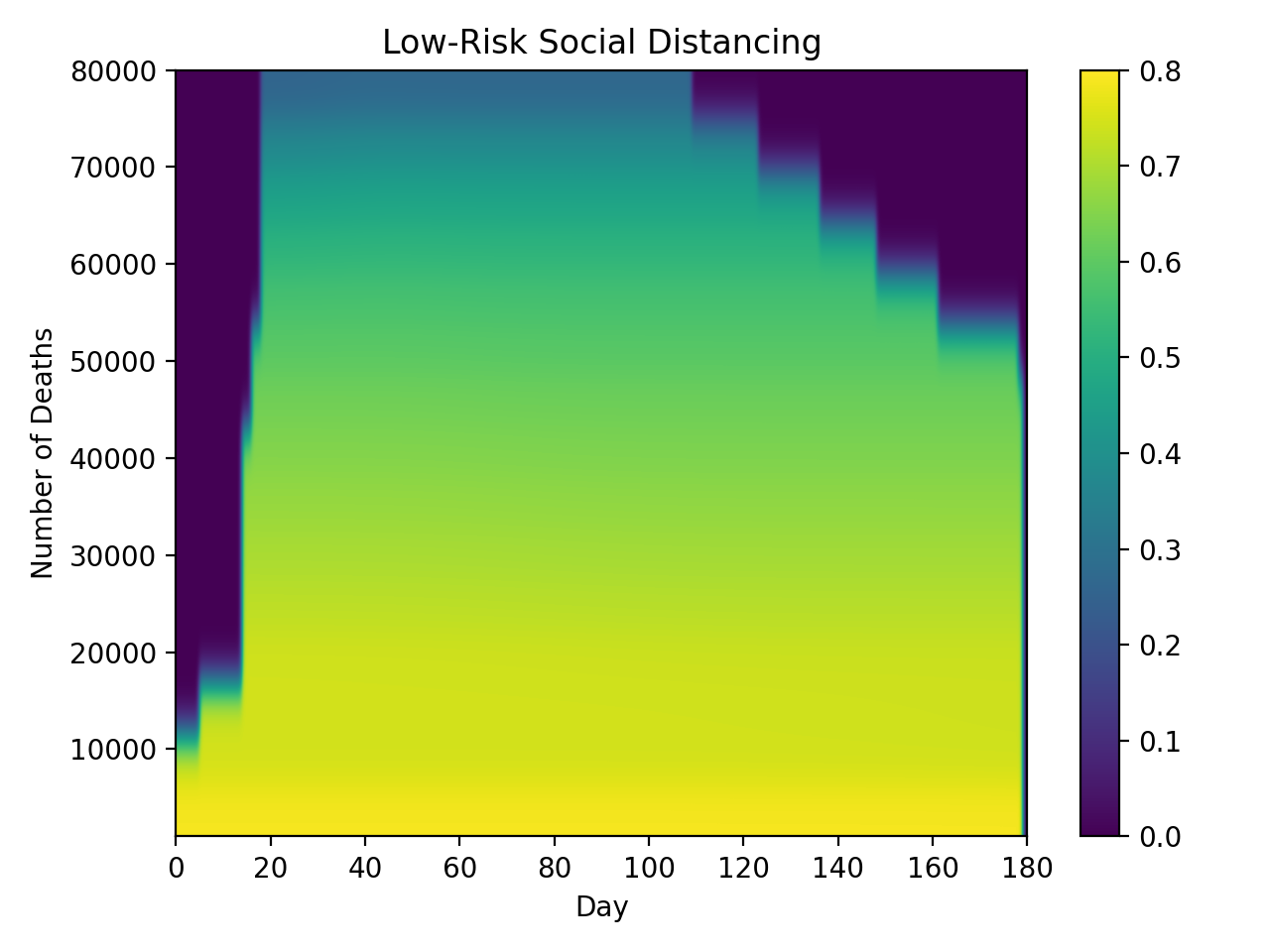}
    \end{subfigure}
    \hfill
    \begin{subfigure}[b]{0.24\columnwidth}
        \includegraphics[width=1.95in]{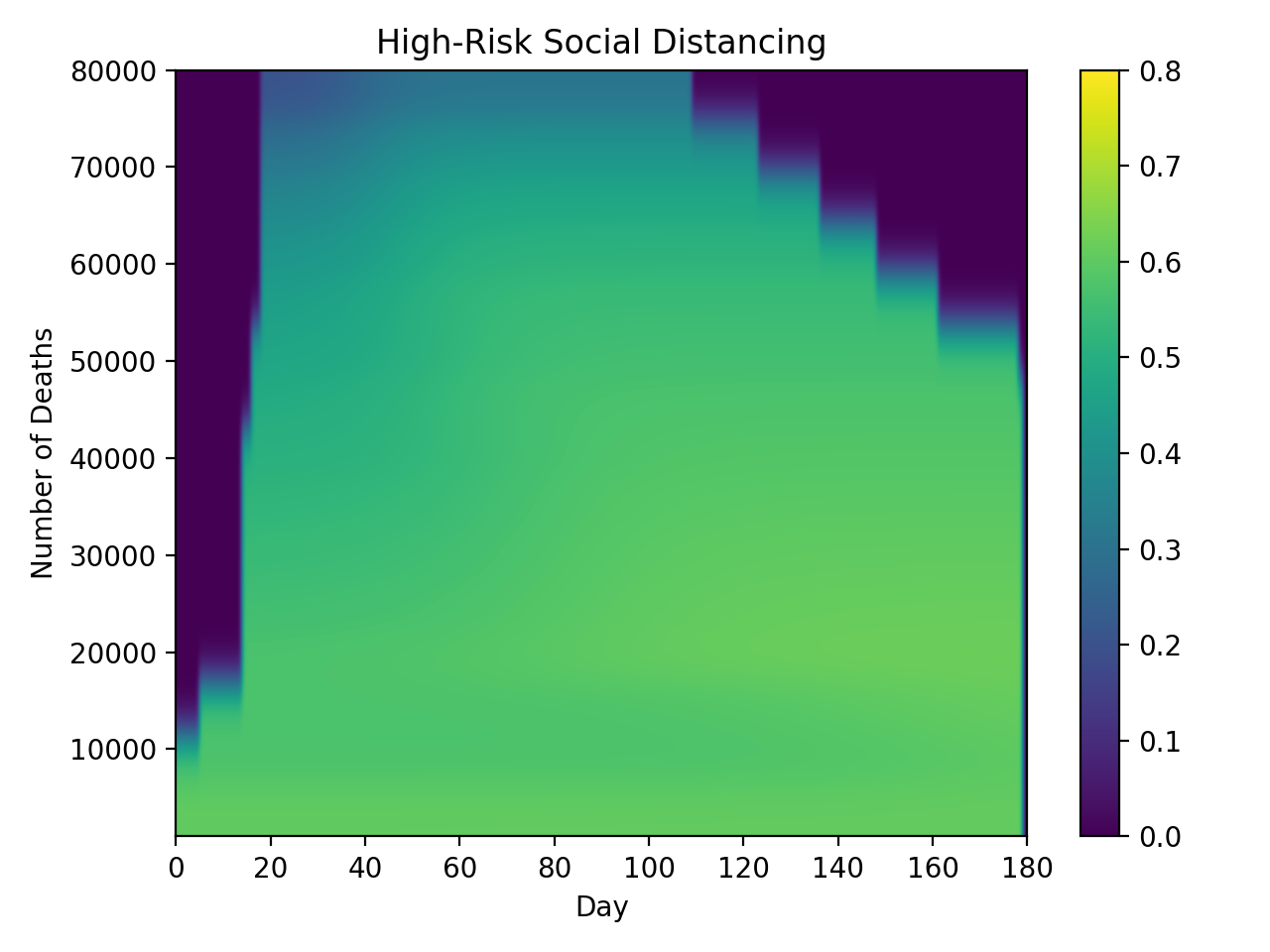}
    \end{subfigure}

    \begin{subfigure}[b]{0.24\columnwidth}    
    \includegraphics[width=1.95in]{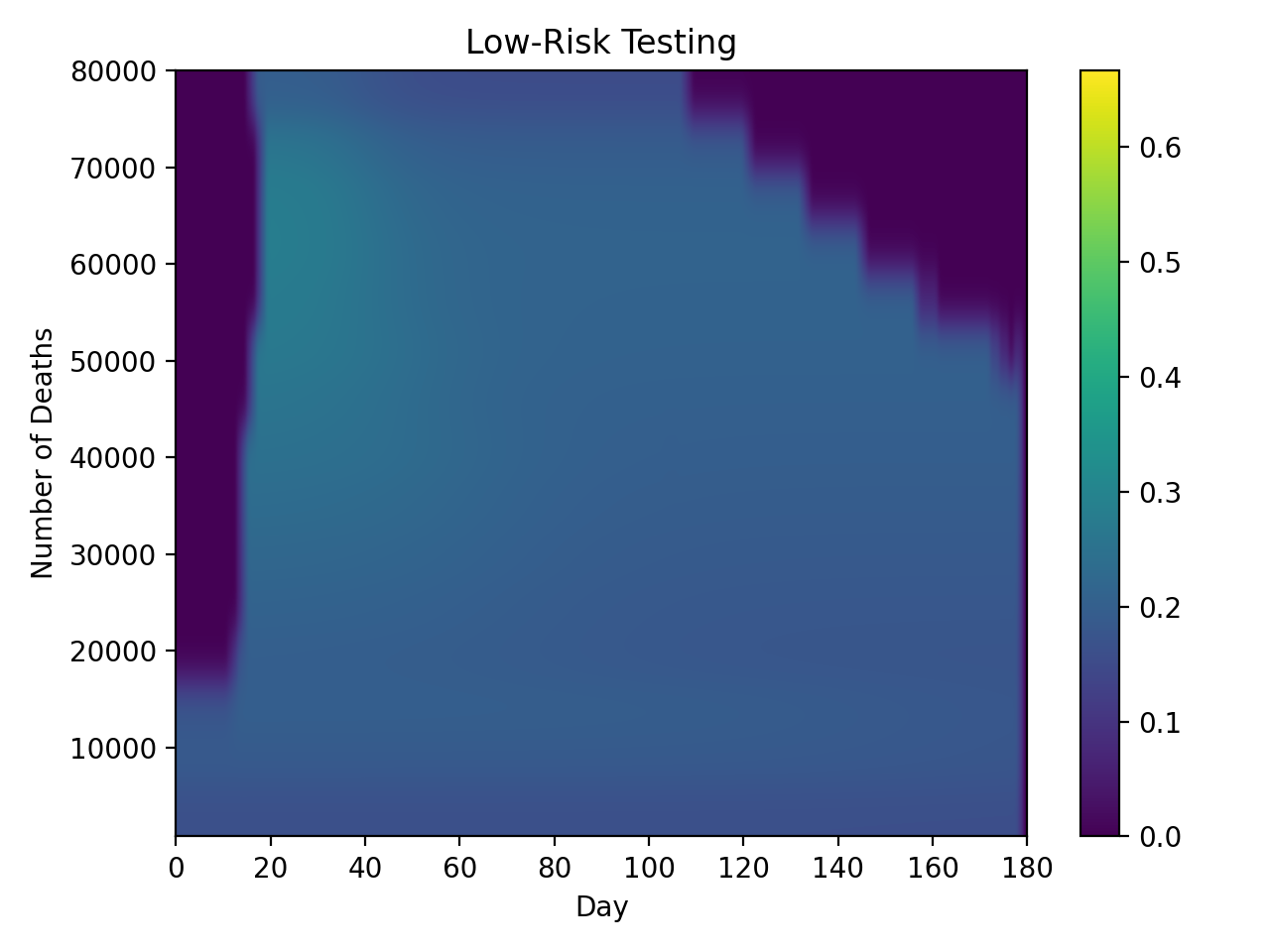}  
    \end{subfigure}
    \hfill
    \begin{subfigure}[b]{0.24\columnwidth}
        \includegraphics[width=1.95in]{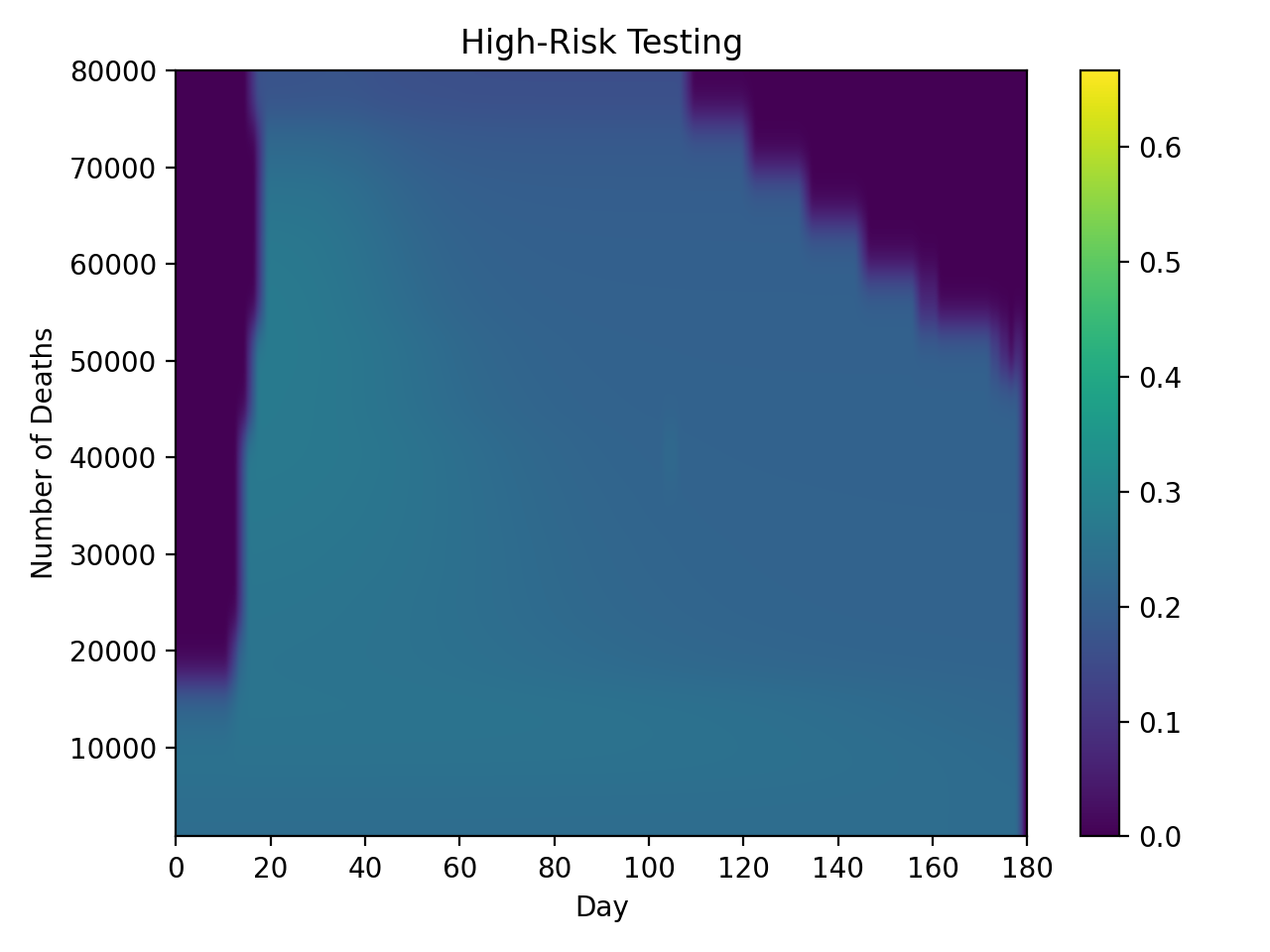}
    \end{subfigure}
 \hfill
    \begin{subfigure}[b]{0.24\columnwidth}
        \includegraphics[width=1.95in]{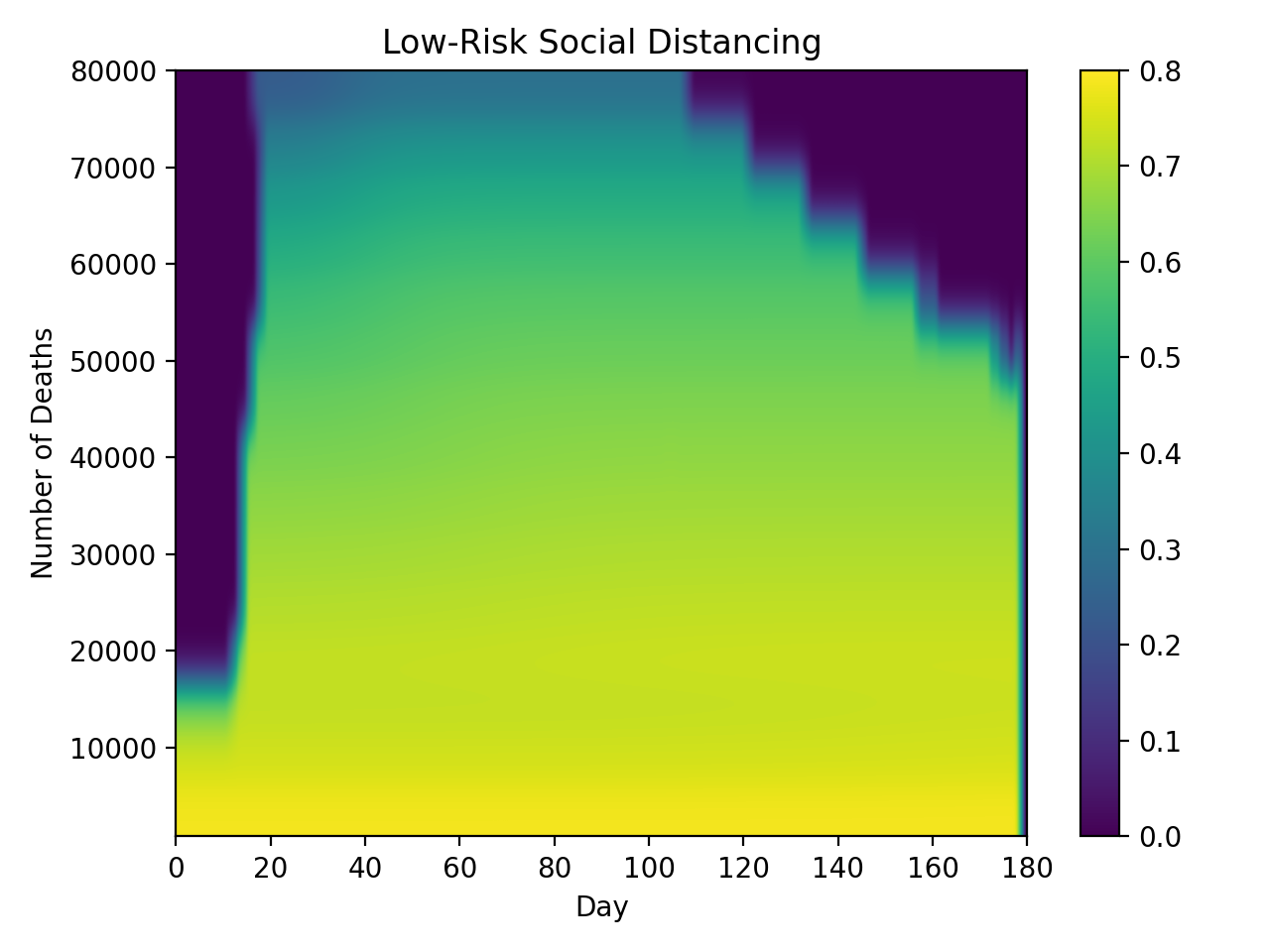}
    \end{subfigure}
    \hfill
    \begin{subfigure}[b]{0.24\columnwidth}
        \includegraphics[width=1.95in]{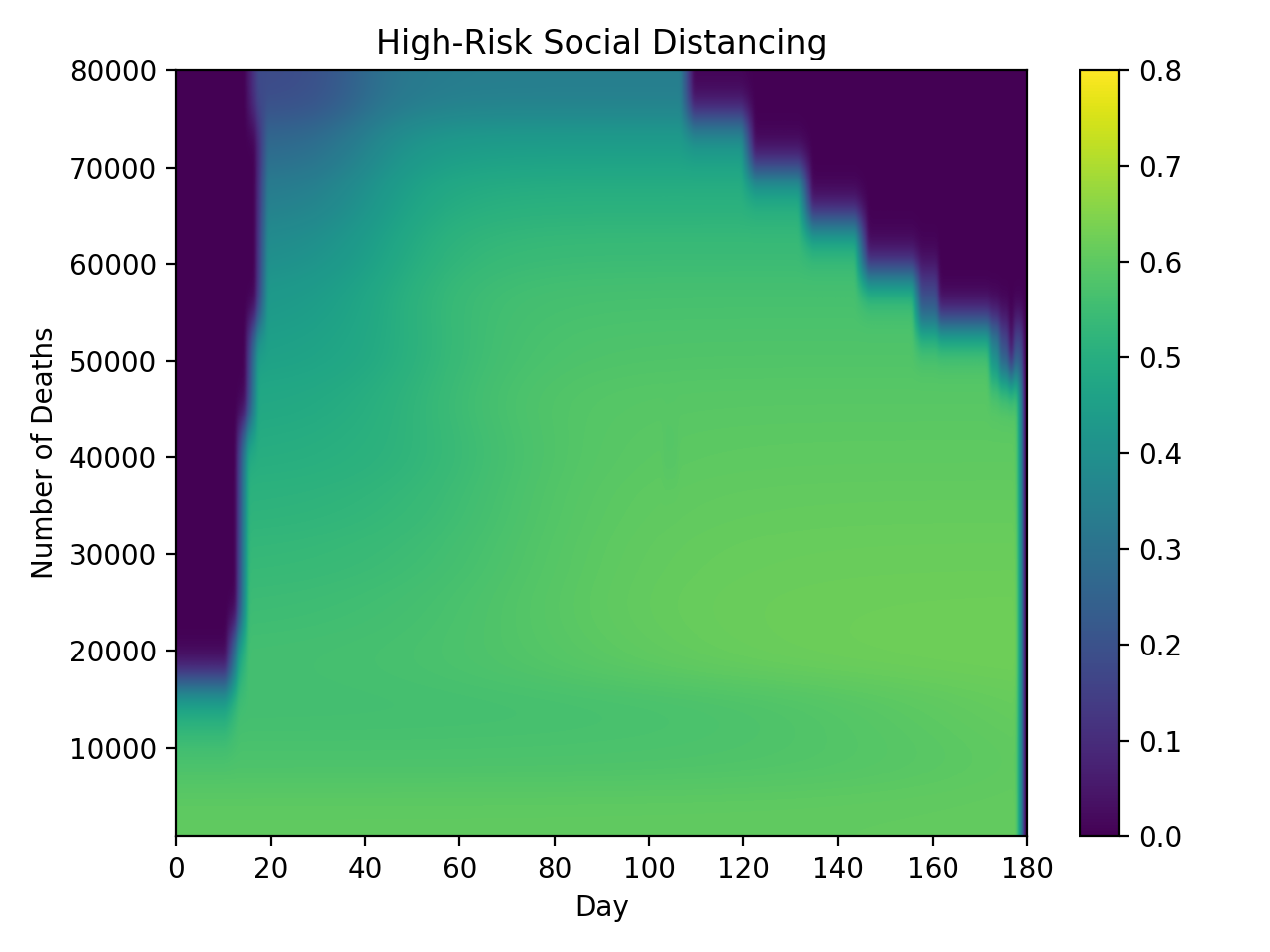}
    \end{subfigure}

    \begin{subfigure}[b]{0.24\columnwidth}    
    \includegraphics[width=1.95in]{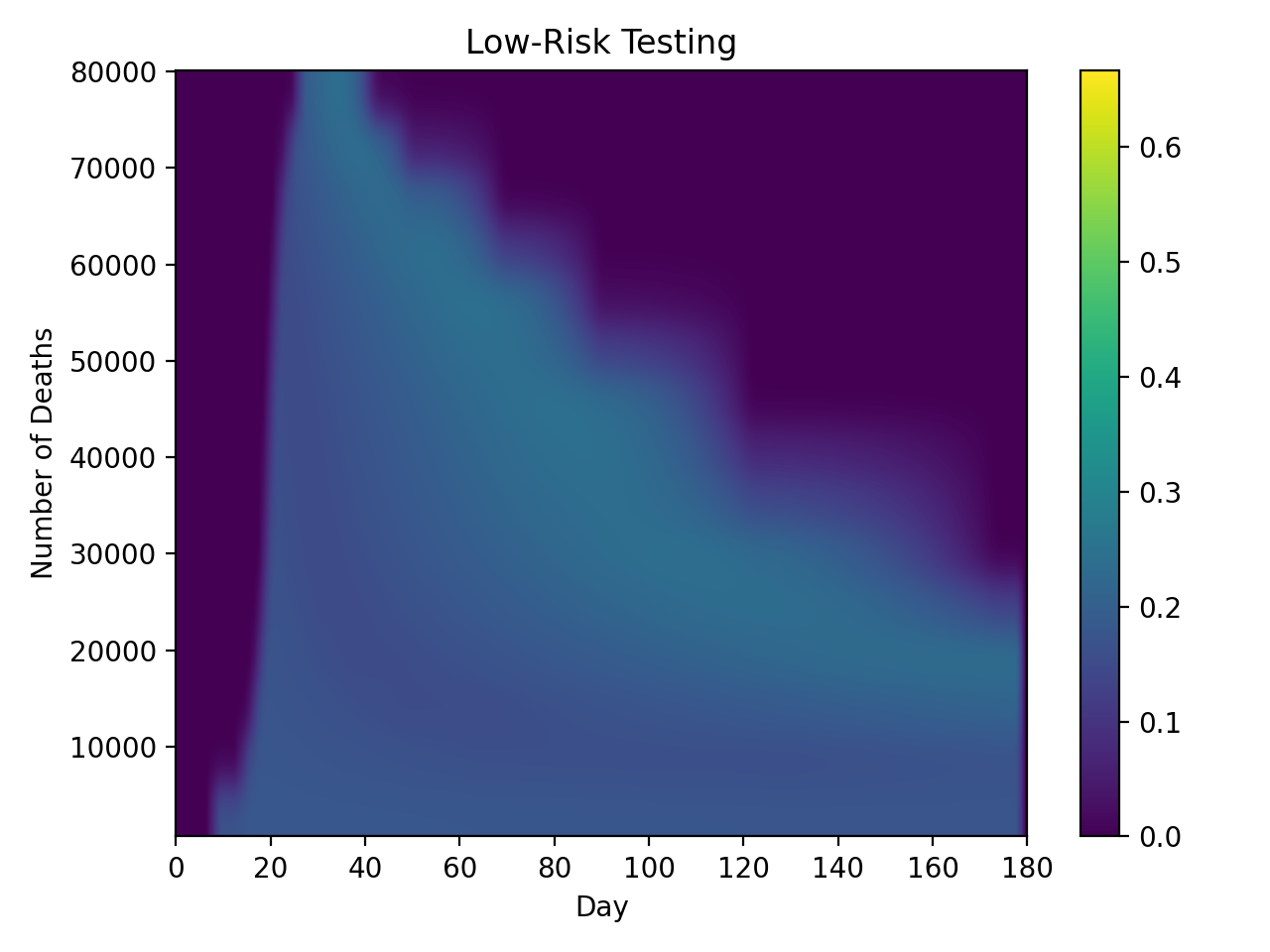}  
    \end{subfigure}
    \hfill
    \begin{subfigure}[b]{0.24\columnwidth}
        \includegraphics[width=1.95in]{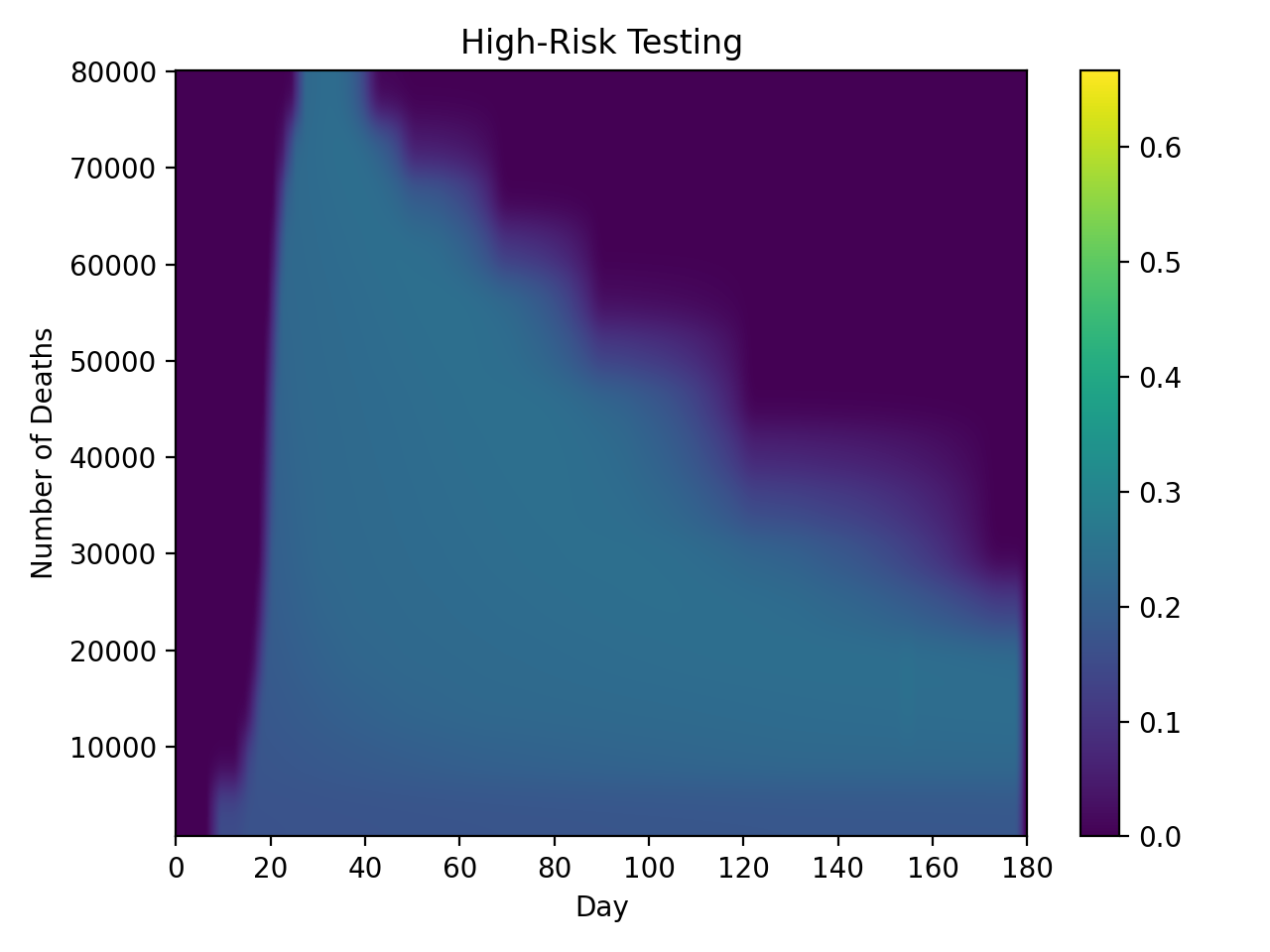}
    \end{subfigure}
 \hfill
    \begin{subfigure}[b]{0.24\columnwidth}
        \includegraphics[width=1.95in]{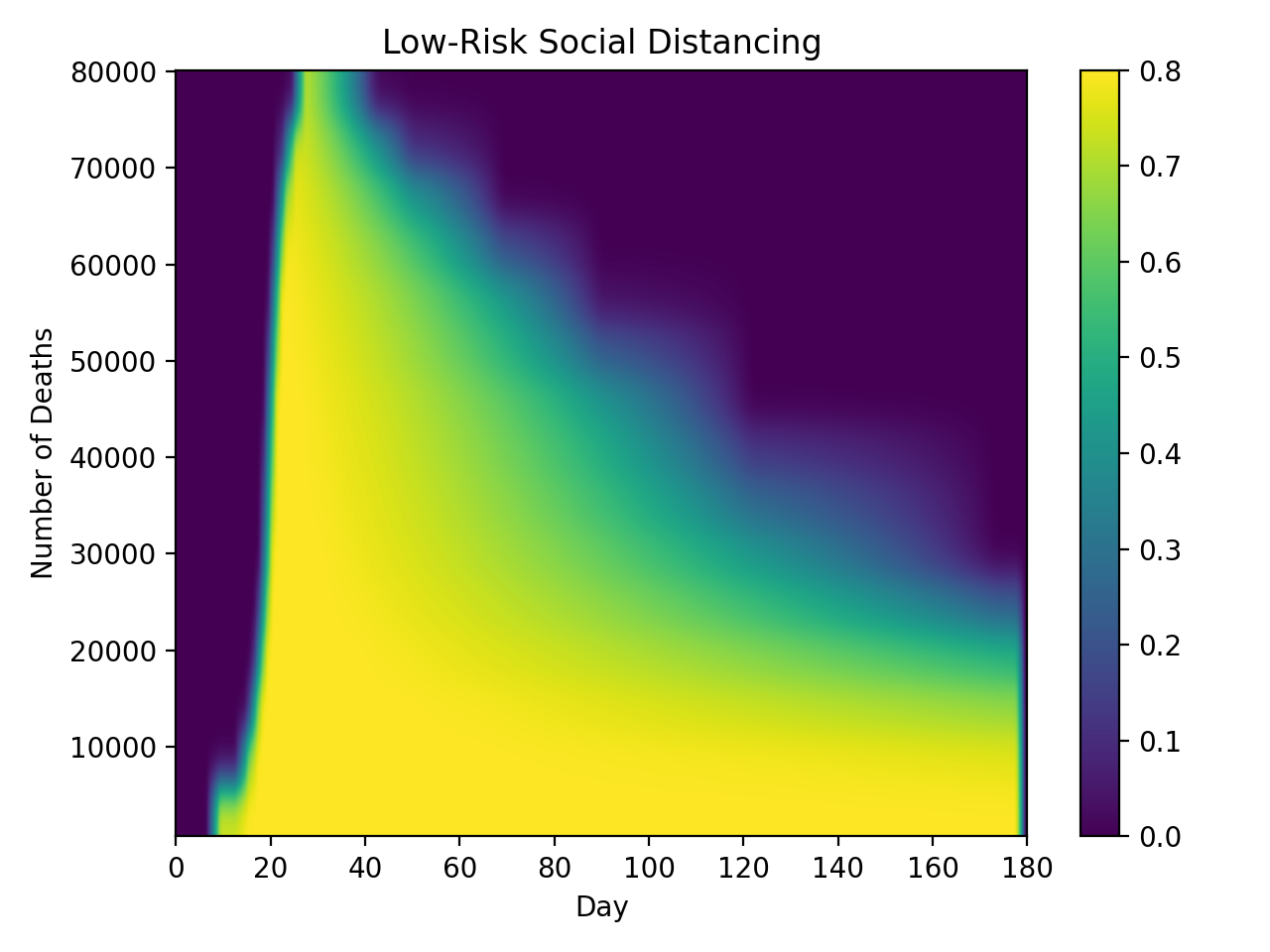}
    \end{subfigure}
    \hfill
    \begin{subfigure}[b]{0.24\columnwidth}
        \includegraphics[width=1.95in]{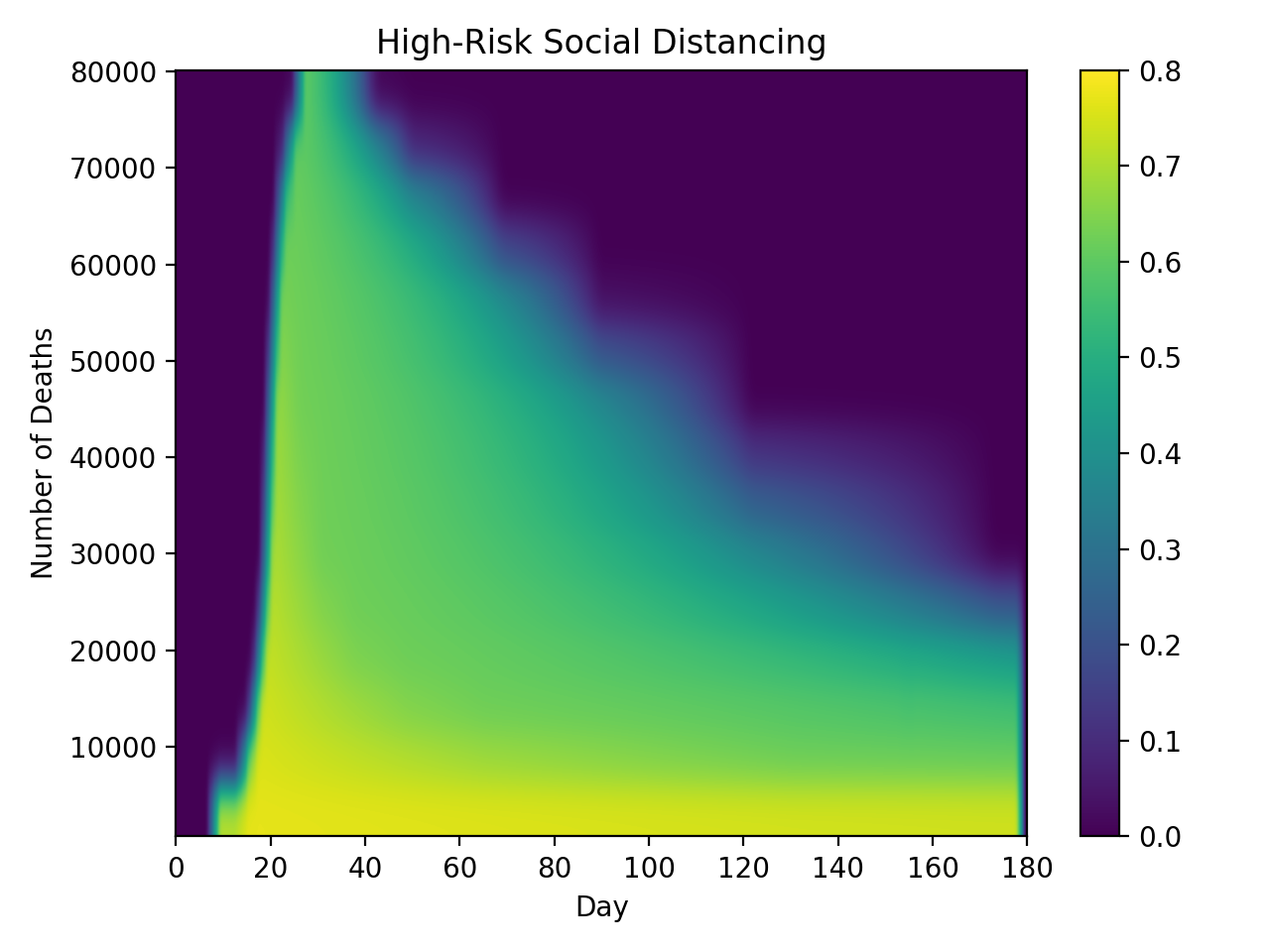}
    \end{subfigure}

\caption{Control levels (low risk testing, high risk testing, low risk distancing, high risk distancing) for constant budget (first row), constant $R_e$ fraction (second row), and constant $R_e$ target (third row) basic control strategies. The color bar indicates control levels (0  to 0.8 for distancing, 0 to 0.66 for testing) over time (horizontal axis) for strategies that achieve total deaths from 0 to 120,000 (vertical axis).}
 \label{fig:four figures}
\end{figure}
\end{landscape}

Figure \ref{fig:four figures} indicates the application of the four different controls  (low and high risk testing and distancing) for the basic versions of  the three different strategy types, for optimal controls associated with different death outcomes. On each figure, the vertical axis indicates the number of deaths acheived by the strategy; the horizontal axis gives the date and the color indicates the level of each control, according to the colorbar accompanying each figure.  For example, for the optimal $R_e$ fraction strategy that achieves  80,000 deaths, the time progression of low risk social distancing may be obtained from the third plot in the second row by looking across the plot at the 80,000 level on the vertical axis.
The $R_e$ target strategies (last row of plots) begin with high levels of low-risk (third plot in the row) and high-risk distancing (fourth plot), and then transition towards testing in the later stages (first two plots). In contrast, the other two strategy types prioritize distancing (especially distancing of the low-risk group)  over testing throughout the period of control.  
 In budget-based strategies, typically each control is applied at a nearly constant level throughout the period of control as evidenced by the  horizontal color patterns in the first row of plots.
For between 20,000-120,000 deaths, the optimal start dates for all three strategy types are very close to day 20.  The figures show that for deaths below about 50,000, the control continues up the end of the period, indicating that the disease is still extant and herd immunity has not yet been reached.  

\begin{landscape}
\begin{figure}
    \begin{subfigure}[b]{0.24\columnwidth}
    \includegraphics[width=1.95in]{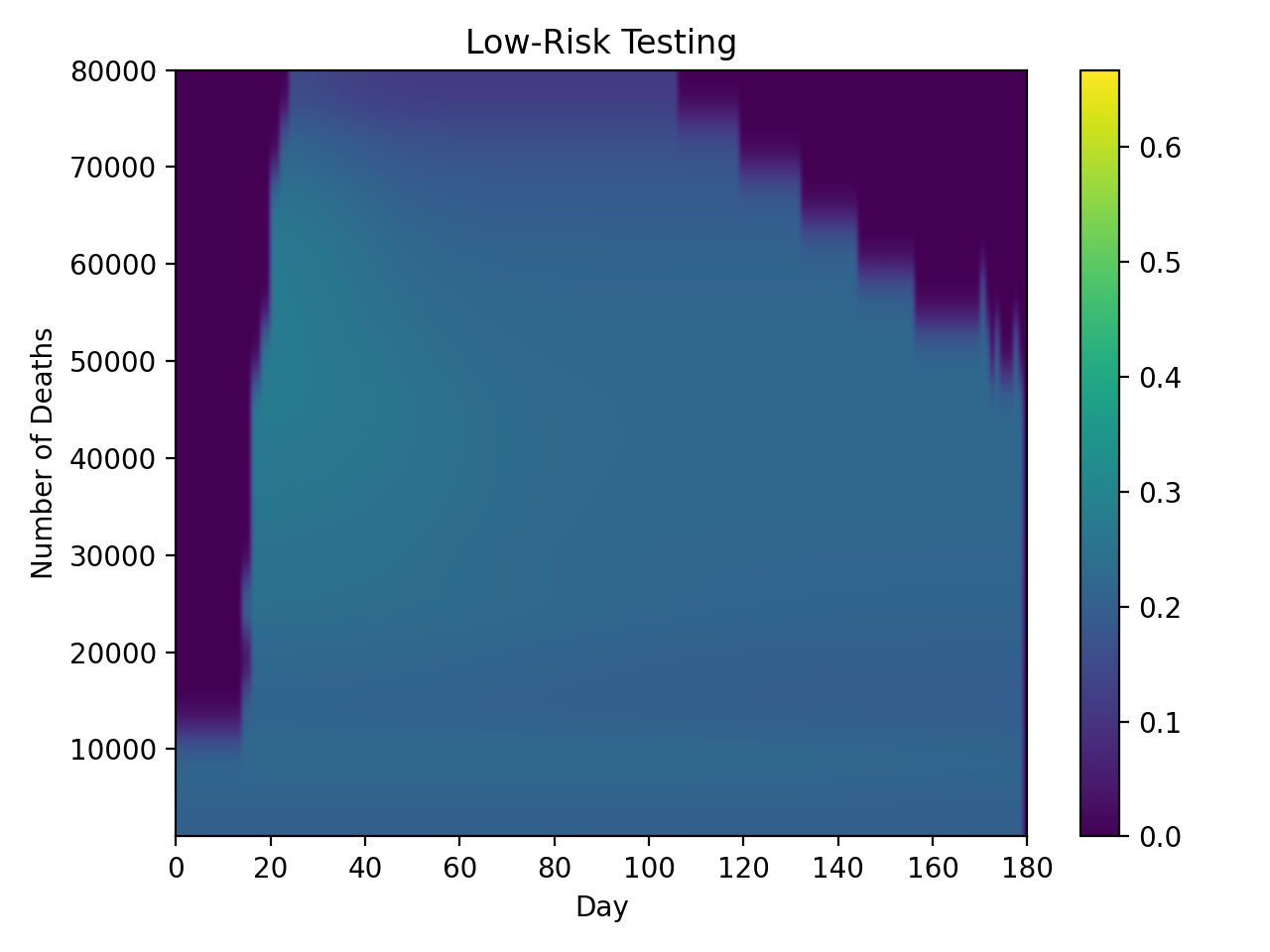}      
    \end{subfigure}
    \hfill
     \begin{subfigure}[b]{0.24\columnwidth}
       \includegraphics[width=1.95in]{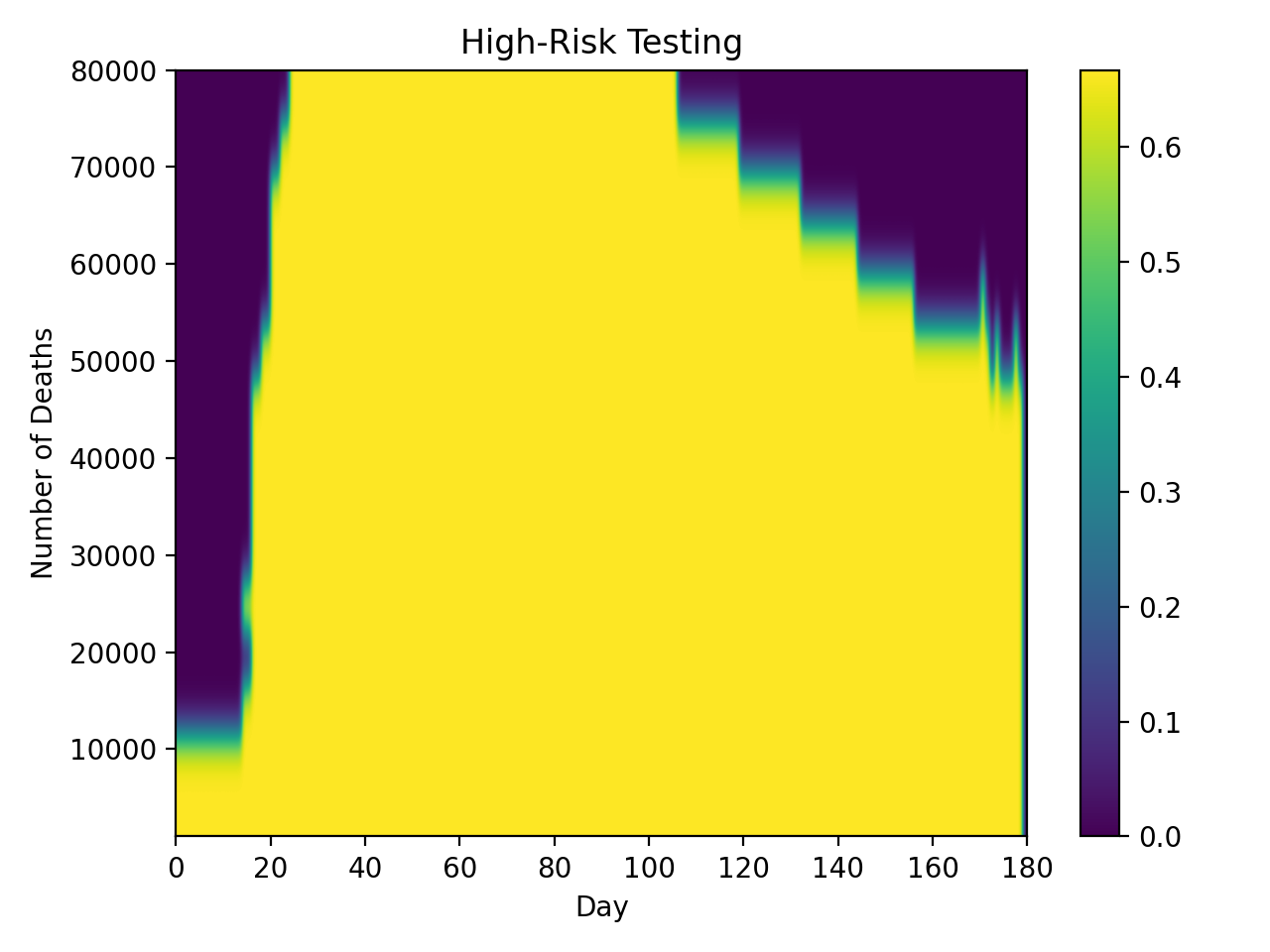}
    \end{subfigure}
    \hfill
    \begin{subfigure}[b]{0.24\columnwidth}
        \includegraphics[width=1.95in]{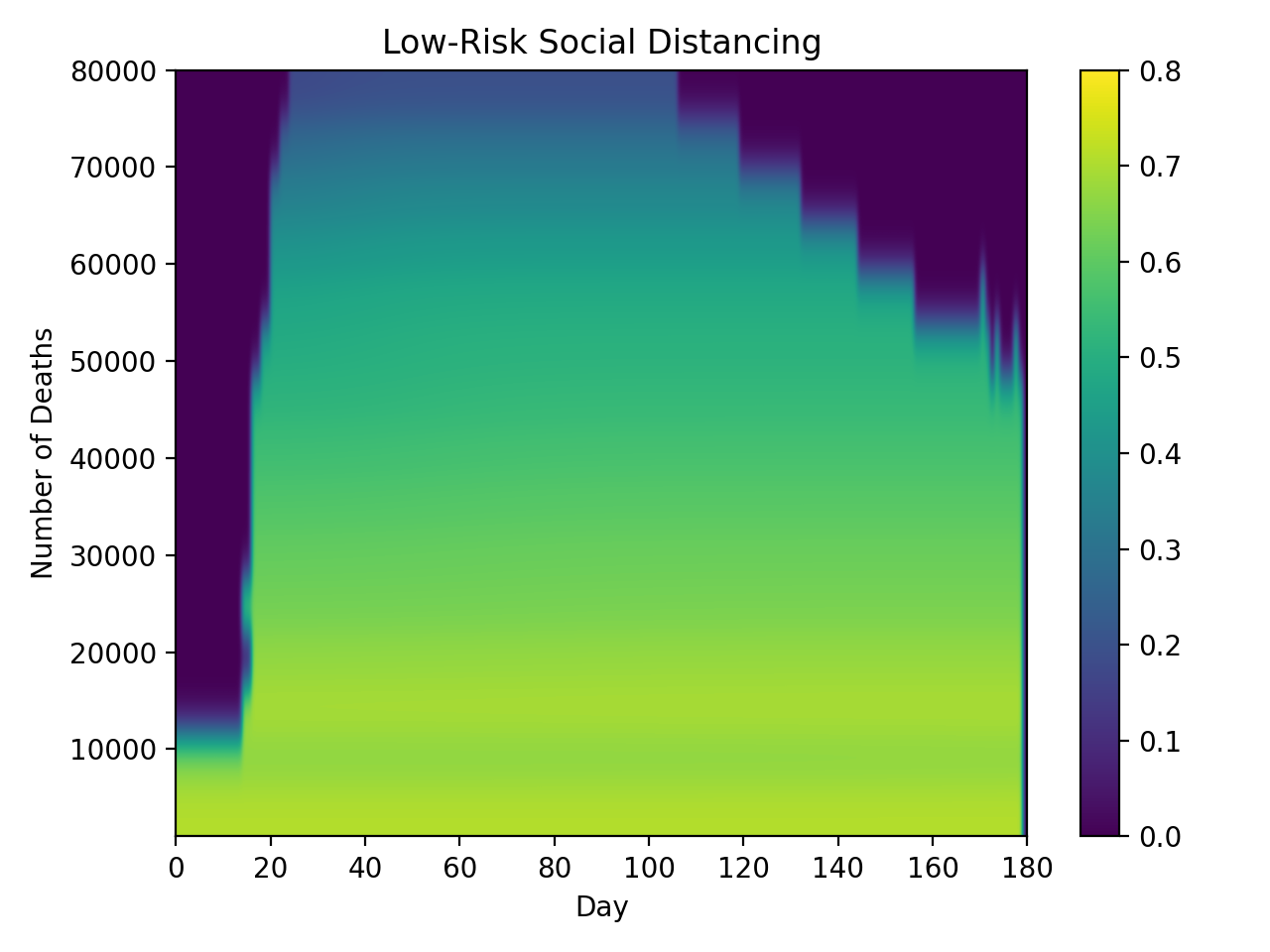}
    \end{subfigure}
    \hfill
    \begin{subfigure}[b]{0.24\columnwidth}
      \includegraphics[width=1.95in]{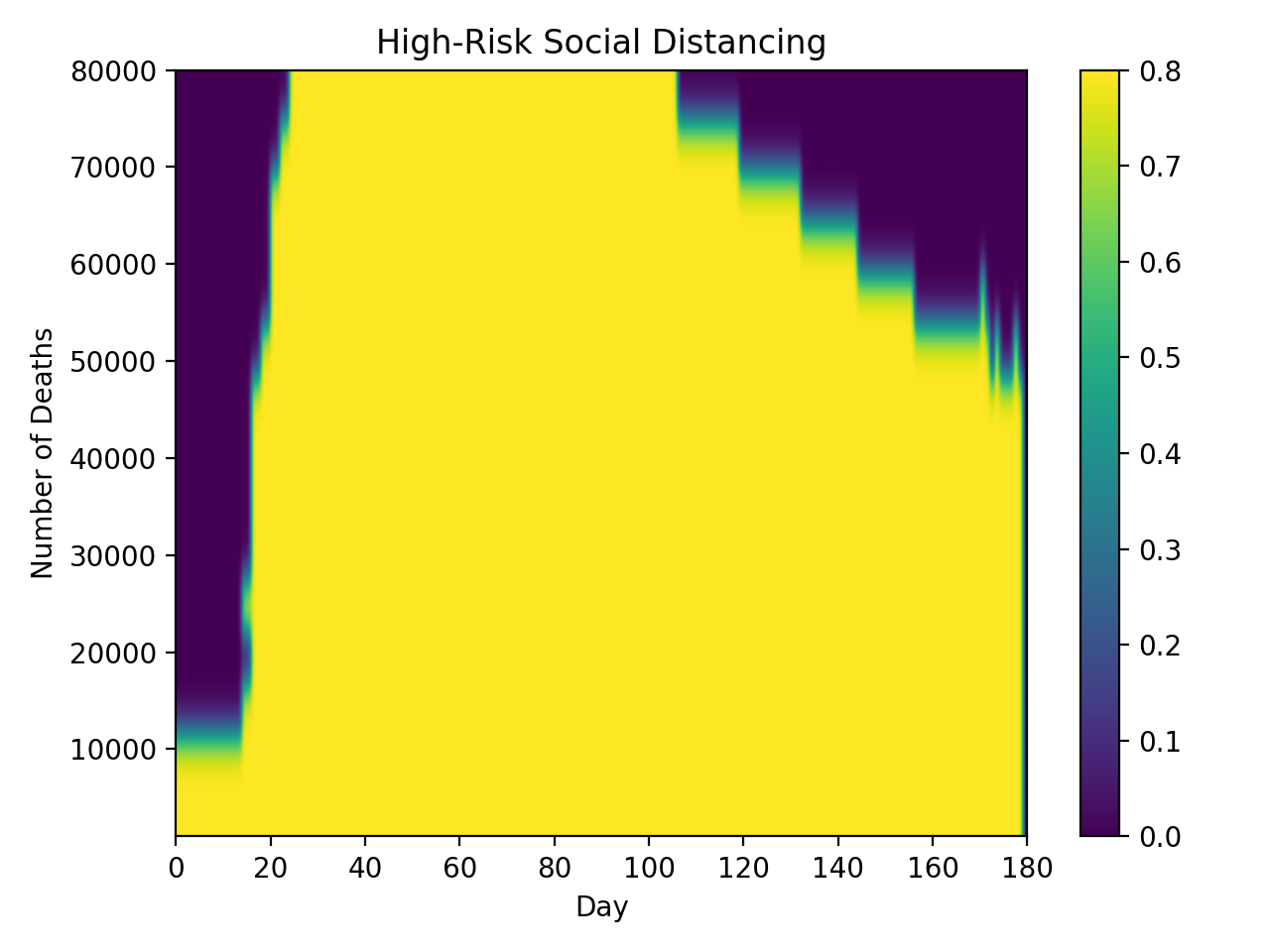}
    \end{subfigure}
  
    \begin{subfigure}[b]{0.24\columnwidth}
    \includegraphics[width=1.95in]{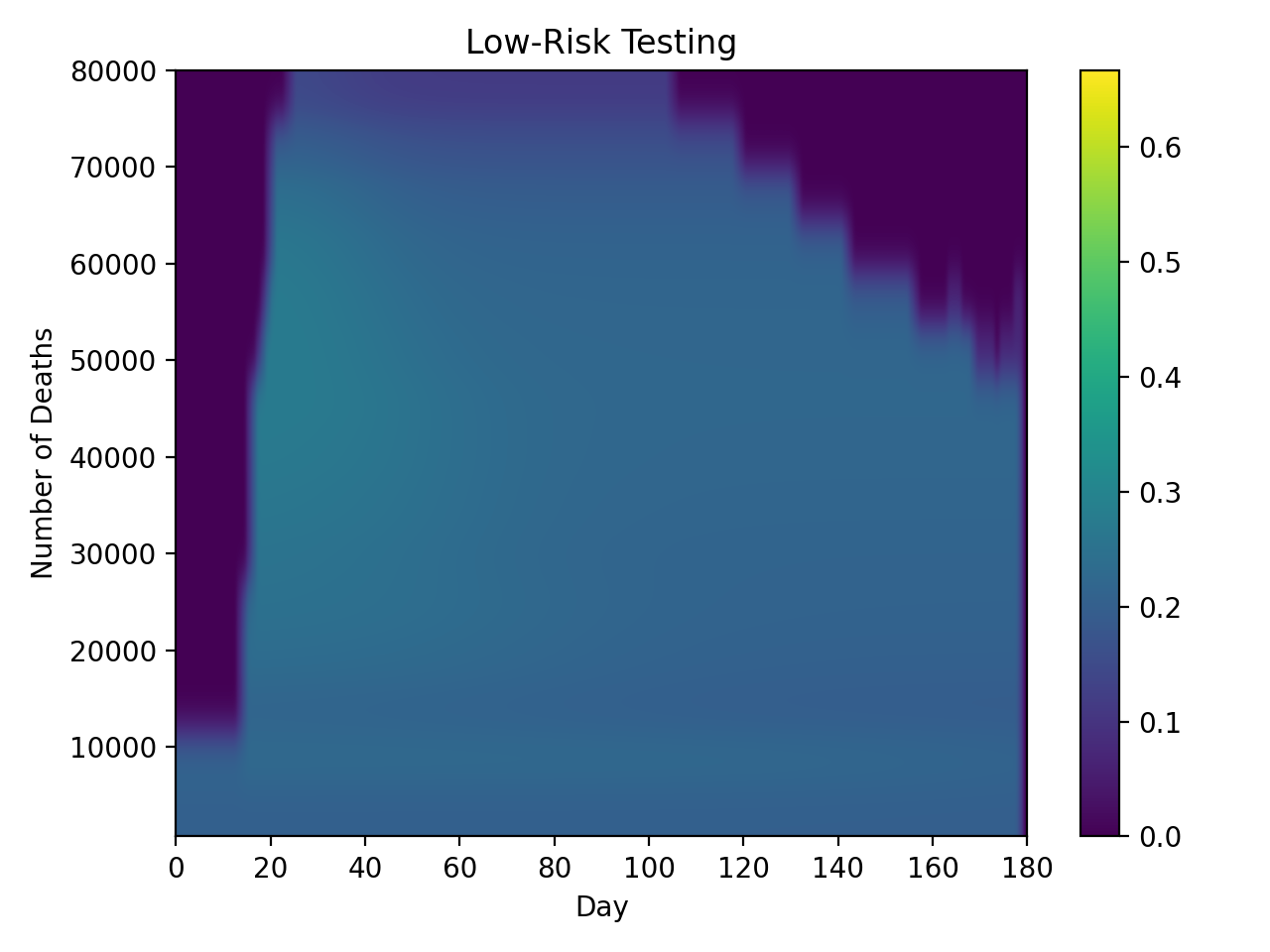}   
    \end{subfigure}
    \hfill
     \begin{subfigure}[b]{0.24\columnwidth}
       \includegraphics[width=1.95in]{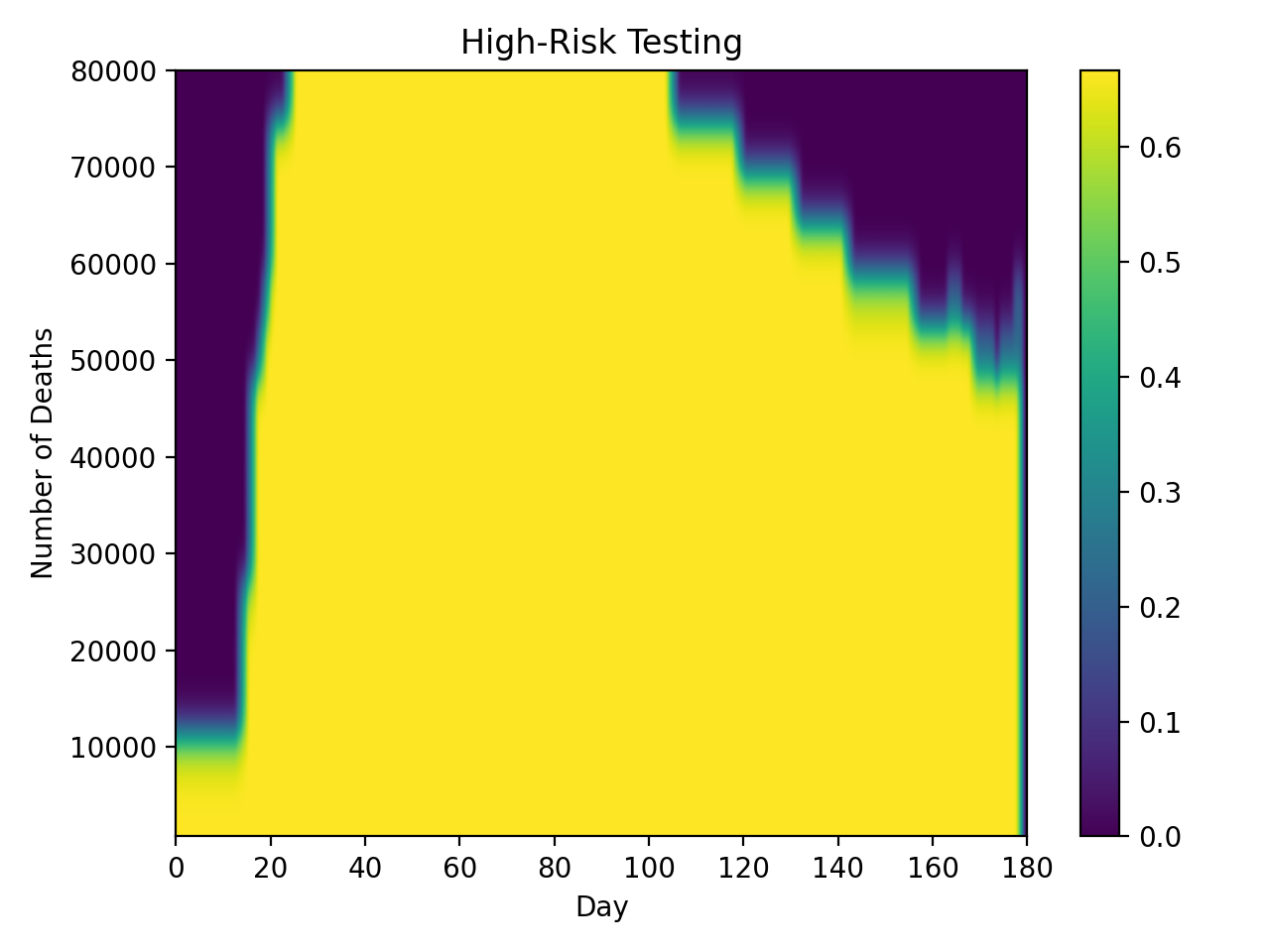}
    \end{subfigure}    
    \hfill
    \begin{subfigure}[b]{0.24\columnwidth}
        \includegraphics[width=1.95in]{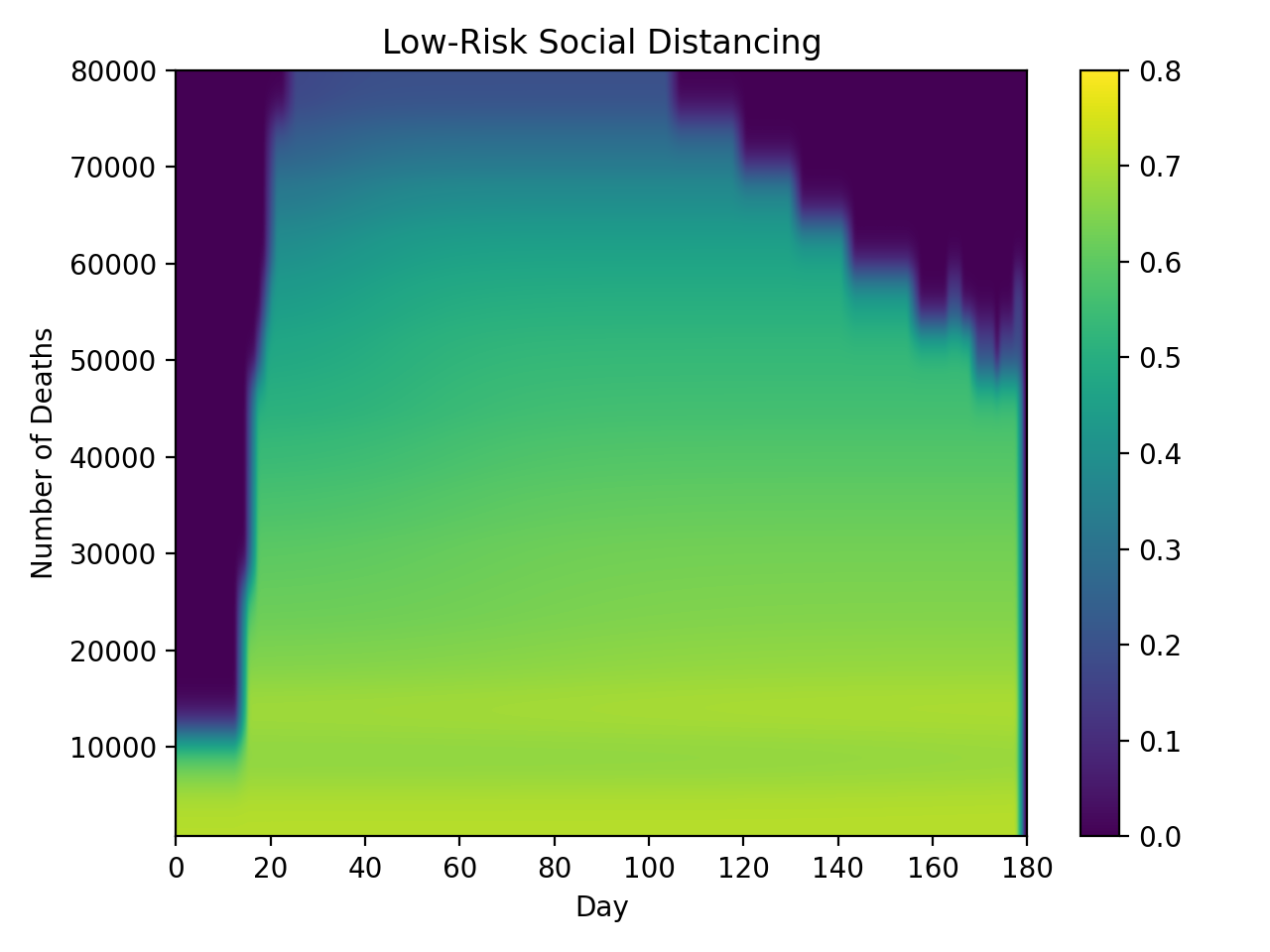}
    \end{subfigure}
    \hfill
    \begin{subfigure}[b]{0.24\columnwidth}
      \includegraphics[width=1.95in]{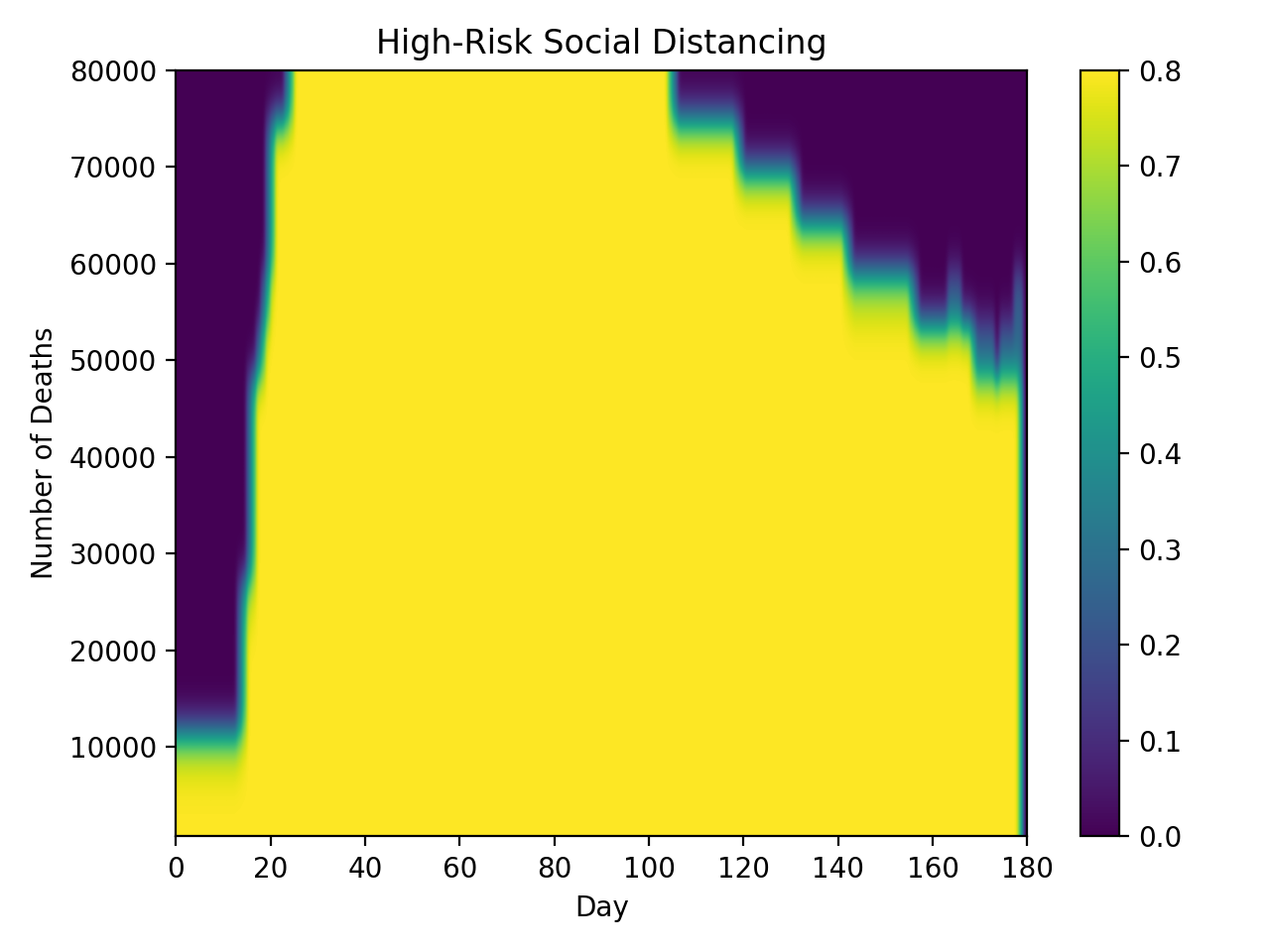}
    \end{subfigure}

    \begin{subfigure}[b]{0.24\columnwidth}
    \includegraphics[width=1.95in]{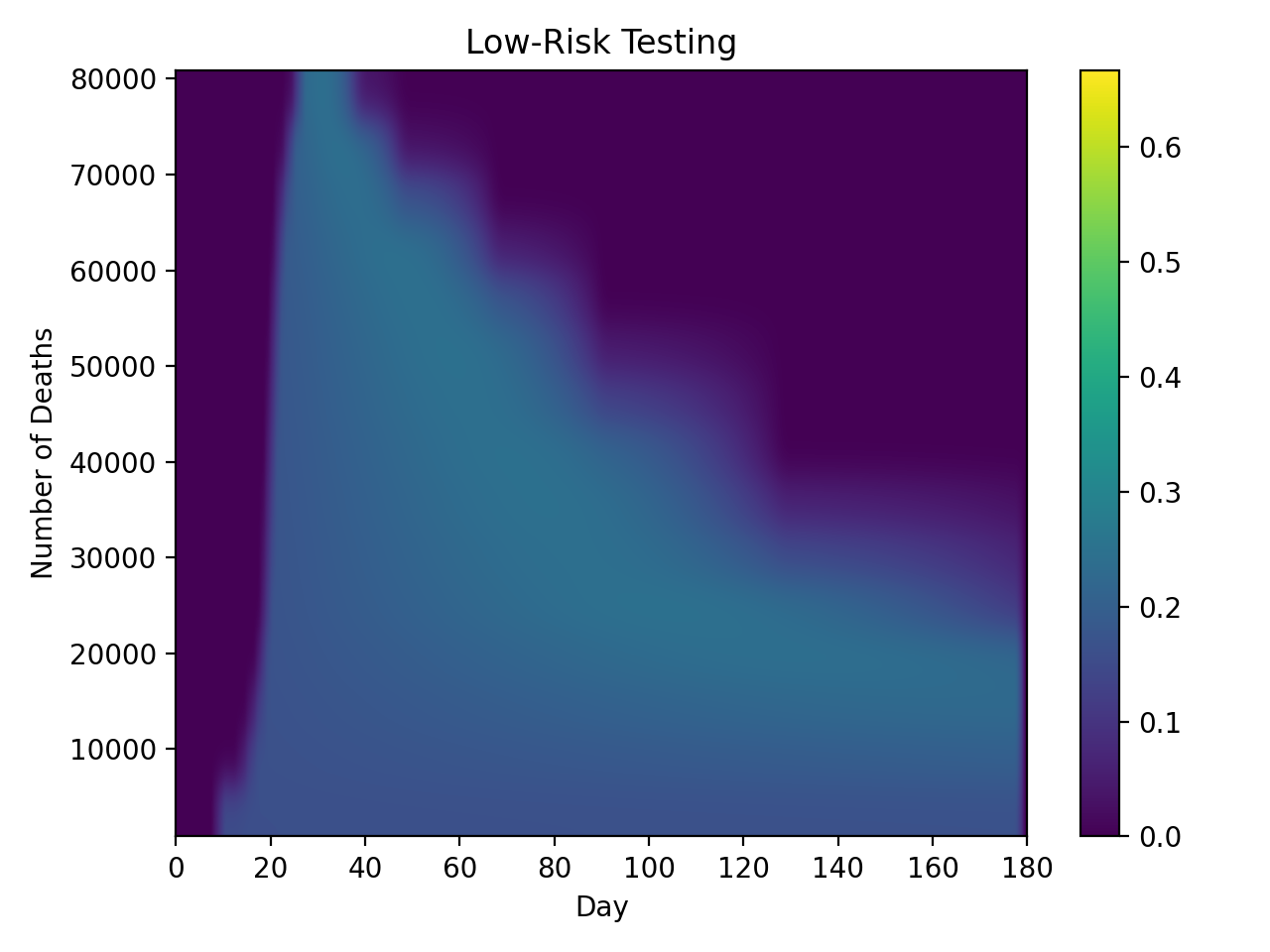}      
    \end{subfigure}
    \hfill
    \begin{subfigure}[b]{0.24\columnwidth}
    \includegraphics[width=1.95in]{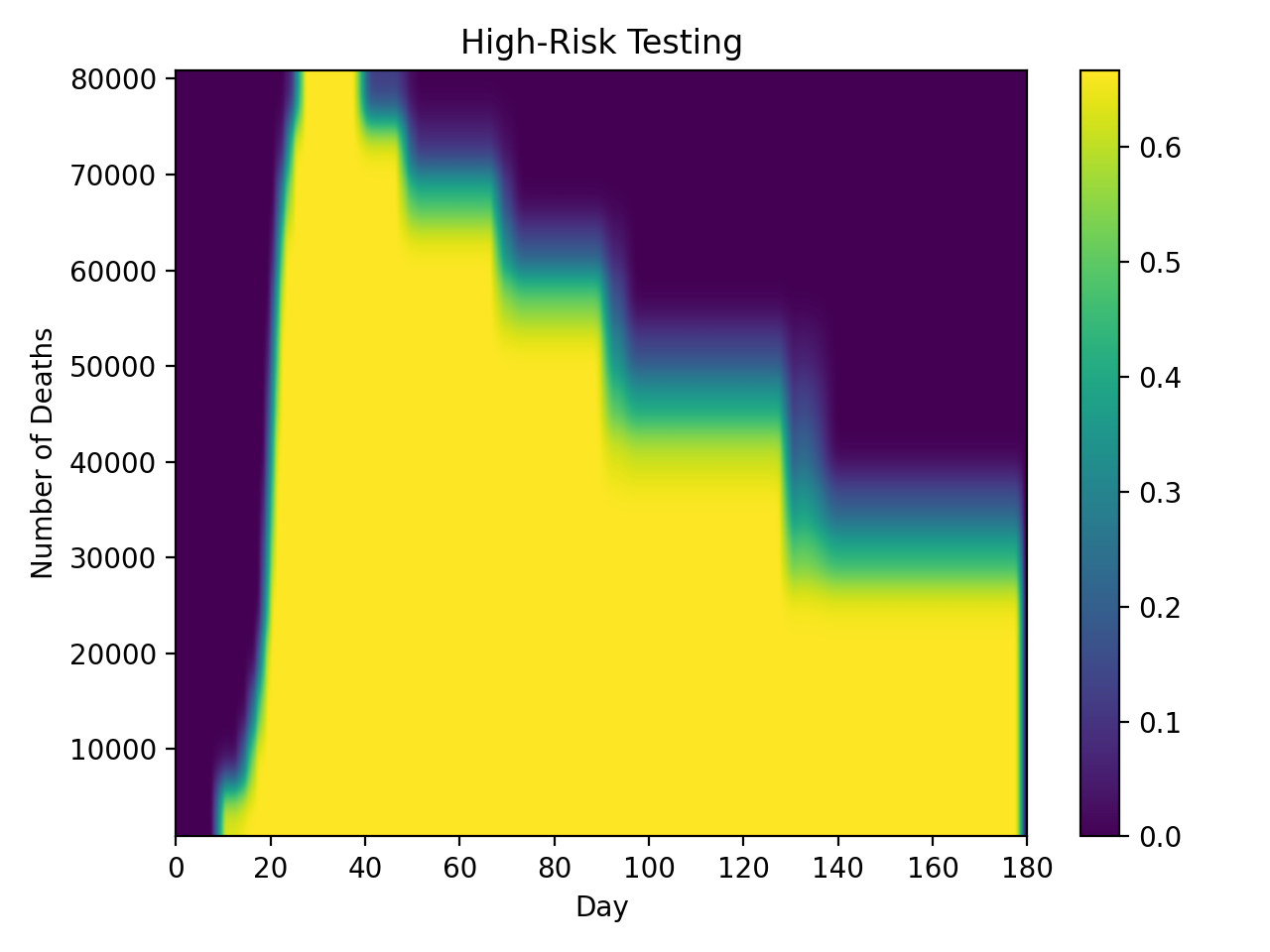}       
    \end{subfigure}
    \hfill
     \begin{subfigure}[b]{0.24\columnwidth}
       \includegraphics[width=1.95in]{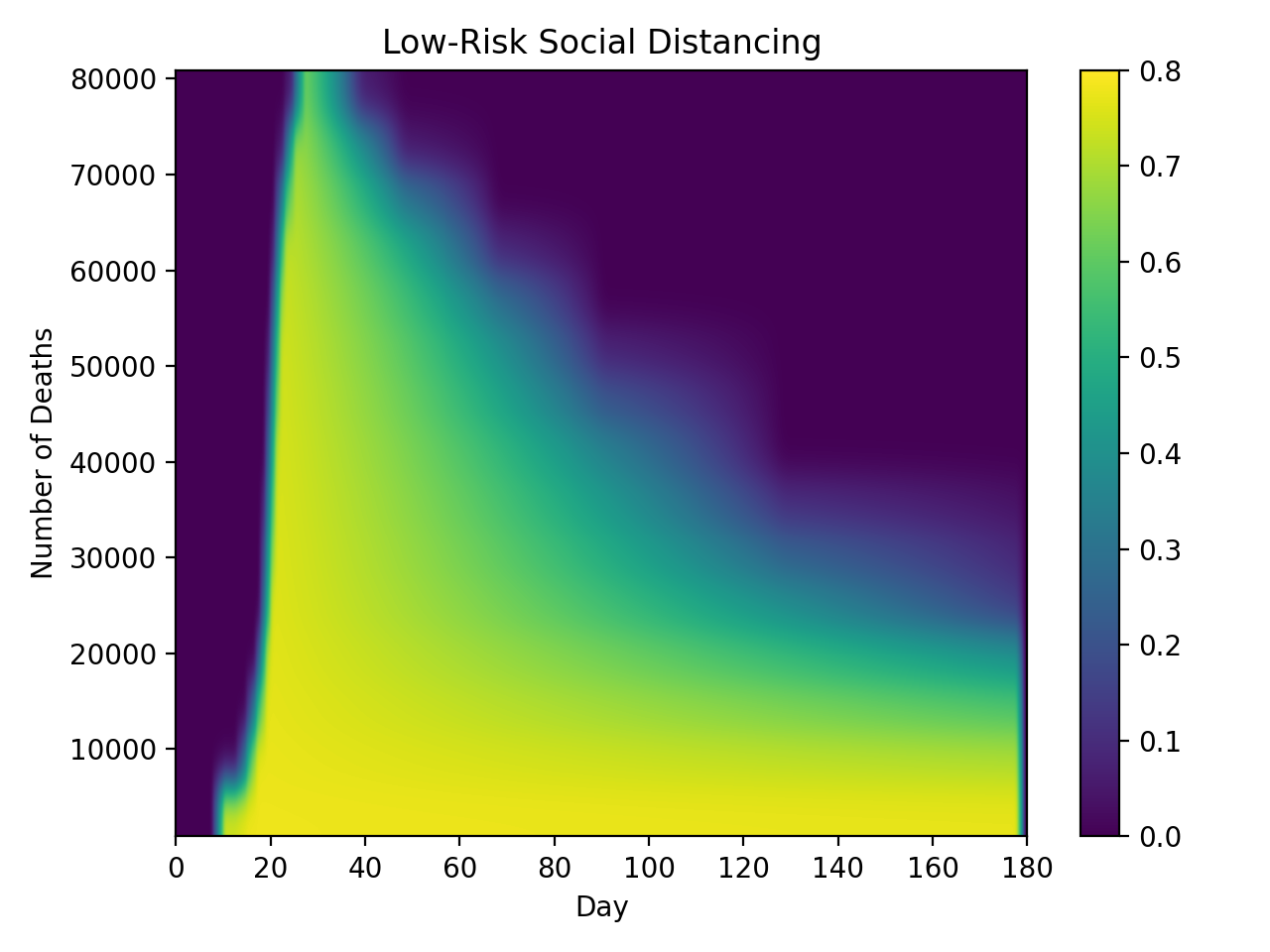}
    \end{subfigure}    
    \hfill
     \begin{subfigure}[b]{0.24\columnwidth}
       \includegraphics[width=1.95in]{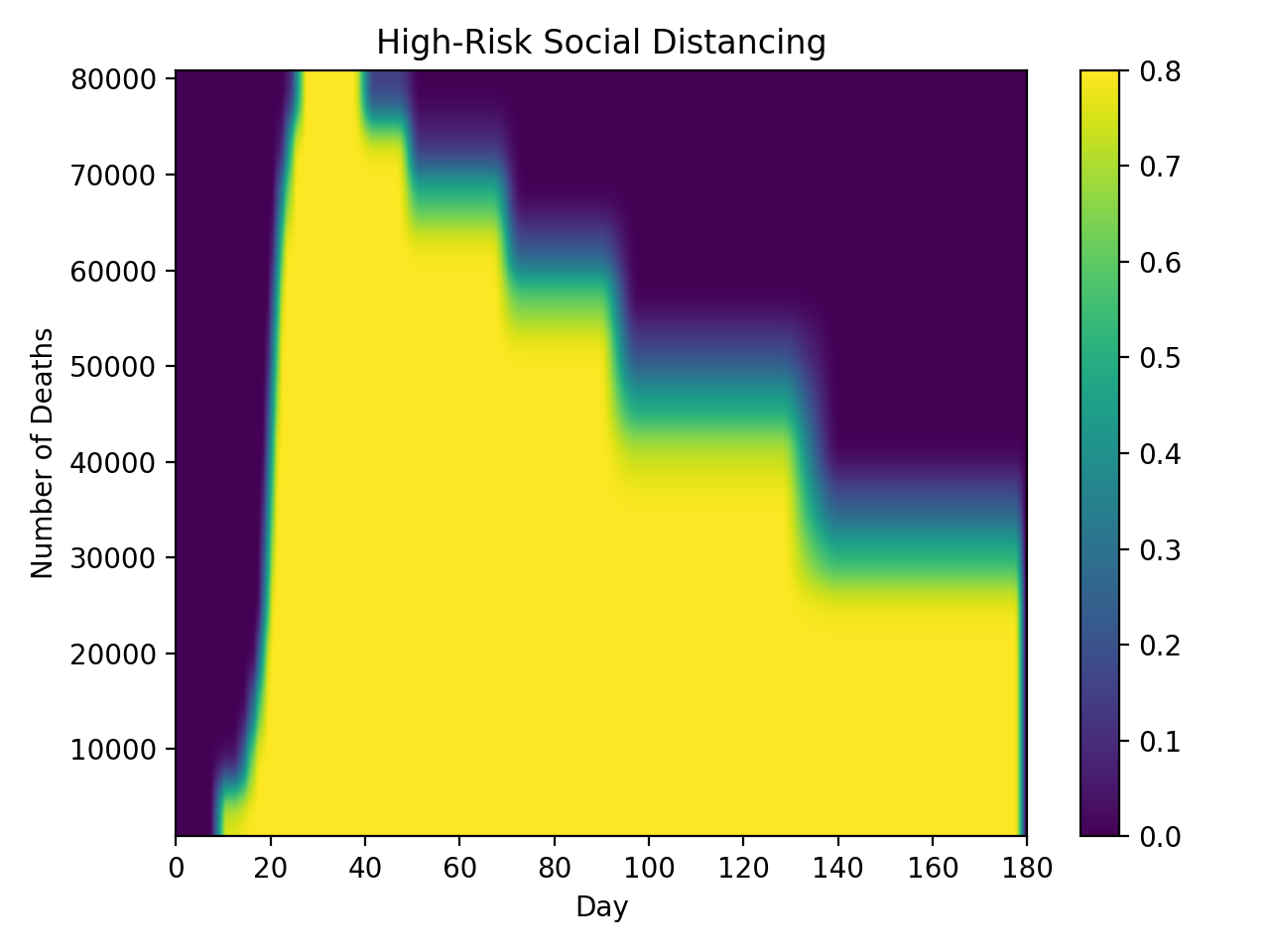}
    \end{subfigure}    
 
\caption{Control levels (low risk testing, high risk testing, low risk distancing, high risk distancing) for constant budget (first row), constant $R_e$ fraction (second row), and constant $R_e$ target (third row) high risk prioritizing control strategies. Plot arrangement and color scales are the same as in Figure~\ref{fig:four figures}}
 \label{fig:four figures2}
\end{figure}
\end{landscape}

Figure \ref{fig:four figures2} is analogous to Figure \ref{fig:four figures}, and represents the control history for the four controls for optimal high-risk prioritizing controls at different death levels. As reflected in the figures, testing and distancing for the high risk group is at maximum level throughout the control period, while testing and distancing for the low risk group resembles Figure \ref{fig:four figures} but at somewhat lower levels.

\begin{landscape}
\begin{figure}
    \begin{subfigure}[b]{0.24\columnwidth}
    \includegraphics[width=1.95in]{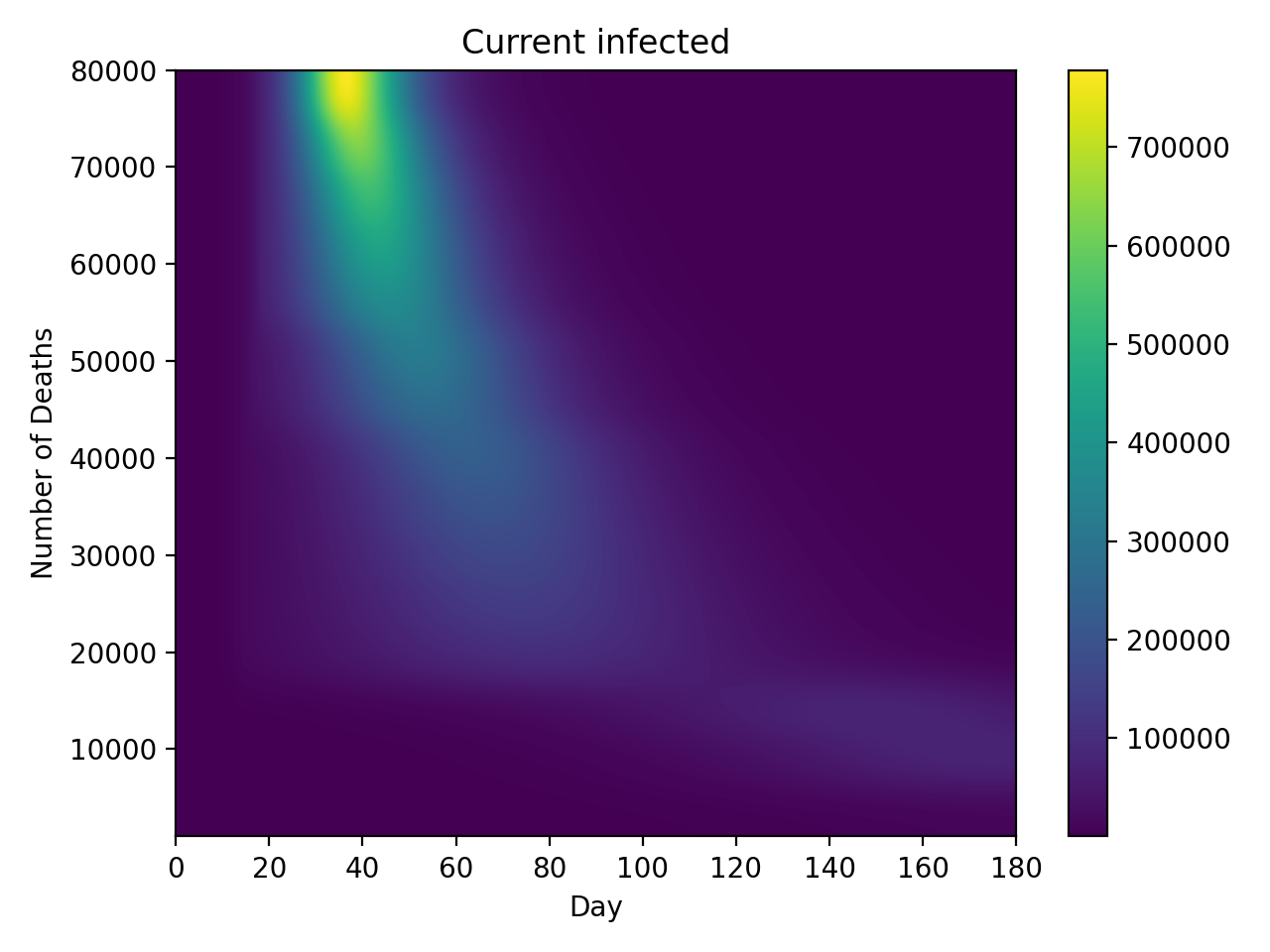}
    \end{subfigure}
     \hfill
    \begin{subfigure}[b]{0.24\columnwidth}
        \includegraphics[width=1.95in]{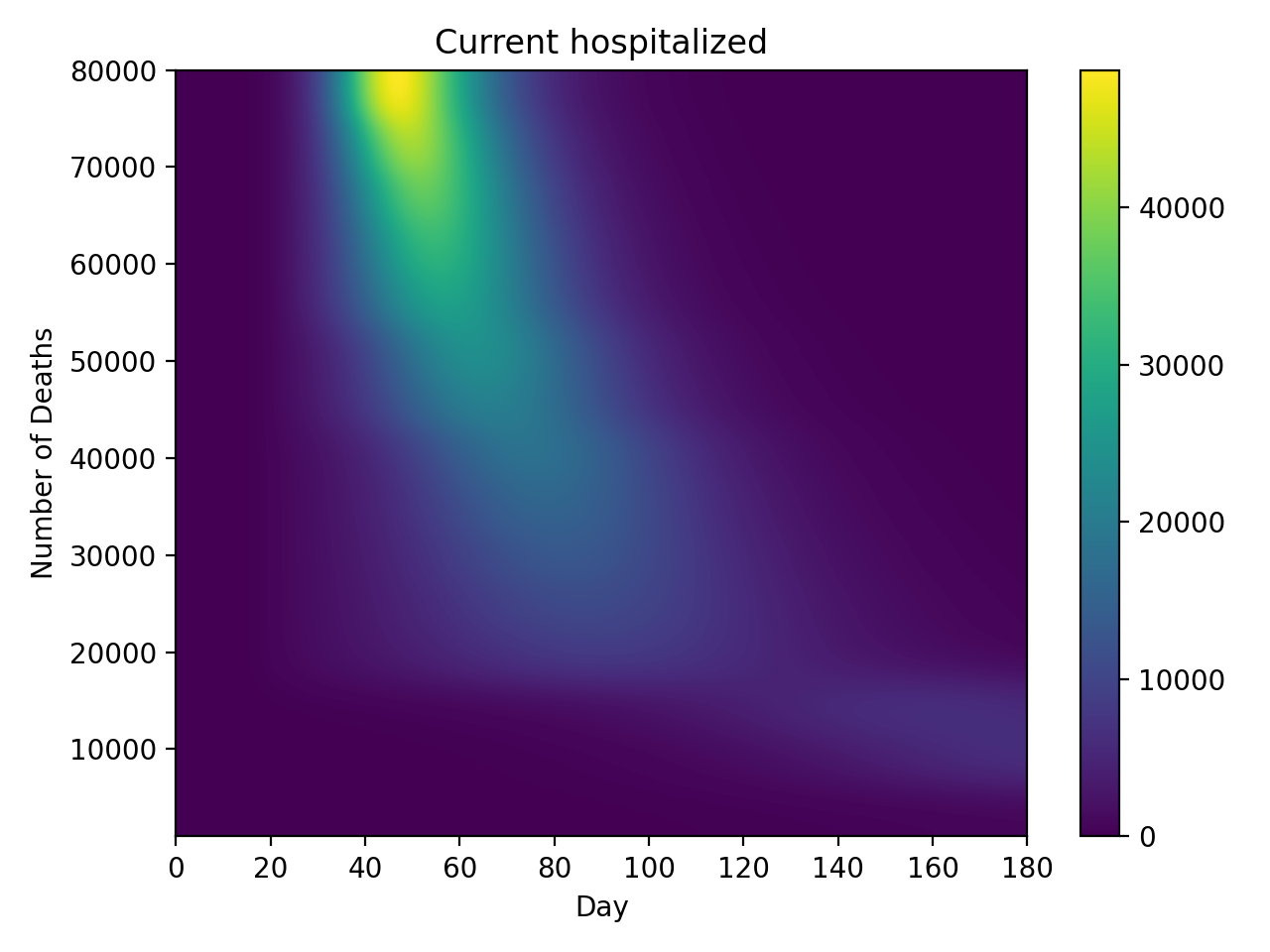}
    \end{subfigure}
    \hfill
    \begin{subfigure}[b]{0.24\columnwidth}
        \includegraphics[width=1.95in]{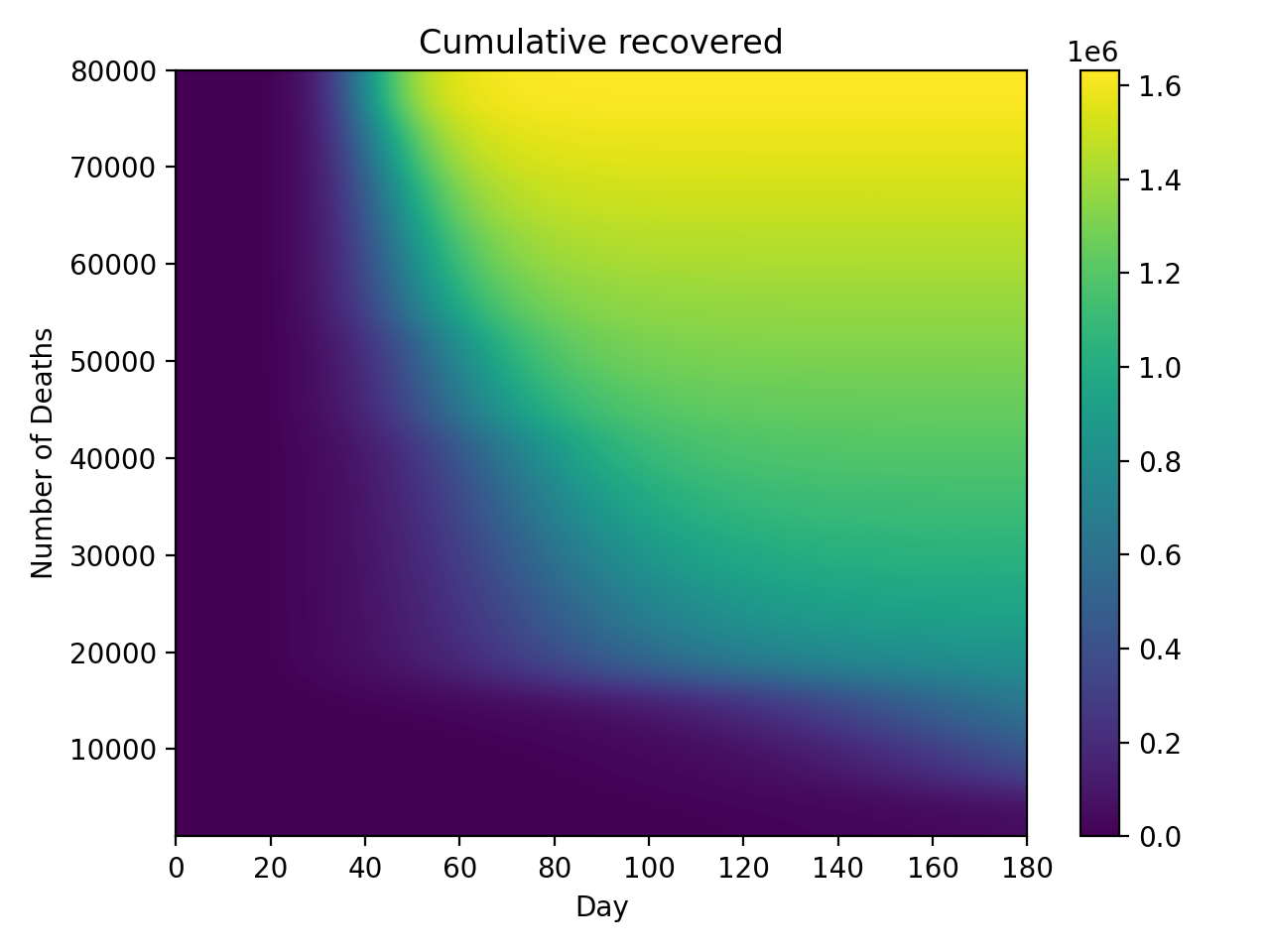}
    \end{subfigure}
 \hfill
    \begin{subfigure}[b]{0.24\columnwidth}
        \includegraphics[width=1.95in]{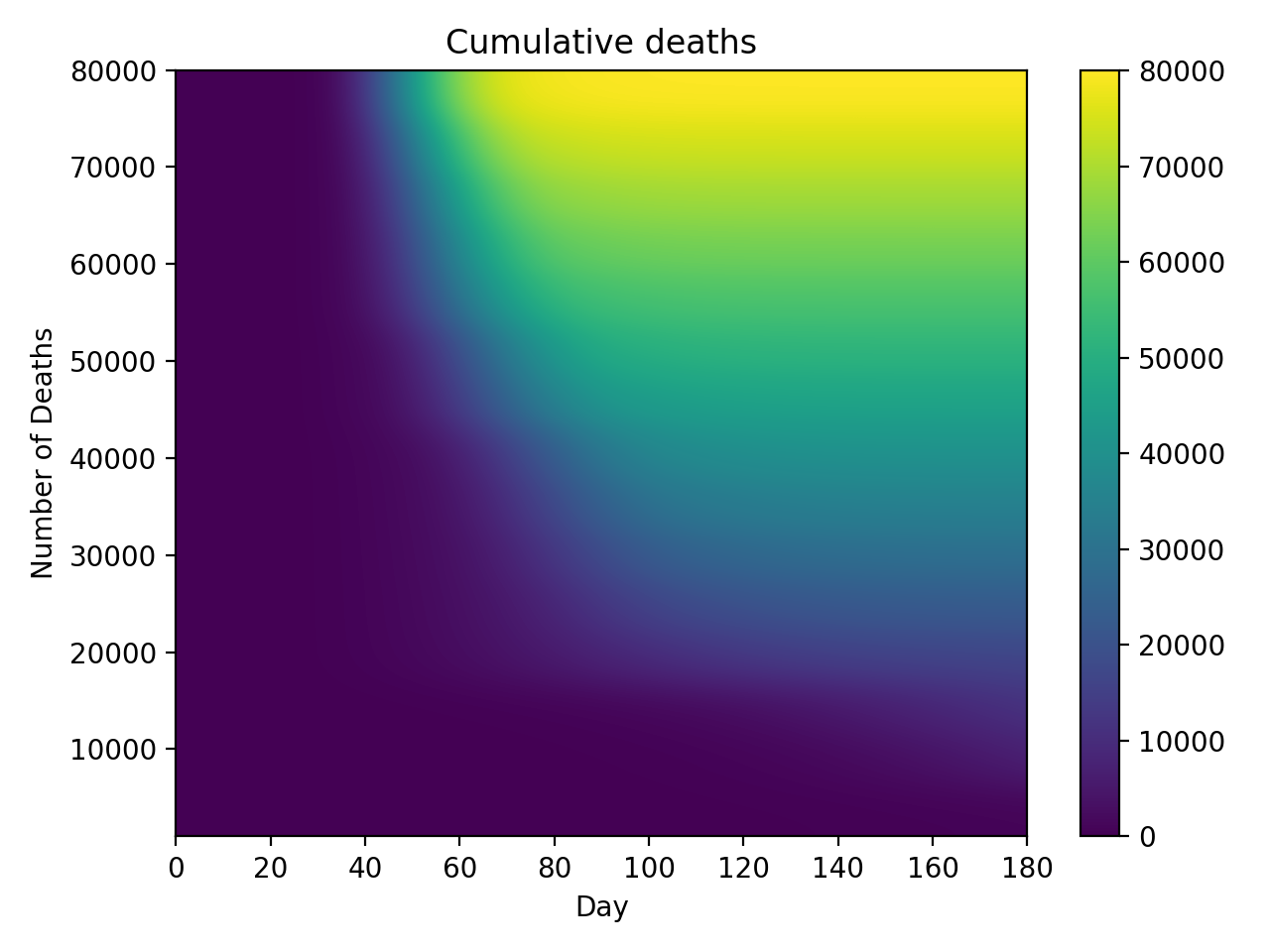}
    \end{subfigure}   
  \begin{subfigure}[b]{0.24\columnwidth}
        \includegraphics[width=1.95in]{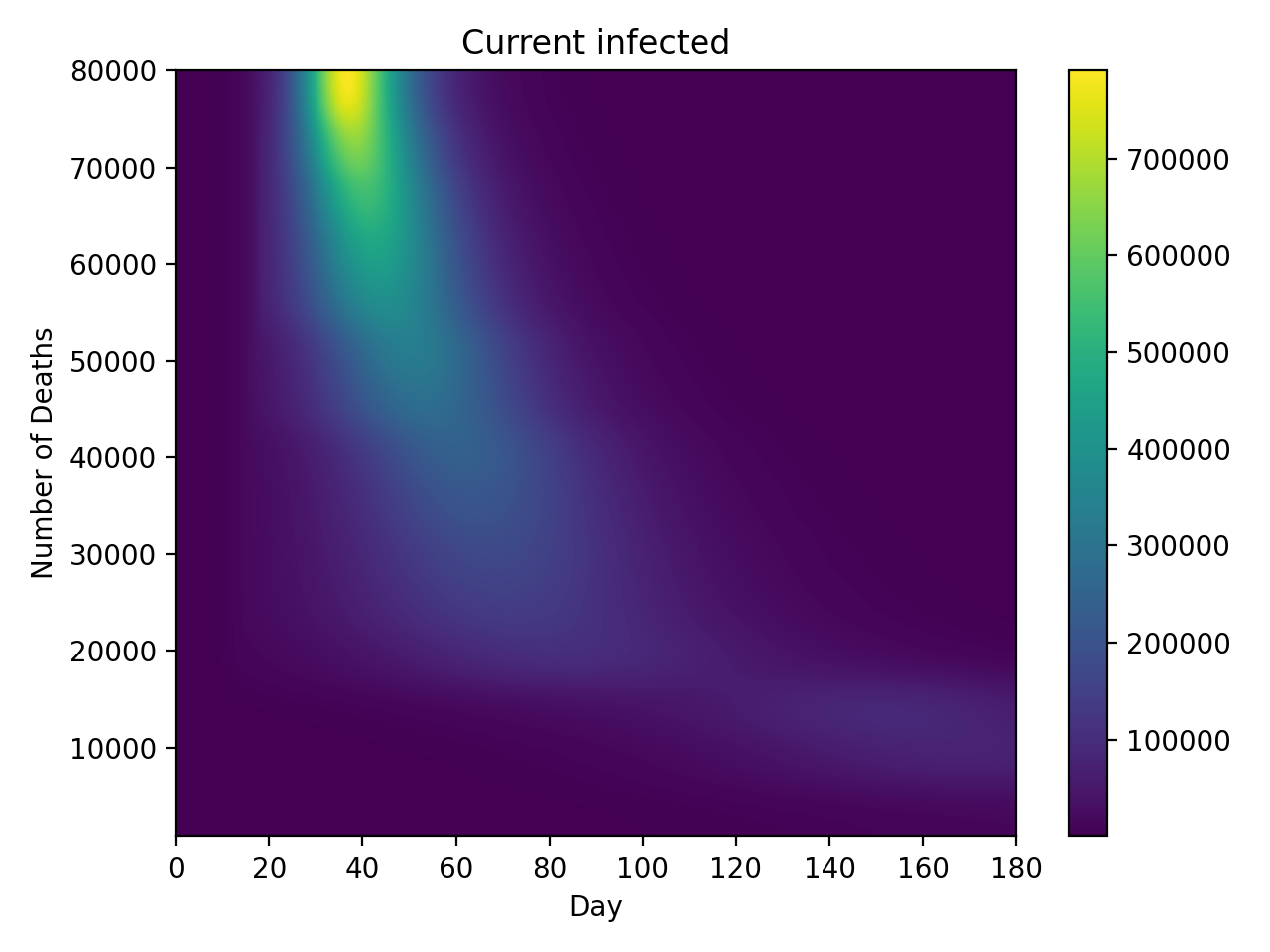}
    \end{subfigure}
    \hfill 
 \begin{subfigure}[b]{0.24\columnwidth}
        \includegraphics[width=1.95in]{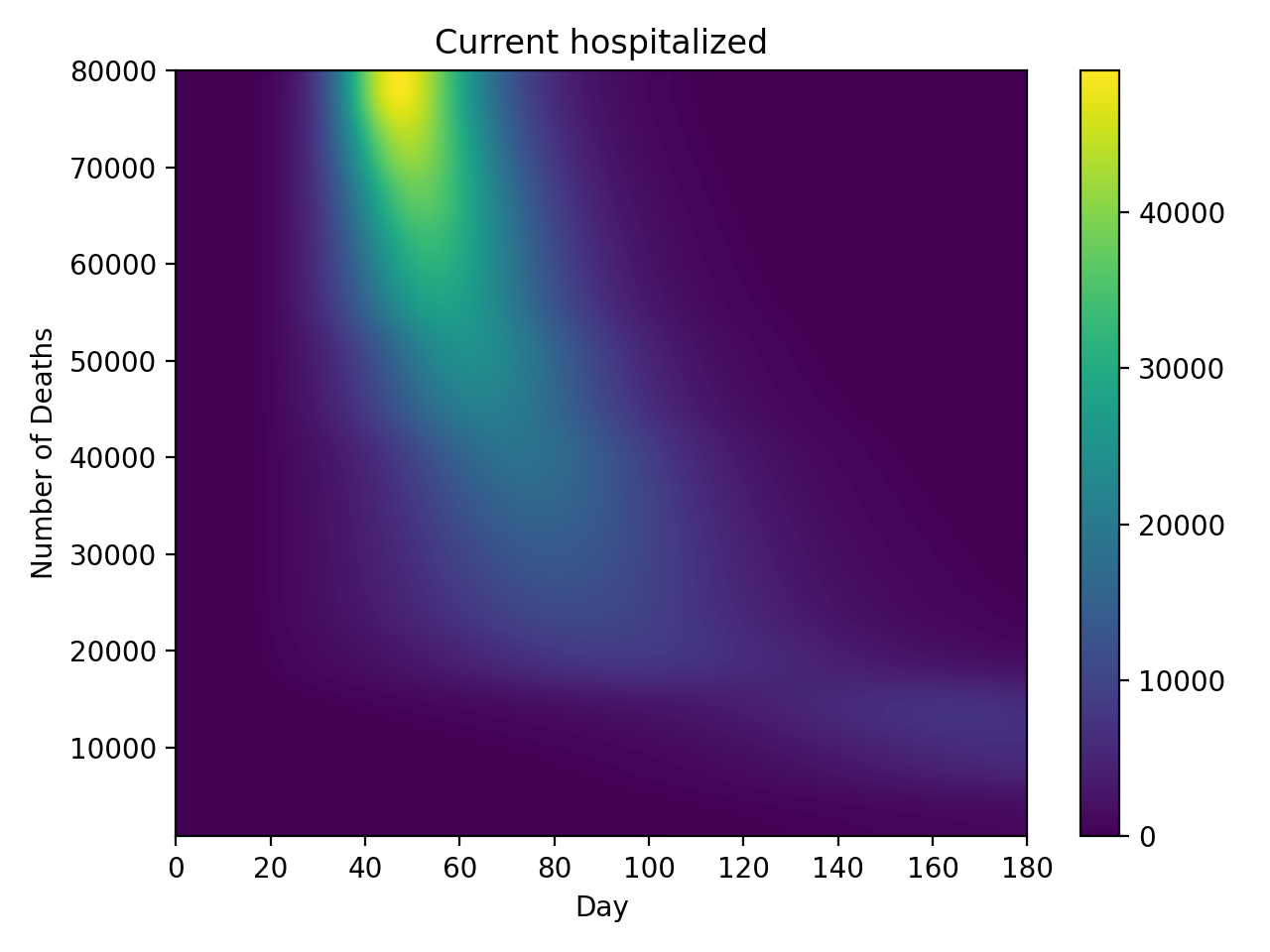}
    \end{subfigure}
    \hfill 
\begin{subfigure}[b]{0.24\columnwidth}
\includegraphics[width=1.95in]{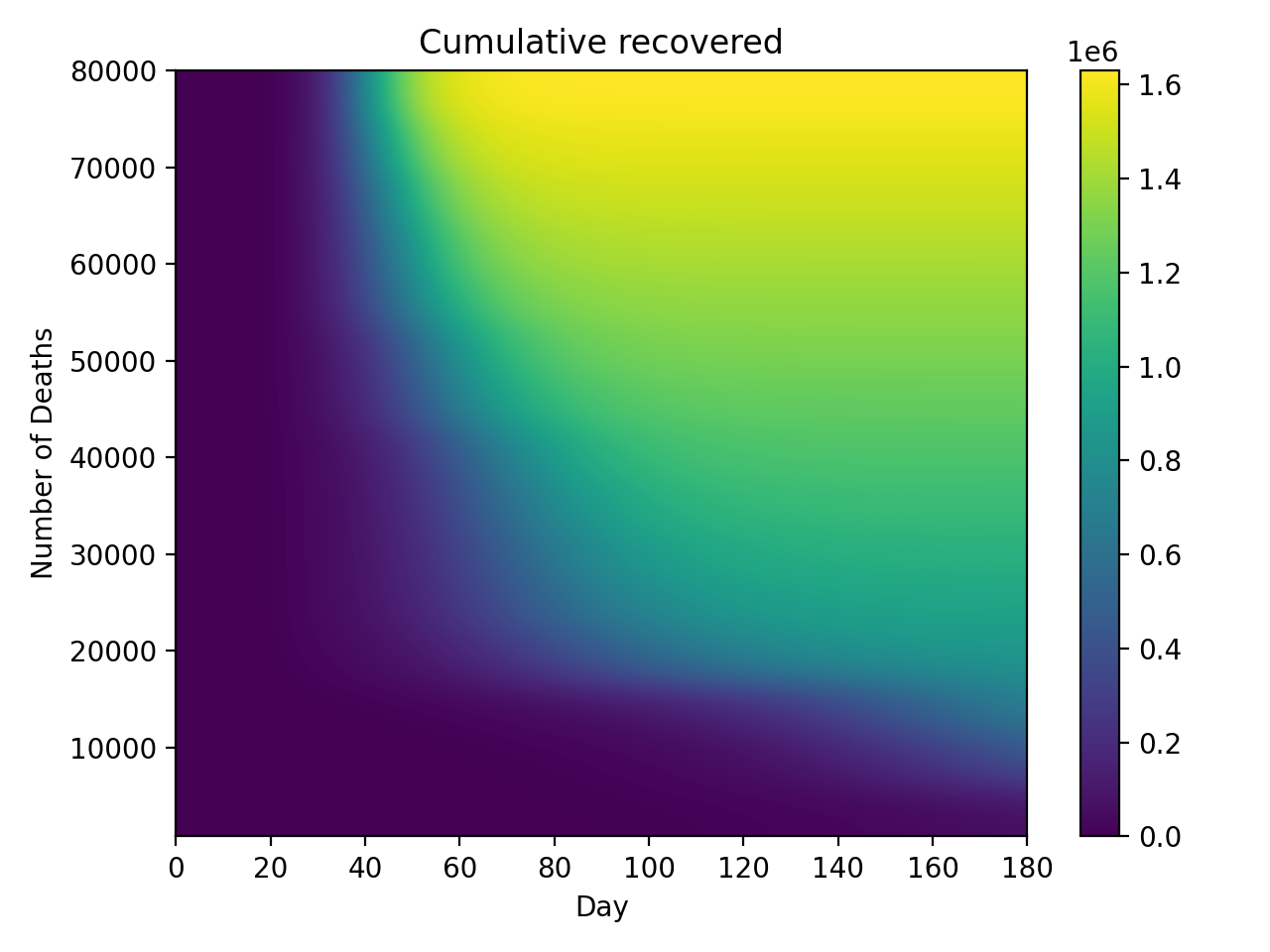}
    \end{subfigure}
    \hfill   
\begin{subfigure}[b]{0.24\columnwidth}
        \includegraphics[width=1.95in]{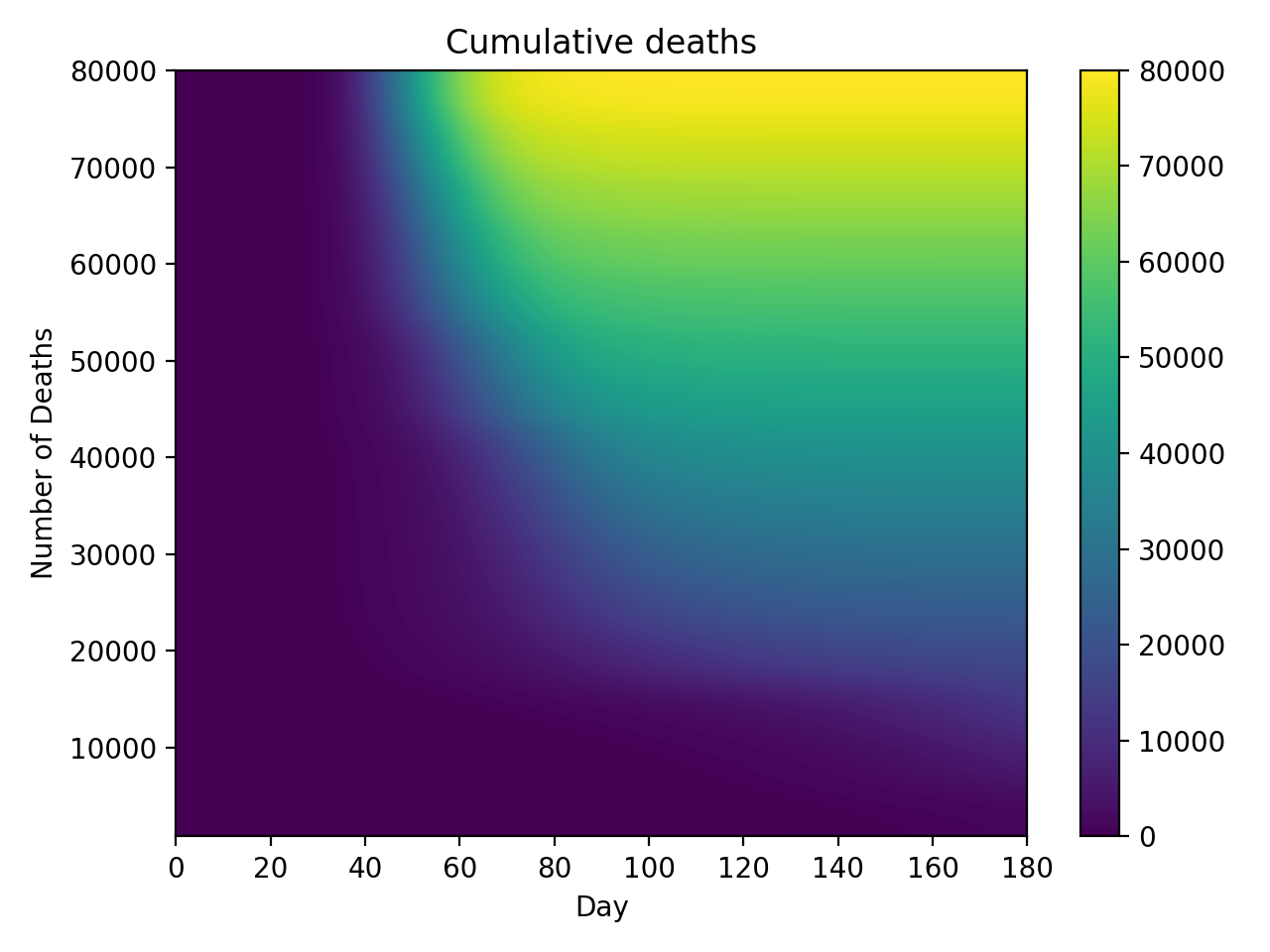}
    \end{subfigure}
    \hfill   
    \begin{subfigure}[b]{0.24\columnwidth}
    \includegraphics[width=1.95in]{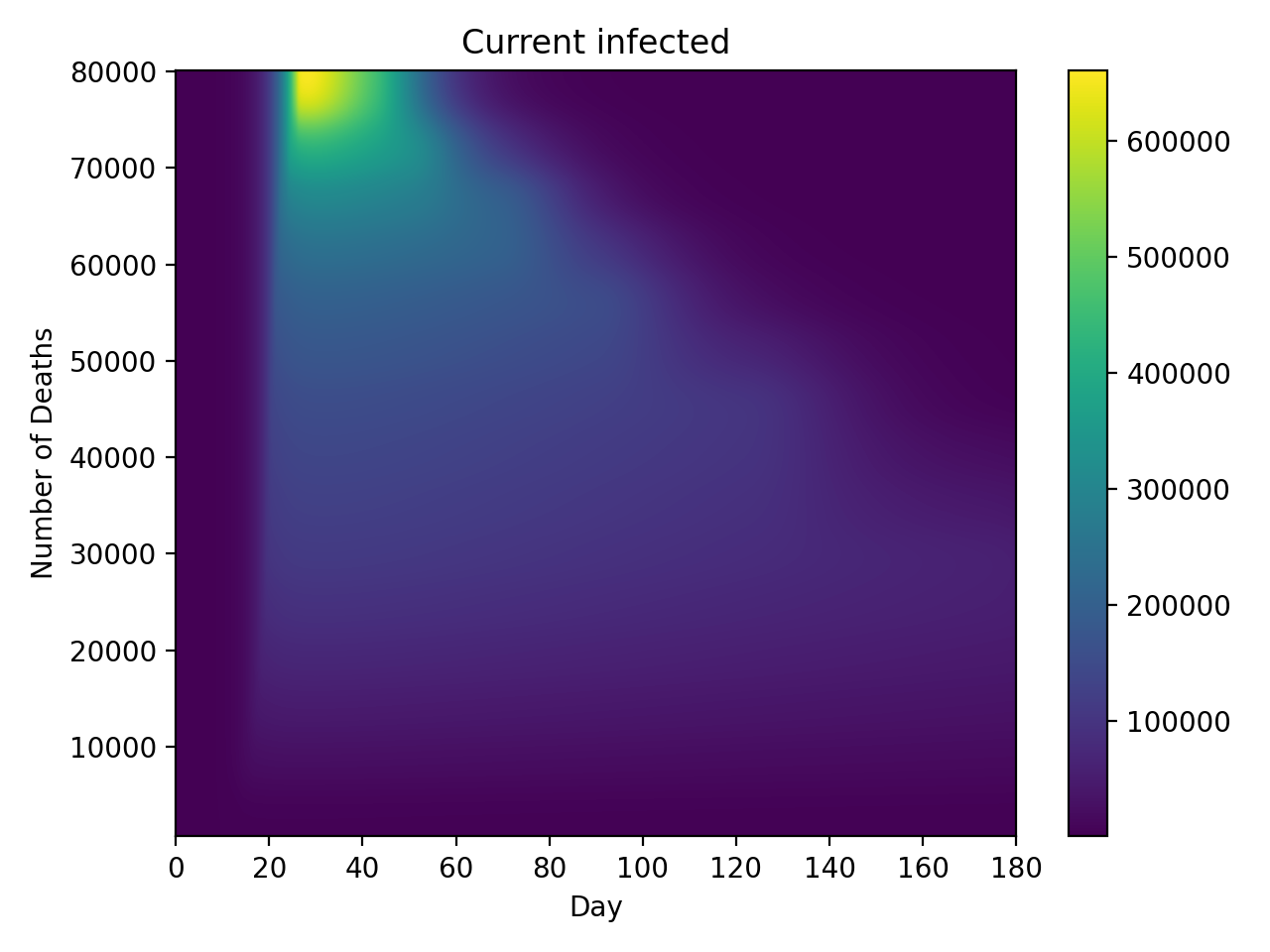}
    \end{subfigure}
   \hfill
 \begin{subfigure}[b]{0.24\columnwidth}
        \includegraphics[width=1.95in]{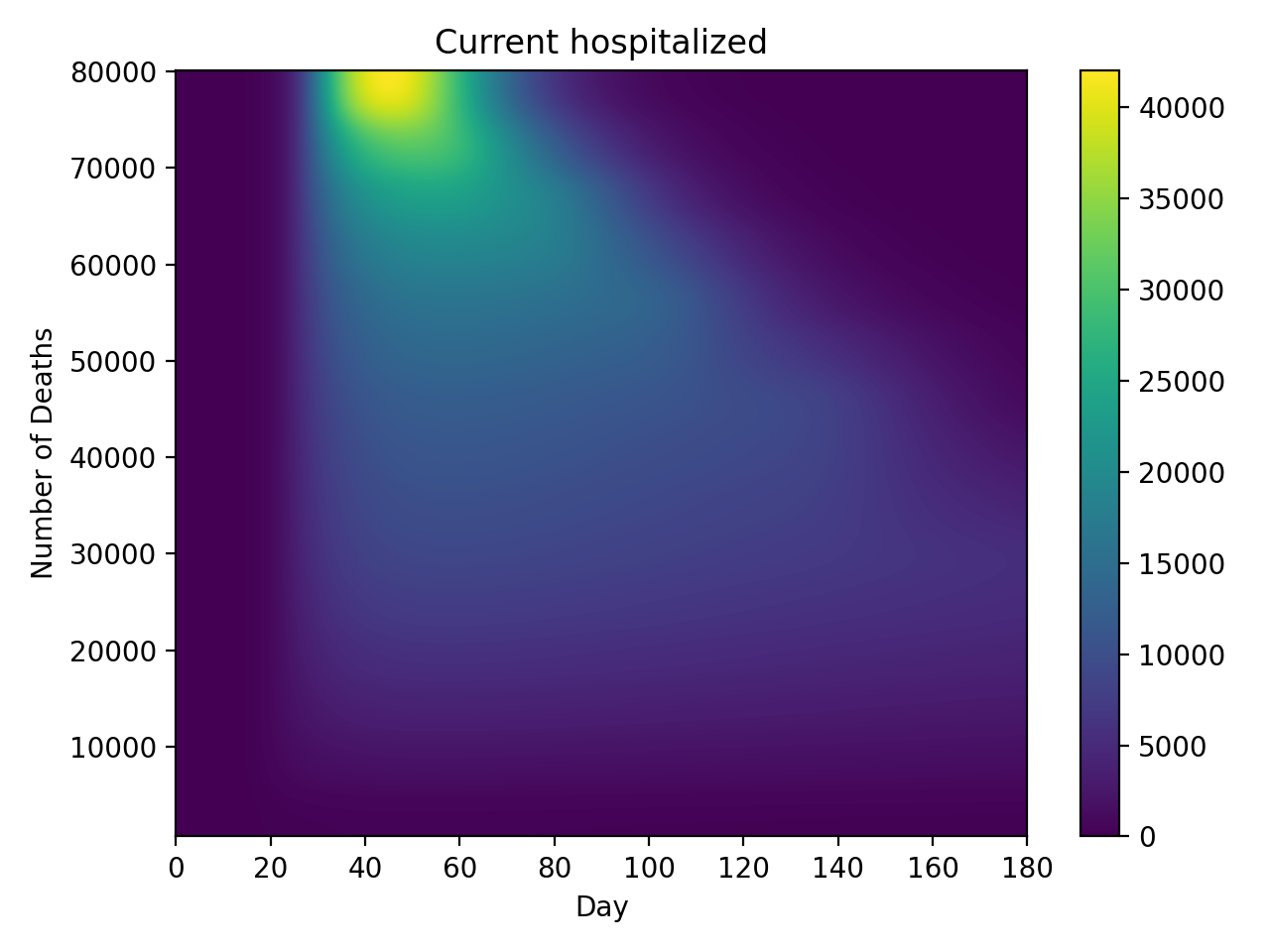}
    \end{subfigure}
    \hfill
\begin{subfigure}[b]{0.24\columnwidth}
    \includegraphics[width=1.95in]{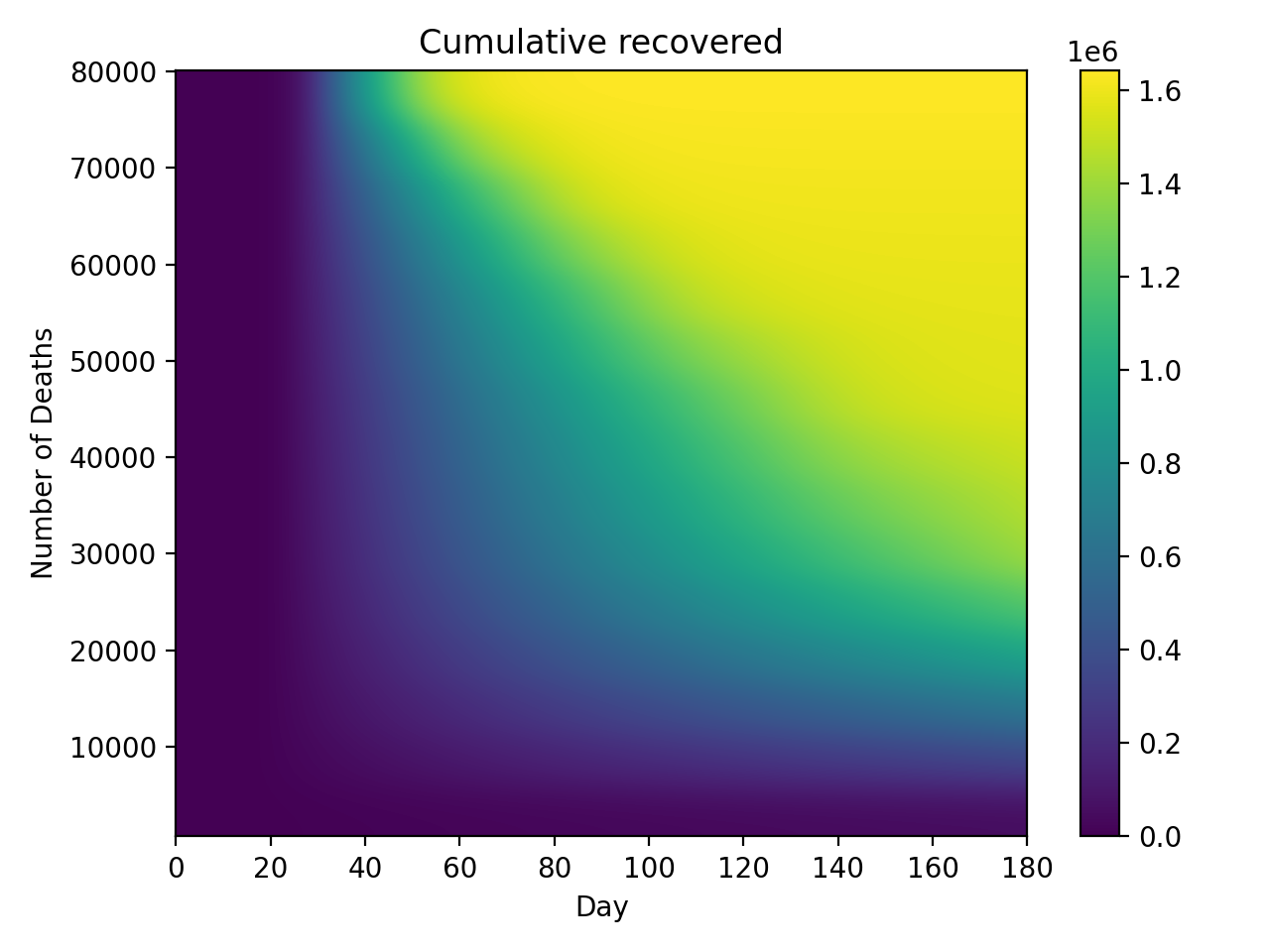}
    \end{subfigure}
   \hfill
 \begin{subfigure}[b]{0.24\columnwidth}
 \includegraphics[width=1.95in]{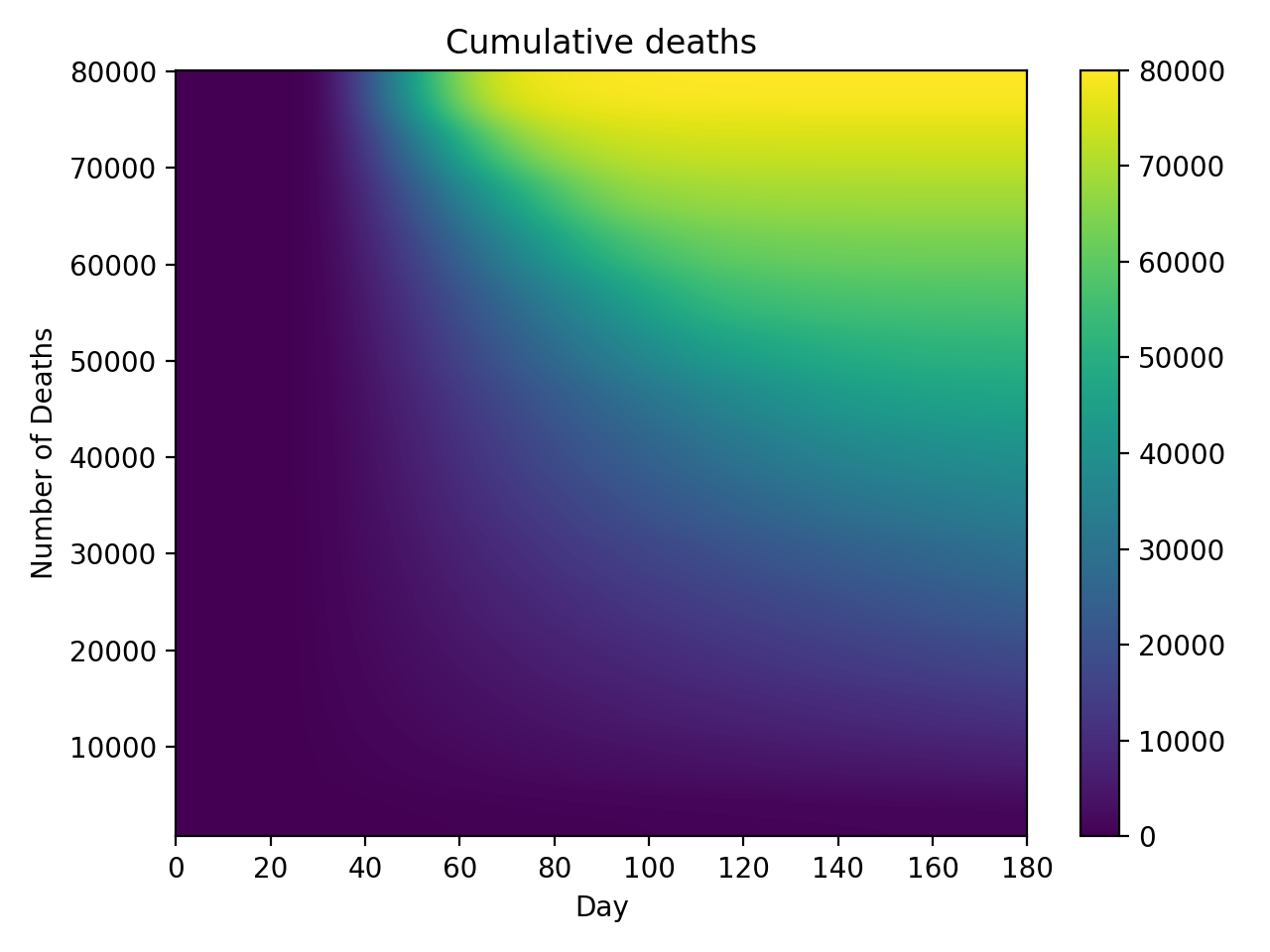}   
\end{subfigure}
\caption{Infections, hospitalizations, recovered, and deaths as a function of time (horizontal axis) and total number of deaths (vertical axis), for constant budget (first row), constant $R_e$ fraction (second row), and constant $R_e$ target (third row) basic control strategies. Color scales for plots correspond to infections (first column), hospitalizations (second column), recovereds (third column) and deaths (fourth column).}\label{fig:BasicInfHosRecDeaths}
\end{figure}
\end{landscape}

\begin{landscape}
 \begin{figure}
     \begin{subfigure}[b]{0.24\columnwidth}
        \includegraphics[width=1.95in]{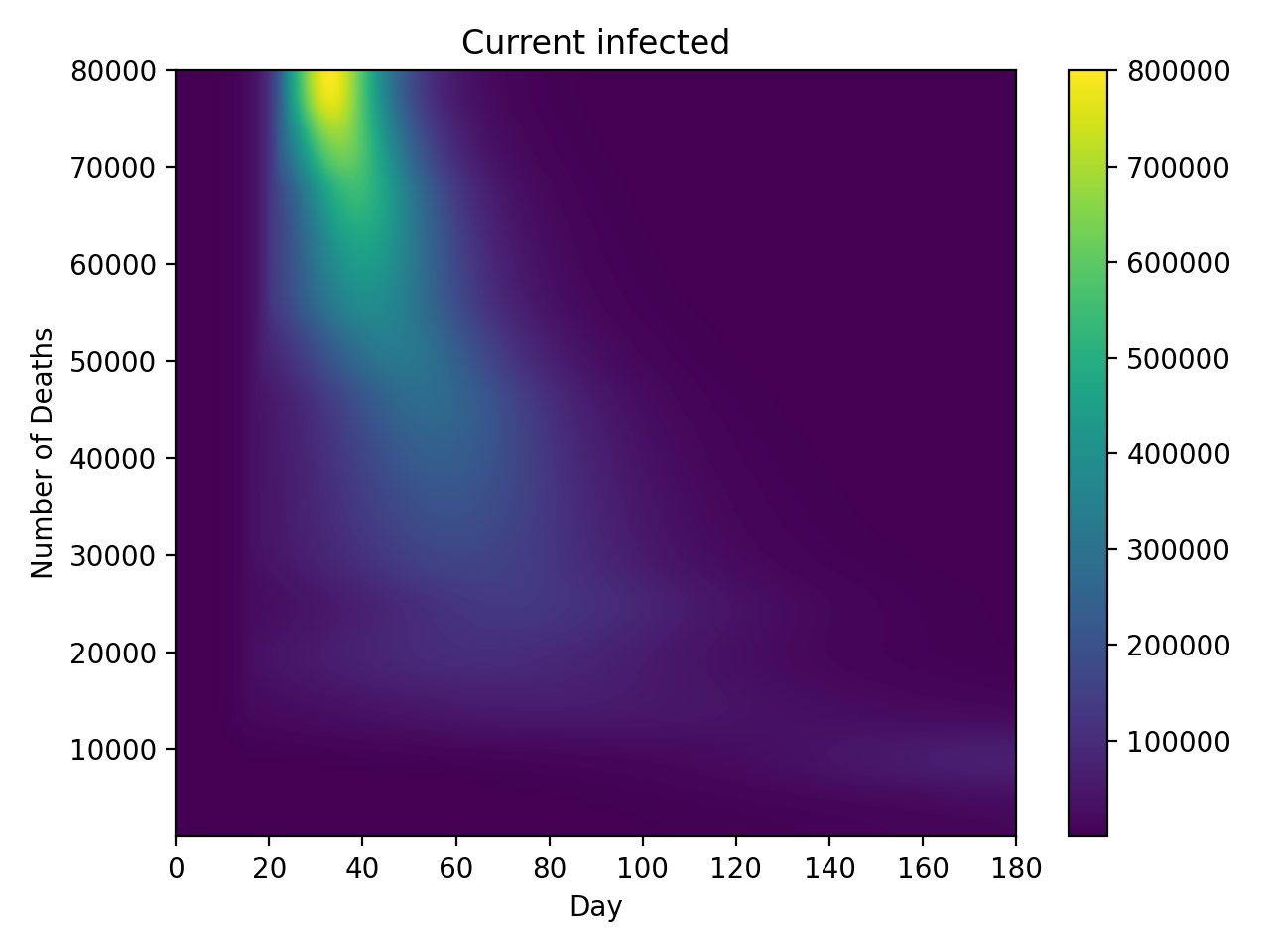}
    \end{subfigure}     
    \hfill
     \begin{subfigure}[b]{0.24\columnwidth}
       \includegraphics[width=1.95in]{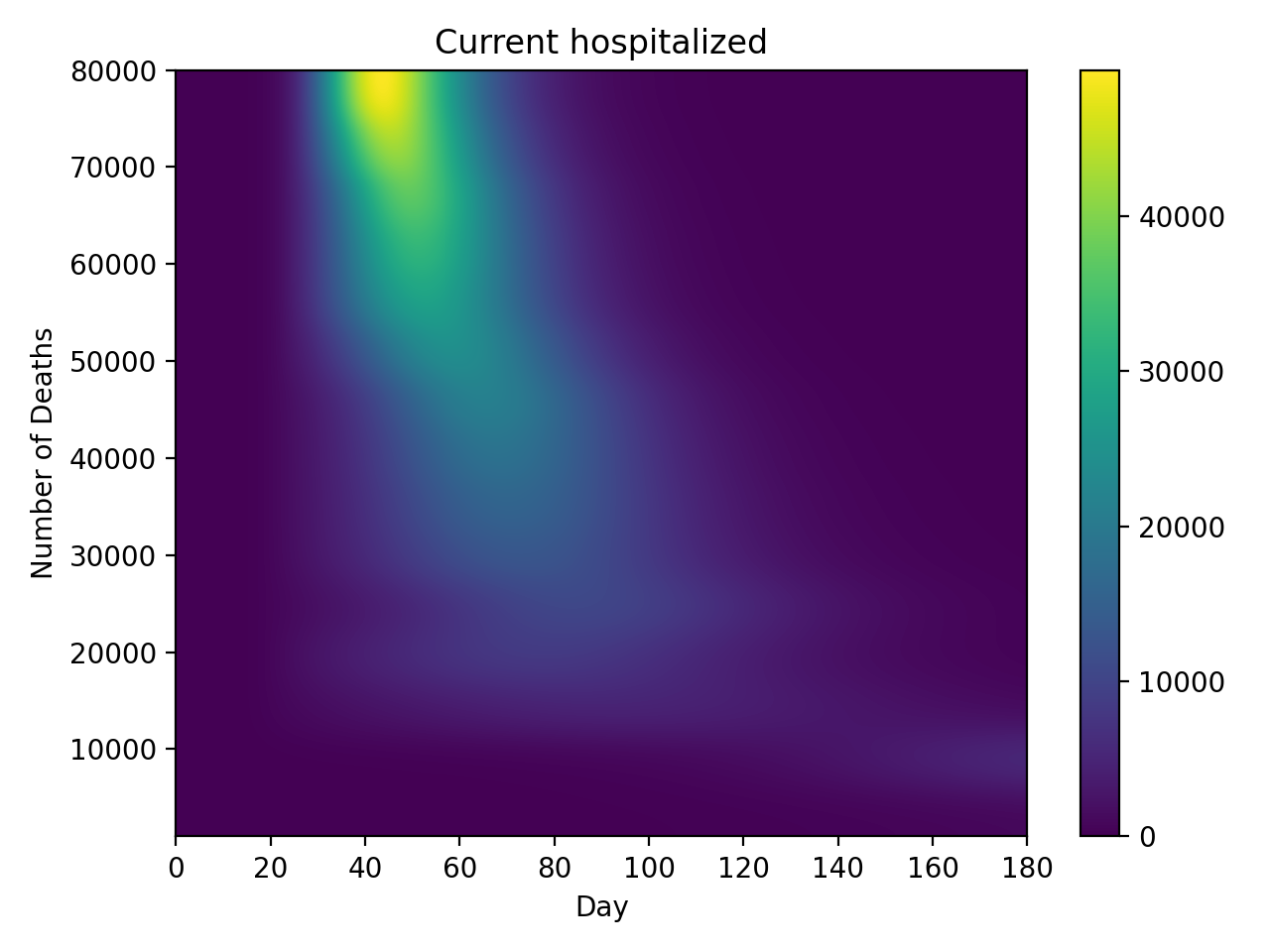}
    \end{subfigure}
 \hfill
    \begin{subfigure}[b]{0.24\columnwidth}
        \includegraphics[width=1.95in]{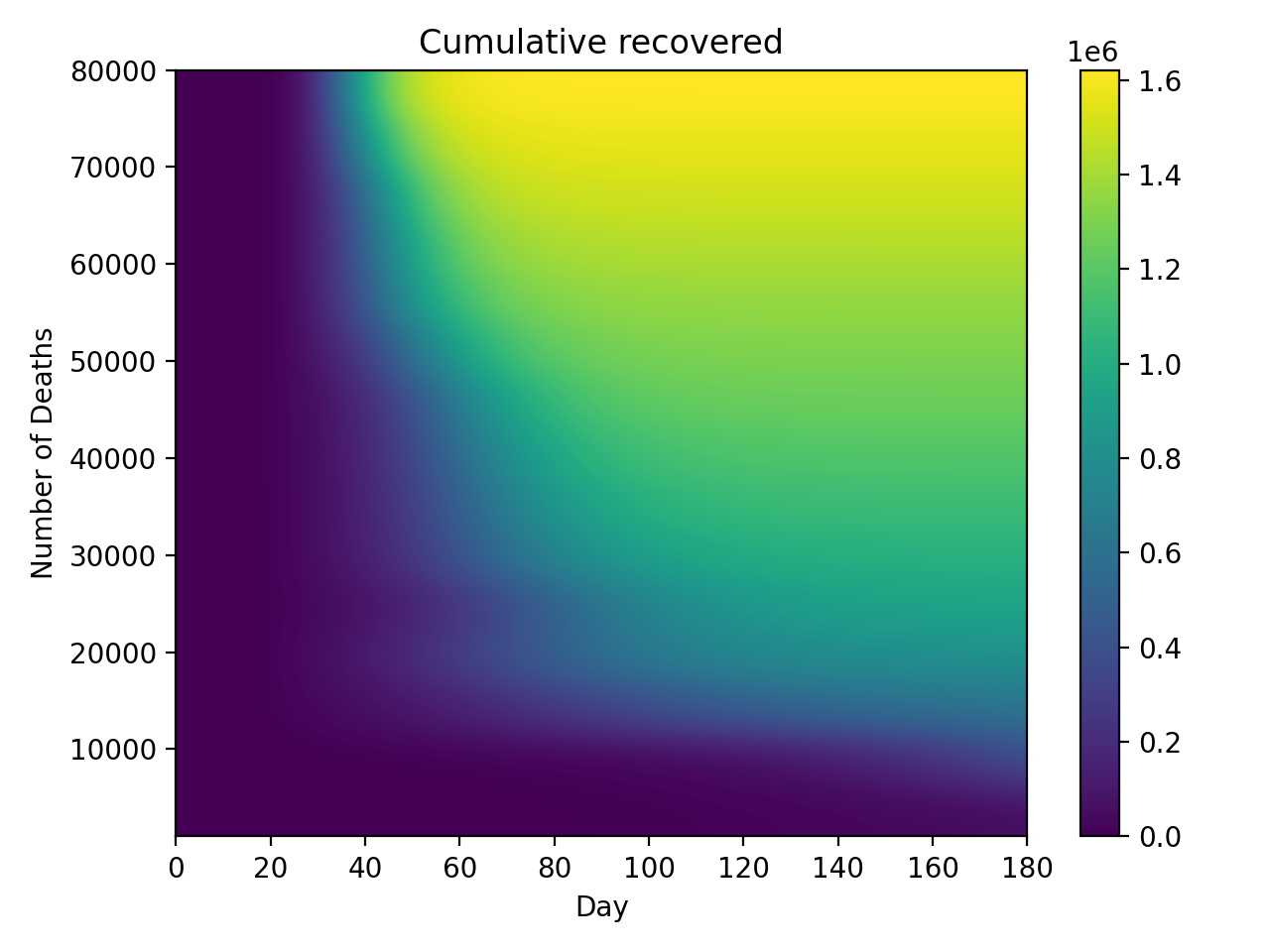}
    \end{subfigure}
    \hfill
    \begin{subfigure}[b]{0.24\columnwidth}
      \includegraphics[width=1.95in]{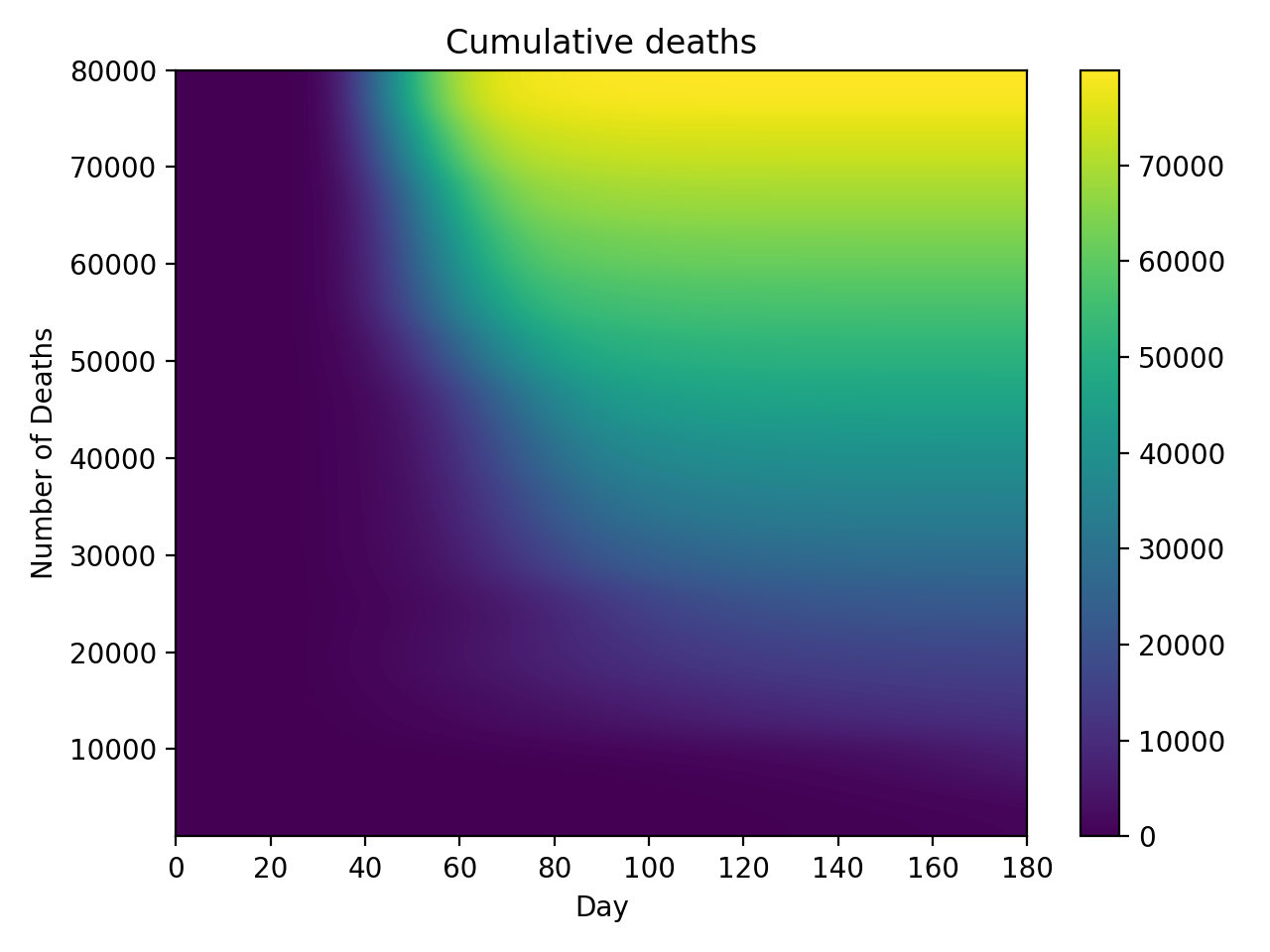}
    \end{subfigure}  
  \begin{subfigure}[b]{0.24\columnwidth}
       \includegraphics[width=1.95in]{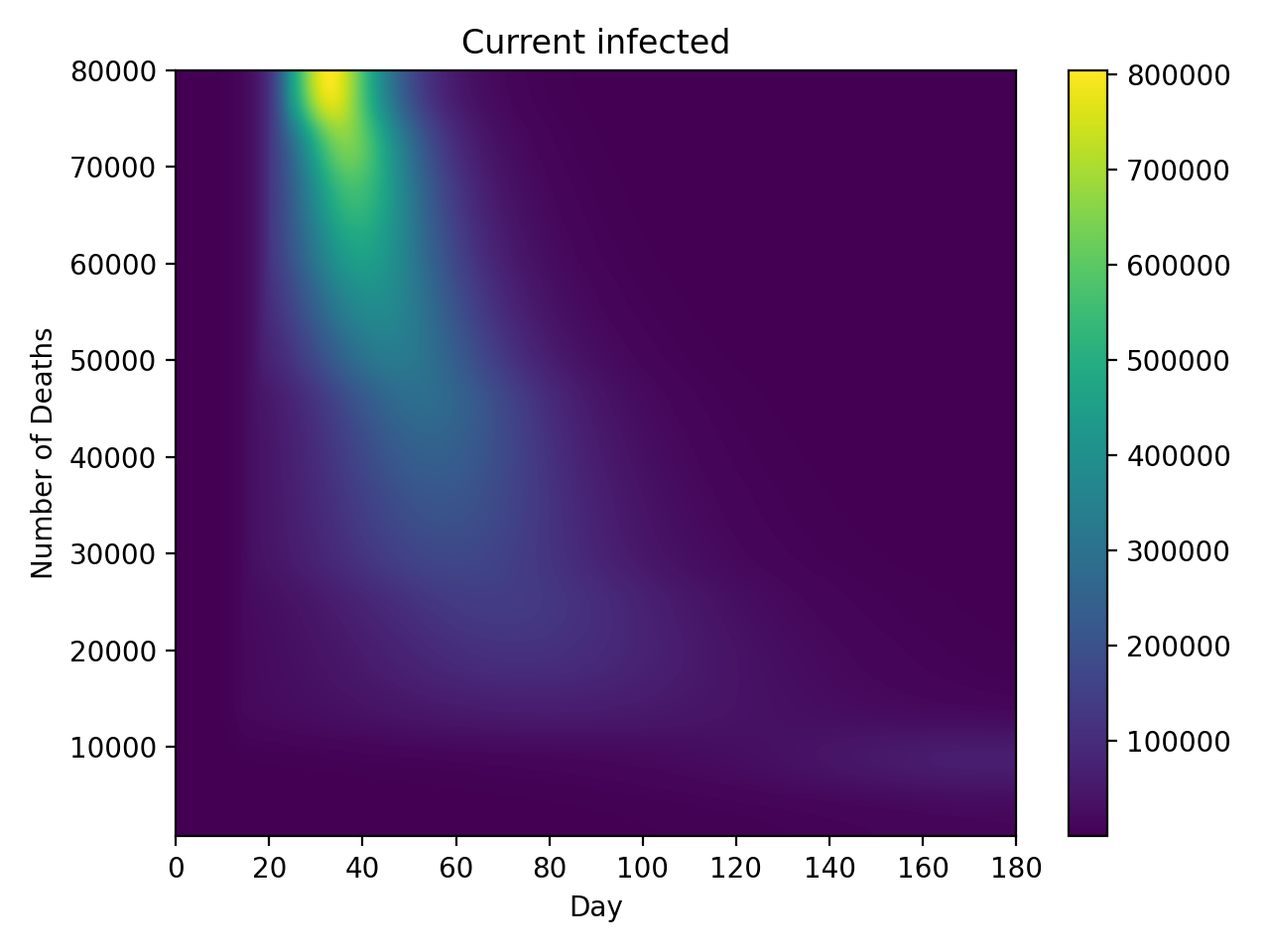}
    \end{subfigure}
 \hfill
 \begin{subfigure}[b]{0.24\columnwidth}
      \includegraphics[width=1.95in]{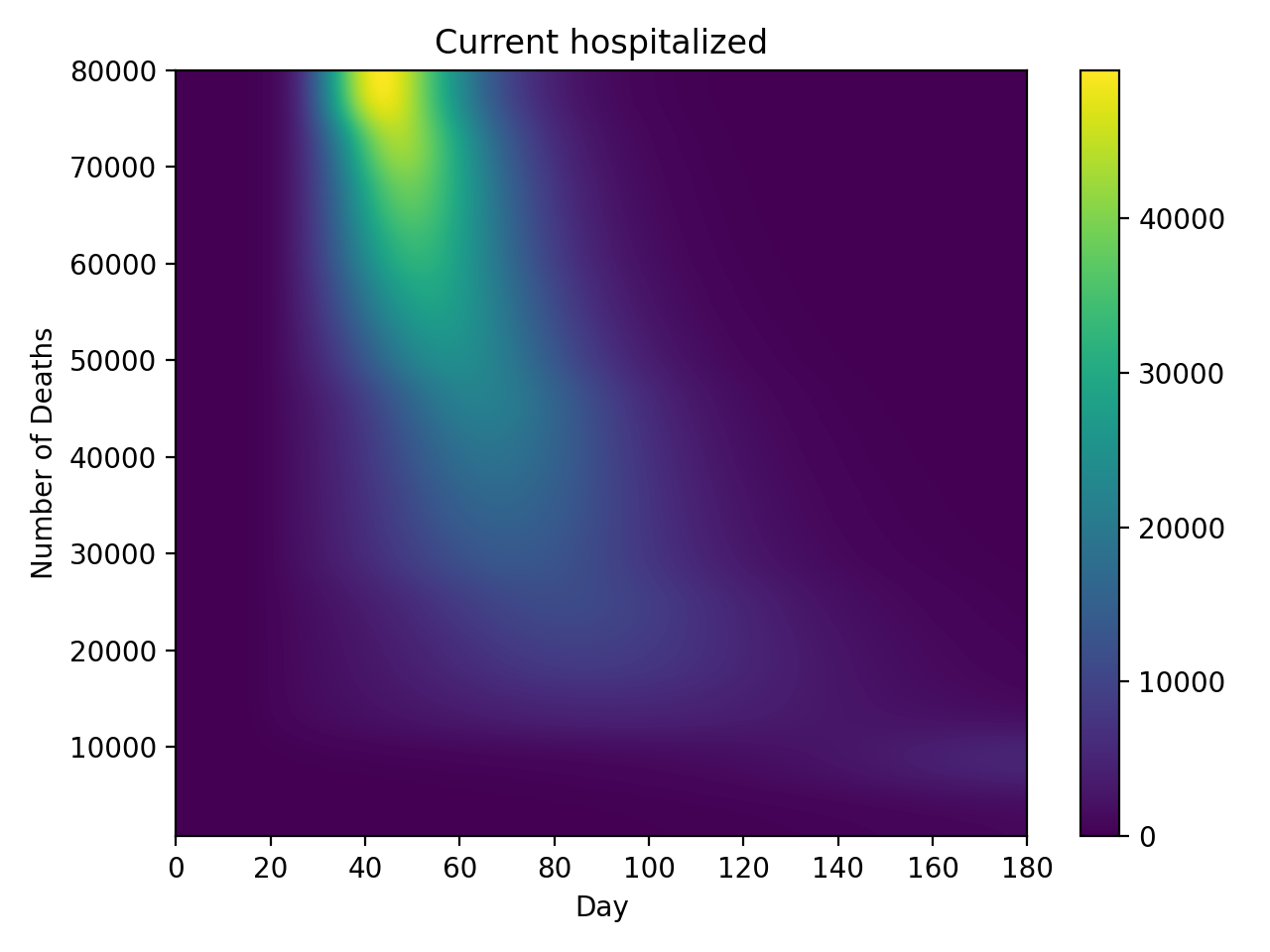}
    \end{subfigure}
    \hfill
 \begin{subfigure}[b]{0.24\columnwidth}
        \includegraphics[width=1.95in]{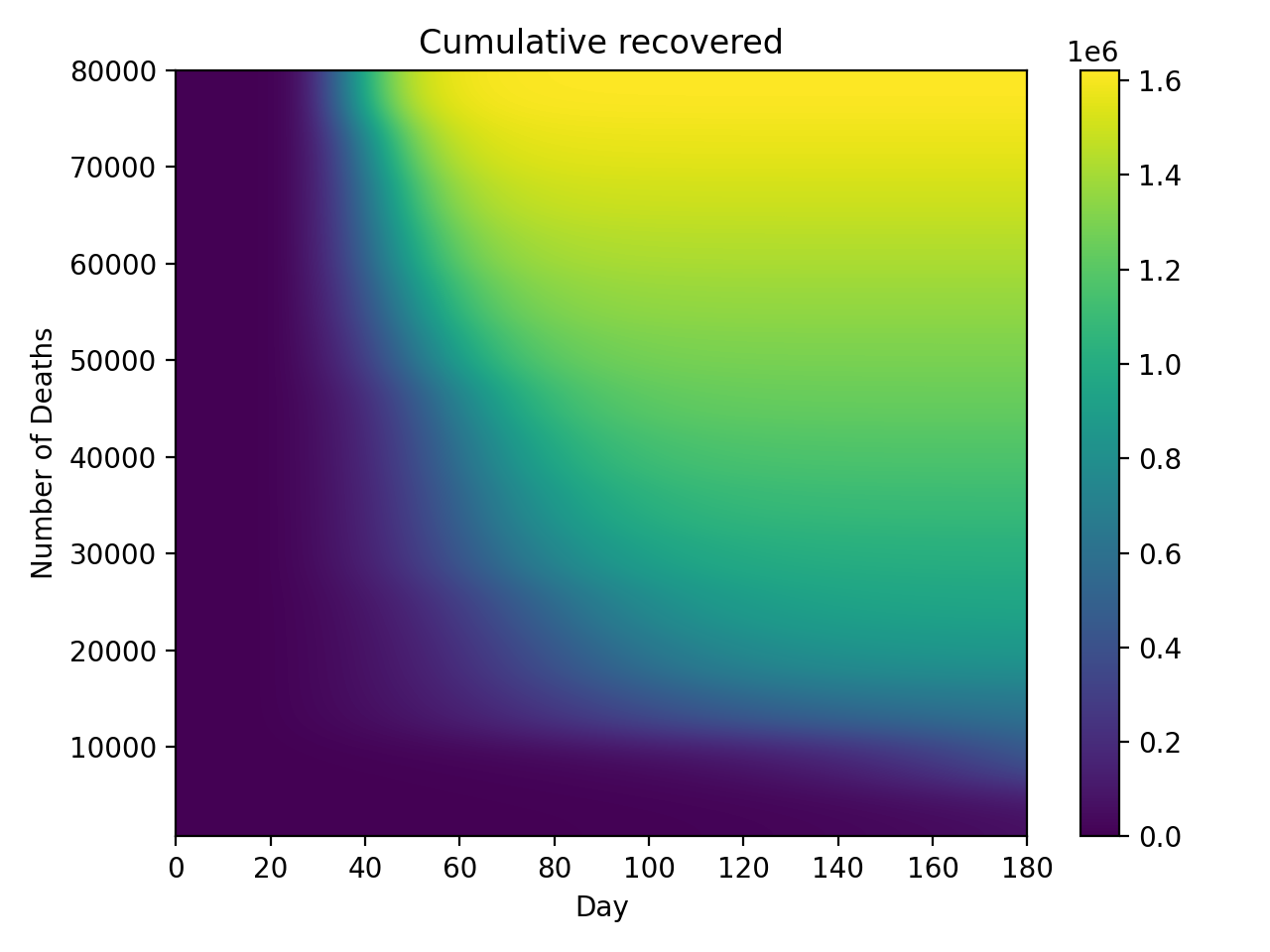}
    \end{subfigure}
    \hfill
    \begin{subfigure}[b]{0.24\columnwidth}
        \includegraphics[width=1.95in]{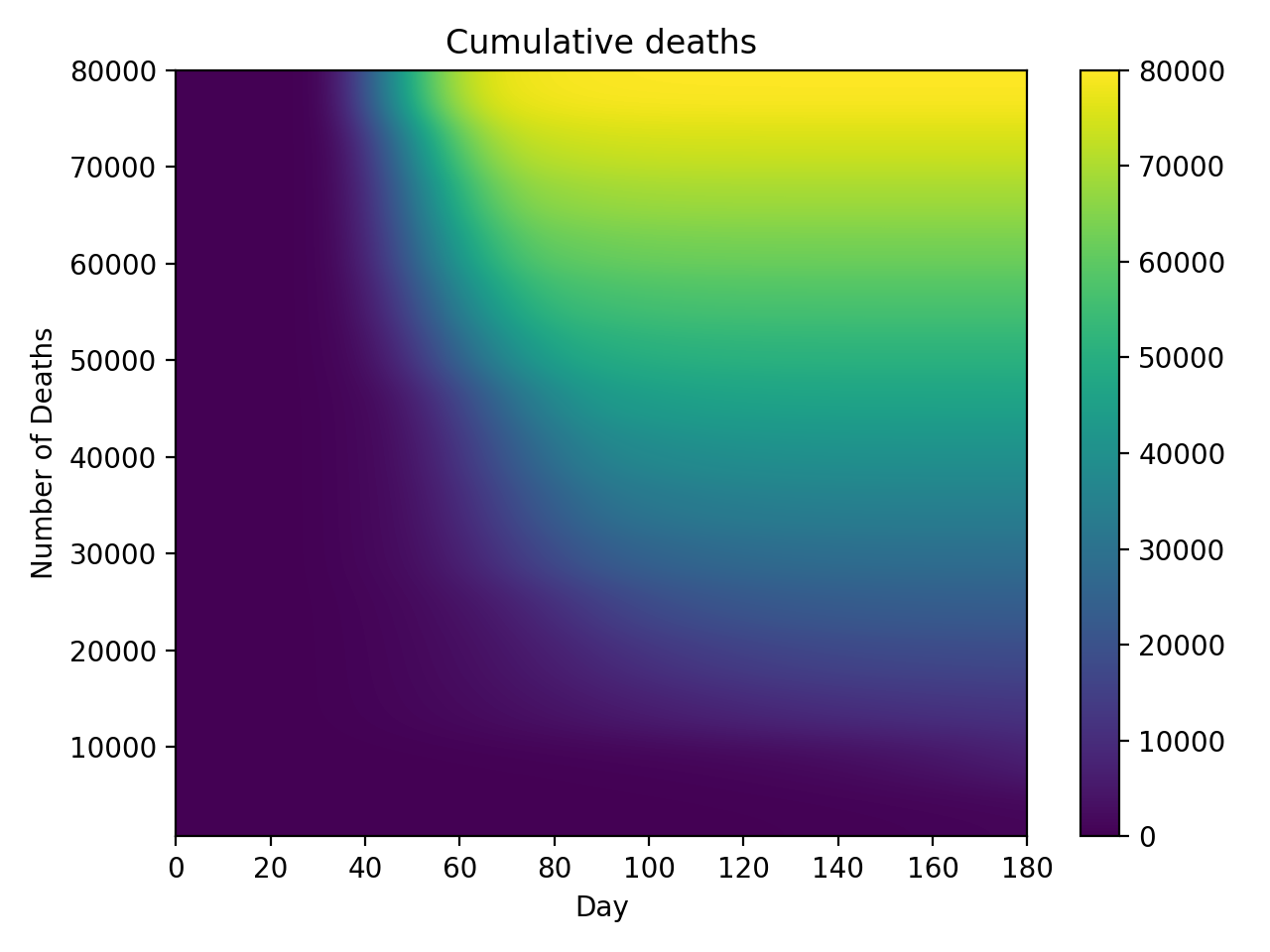}
    \end{subfigure}
    \hfill
\begin{subfigure}[b]{0.24\columnwidth}
        \includegraphics[width=1.95in]{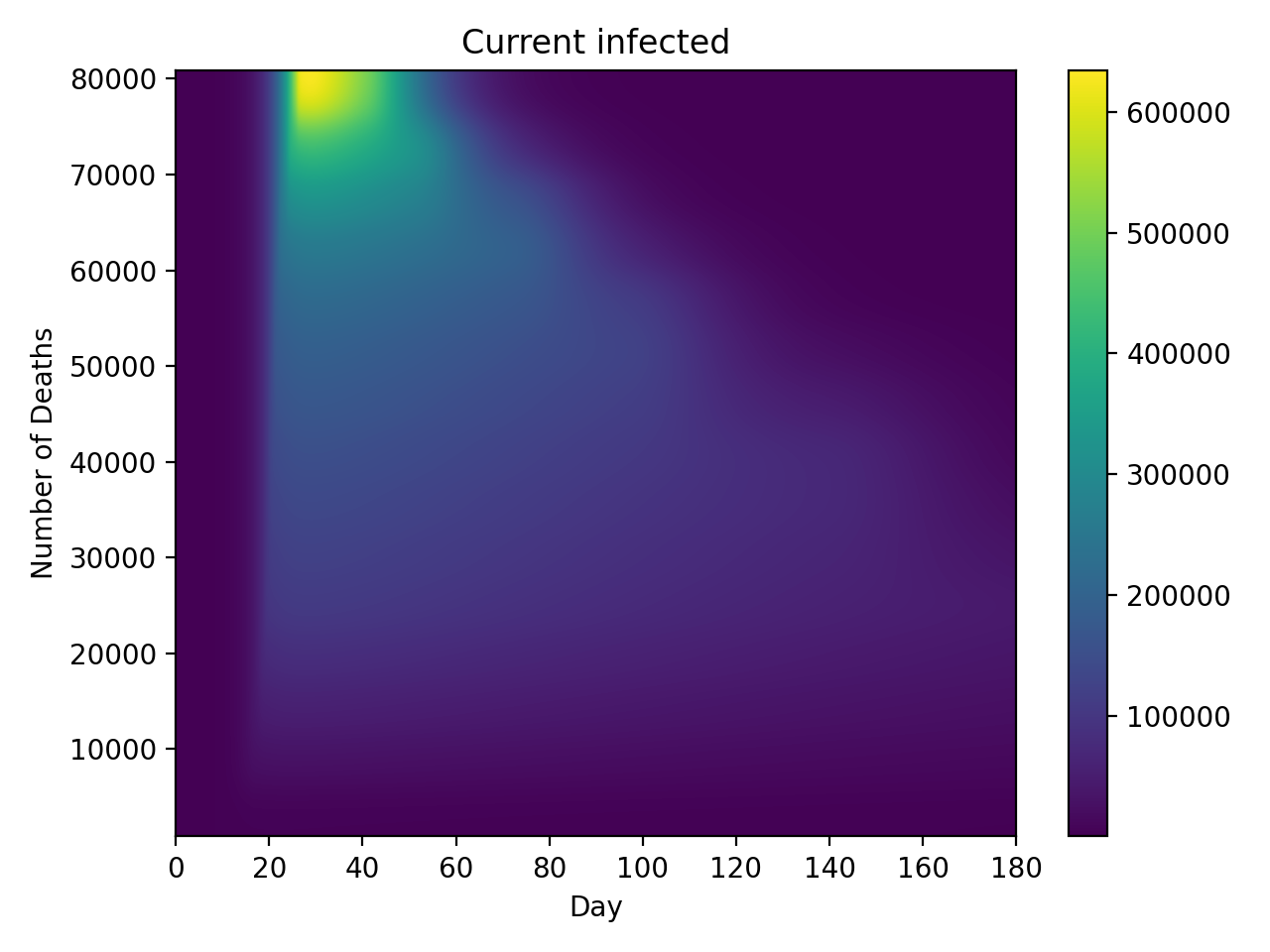}
    \end{subfigure}
    \hfill  
    \begin{subfigure}[b]{0.24\columnwidth}
        \includegraphics[width=1.95in]{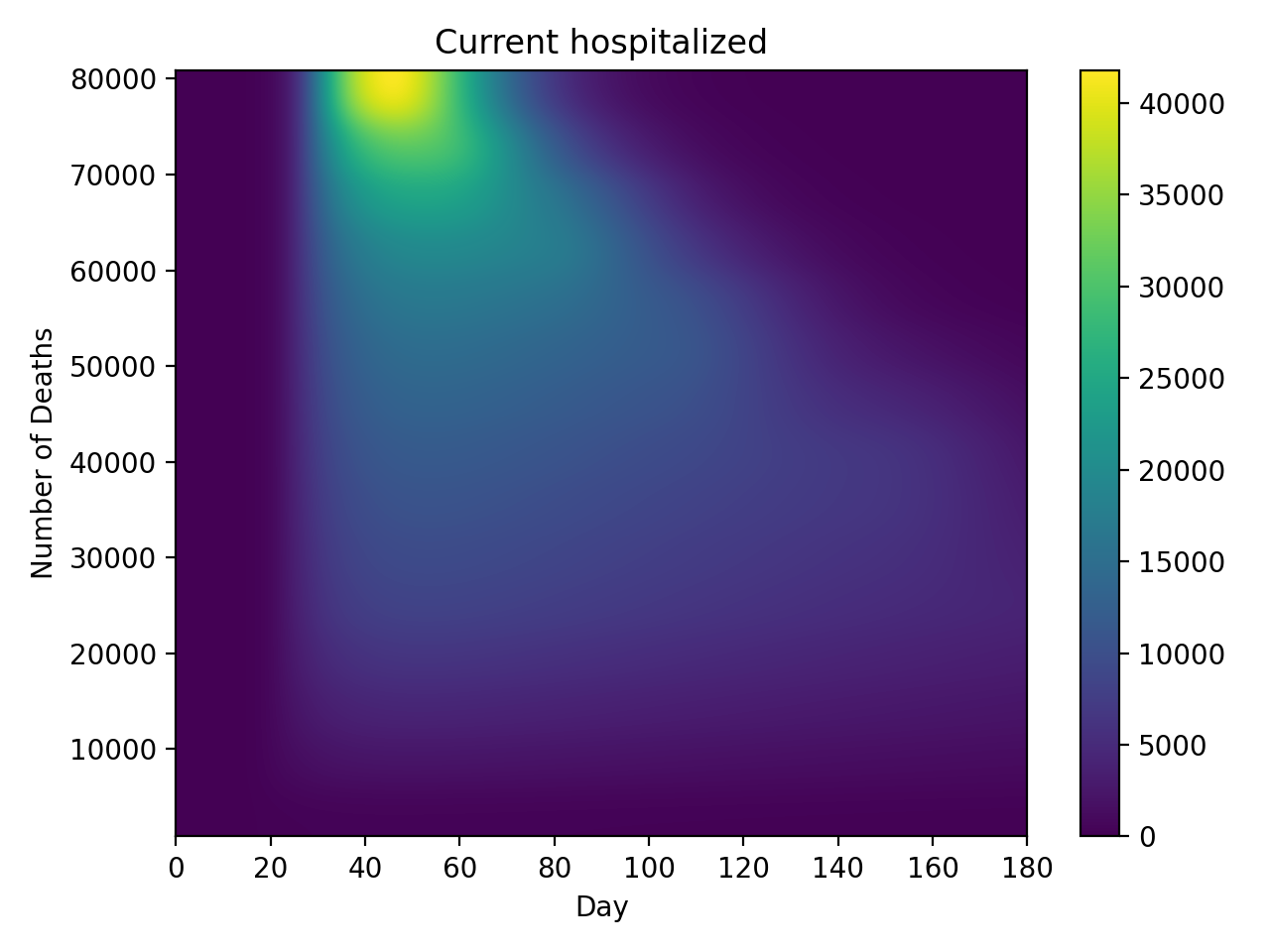}
    \end{subfigure}
    \hfill
    \begin{subfigure}[b]{0.24\columnwidth}
        \includegraphics[width=1.95in]{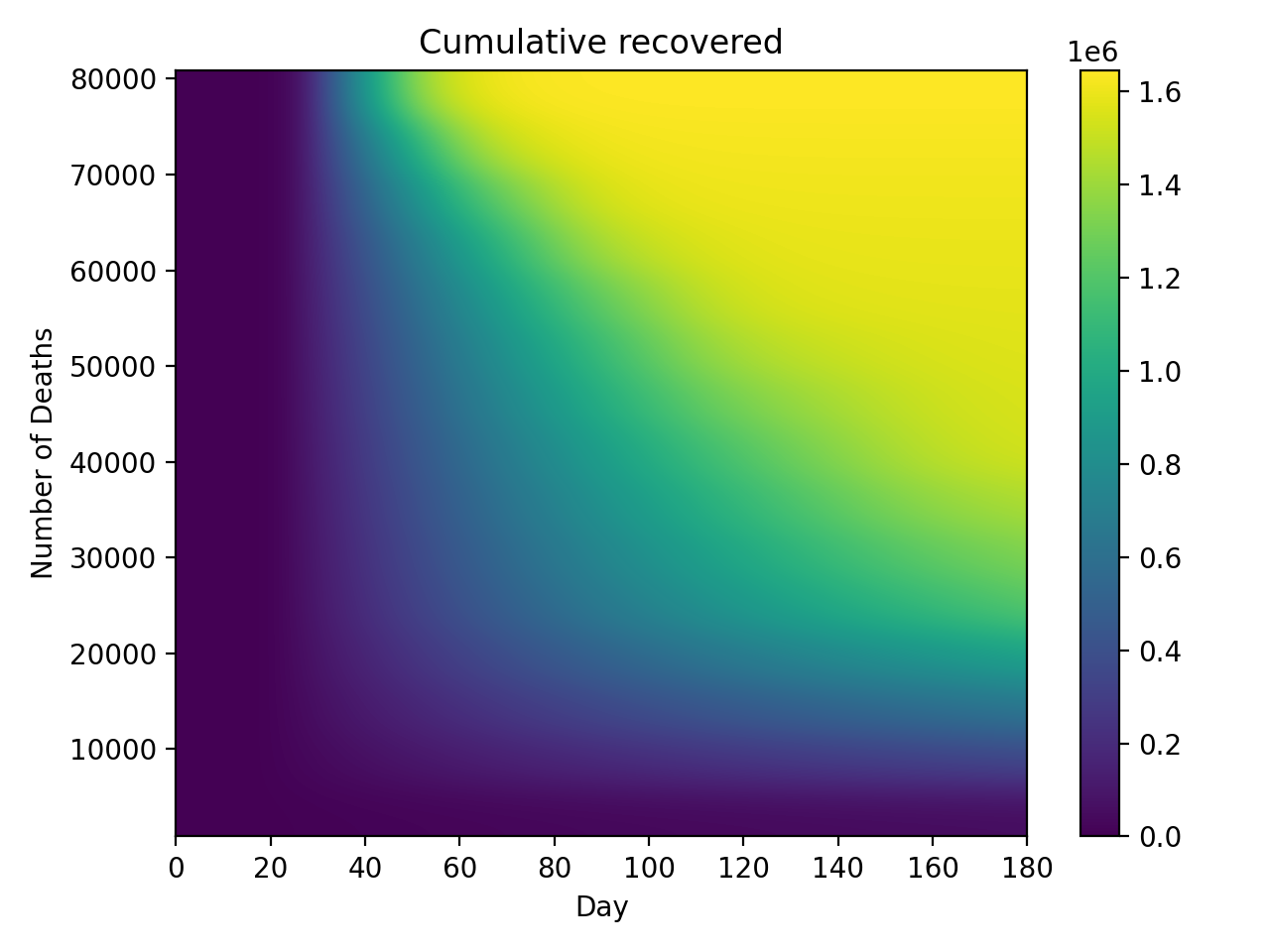}
    \end{subfigure}
    \hfill  
     \begin{subfigure}[b]{0.24\columnwidth}
       \includegraphics[width=1.95in]{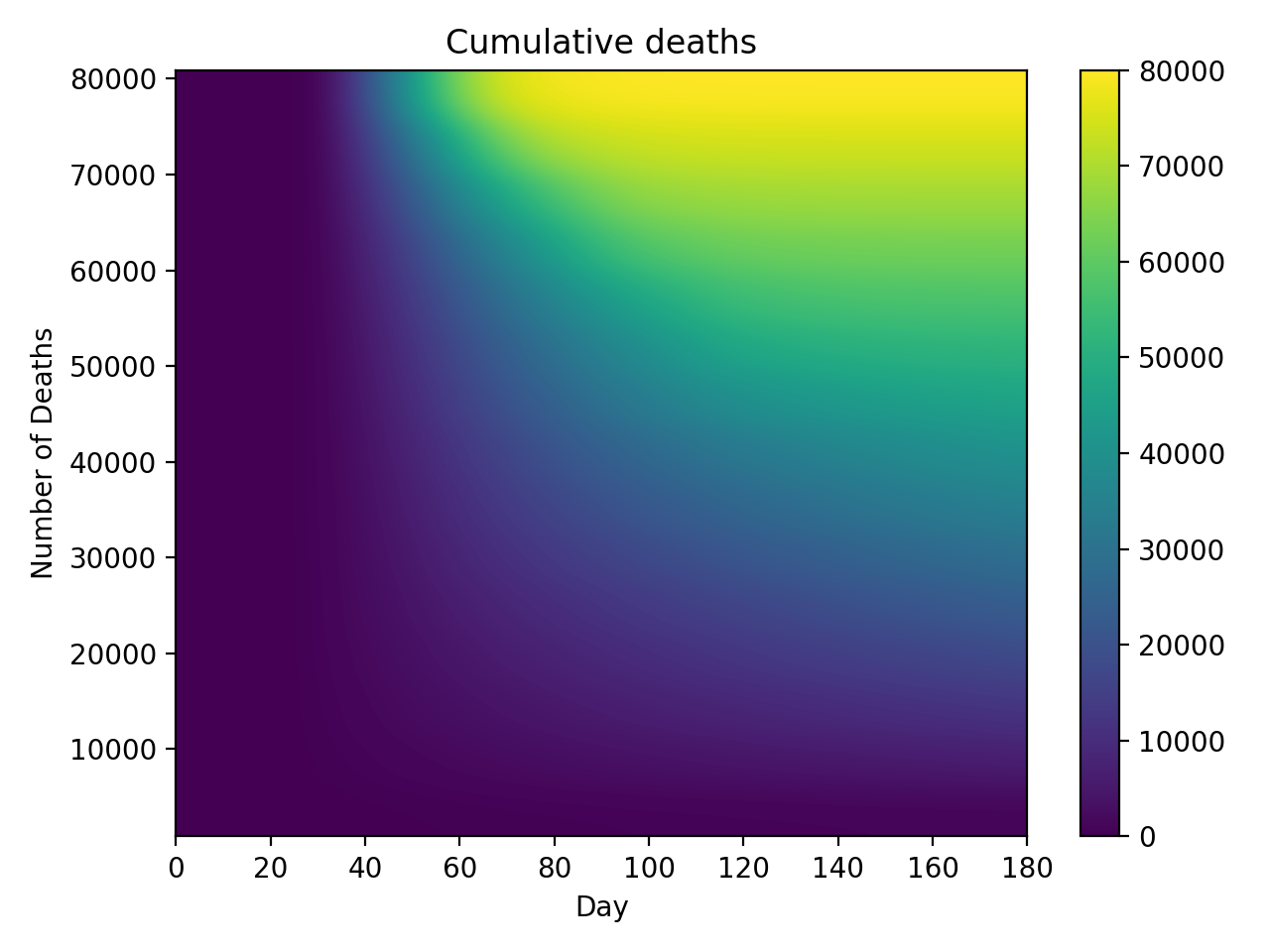}
    \end{subfigure}
\caption{Infections, hospitalizations, recovered, and deaths as a function of time and total number of deaths, for constant budget (first row), constant $R_e$ fraction (second row), and constant $R_e$ target (third row) control strategies that maximize controls on the high risk subpopulation. The plots are arranged as in Figure~\ref{fig:BasicInfHosRecDeaths}. }\label{fig:MaxInfHosRecDeaths}
 \end{figure}
 \end{landscape}

Figure \ref{fig:BasicInfHosRecDeaths} shows the progression over time of current infected, current hospitalized, cumulative recovered, and cumulative deaths corresponding to the three basic strategies shown in Figure~\ref{fig:four figures}. Both deaths and recovered increase with decreasing intensity of control. All three strategies show a peak of infected around 30 days, and a hospitalized peak around 40 days: both peaks are flatter with the $R_e$ target strategy compared to the constant budget and $R_e$ fraction strategies. For strategies with high levels of control, infections and hospitalizations persist up to the end of the 180-period.  For example, for constant $R_e$ strategies that reduce deaths below 60,000, even at 180 days, current infection and hospitalizations up to about 50,000 and 5,000 respectively may be experienced.   
Figure~\ref{fig:MaxInfHosRecDeaths} is analogous to Figure \ref{fig:BasicInfHosRecDeaths}, except that current infected, current hospitalized, cumulative recovered, and cumulative deaths are shown for the three strategy types where high risk individuals are subjected to maximum control. The above observations made for Figure \ref{fig:BasicInfHosRecDeaths} also apply to Figure \ref{fig:MaxInfHosRecDeaths}.

\FloatBarrier
\section{Conclusion}\label{sec:4}
In this paper,  an SEIR epidemic model of COVID-19 in the city of Houston, TX USA is presented under testing and social distancing controls with low and high risk population groups. The basic and effective reproduction numbers  for the model have been calculated, and the effective reproduction number has been explored as a key parameter in understanding the dynamics of disease and its relationship with different control measures and strategies. Comprehensive graphical representations of the dependence of effective reproduction number $R_e$ and control cost on control levels have been presented under different levels of population immunity (Figures~\ref{fig:RevsCtrl}-\ref{fig:DistCostTestLevelCost}). 
Restricted strategies that used only one control (either distancing or testing) and/or targeted only part of the population (either high or low risk) were incapable of reducing $R_e$ below 1, implying that such strategies are not sufficient to prevent disease spreading except in cases of high levels of population immunity. 

A sensitivity analysis was performed, which showed that both costs and $R_e$ as well as optimal distancing levels are highly sensitive to the baseline transmission rate, and less sensitive to symptomatic proportion and the relative infectiousness of asymptomatic individuals (Figures~\ref{fig:Levelpara00}-\ref{fig:Levelpara66}). Hence given the difficulty in obtaining exact values for baseline transmission rate, it may be difficult to determine precisely the best distancing policy for given conditions. Overall costs were also found to be highly sensitive to model cost parameters, particularly the increasing costs associated with diminishing returns from social distancing (Figures~\ref{fig:Levelimm00}-\ref{fig:Levelimm66}).  

Optimal instantaneous strategies which combined distancing and testing were computed (Figures~\ref{fig:Optmix0}-\ref{fig:Optmix66}).  The results showed that optimal strategies utilized distancing primarily (especially at high control levels), and were applied nearly equally to both population groups. 
Three different types of long-term control strategies based on $R_e$ were simulated.
The simulations confirm that the starting date of the control has an enormous effect on the effectiveness of the strategies  in preventing deaths, and the minimum number of deaths increases rapidly if controls are delayed past a certain point (Figures~\ref{fig:budget}-\ref{fig:Re_tgt}). However, the results showed that it is not most cost-effective to begin serious controls too soon: for example, it was found that the best $R_e$ targeting strategy that can reduce deaths to 30,000 was begun on day 9 (Figure~\ref{fig:StartDateImpactOnDeaths}). Although the more intensive (and more costly) strategies  reduced the number of deaths, they also do not entirely eliminate the infection, and it was found that all strategies which reduced deaths below 60,000 required continuing control past the 180 day period of the simulation (Figures~\ref{fig:four figures}-\ref{fig:MaxInfHosRecDeaths}).
 Strategies that set a target value for $R_e$ were found to be most cost-effective, even when started later than  other strategies (Figure~\ref{fig:StartDateImpactOnDeaths}). These strategies are characterized by an initial very high level of distancing, which is later reduced and replaced by higher levels of testing (\ref{fig:four figures}-\ref{fig:four figures2}). Strategies that focused primarily on applying controls to the high-risk population were found to be less cost-effective than strategies that were applied evenly across the entire population (Figure~\ref{fig:StartDateImpactOnDeaths}). 
 
The situation with the COVID epidemic, as with previous epidemics, is continuously changing. The development of vaccines introduces new possibilities for control. The baseline model we have developed in this research can readily be modified to accommodate such changes. 
\section*{Abbreviations}
\begin{enumerate}
\item COVID-19 : Coronavirus disease 2019,\\
\item SEIR: Susceptible, Exposed, Infectious, Recovered,\\
\item TX, USA: Texas, United States of America,\\
\item WHO: World Health Organization,\\
\item SIR: Susceptible, Infectious, Recovered,\\
\item SARS-CoV-2: Severe Acute Respiratory Syndrome Coronavirus 2,\\
\item PCR: Polymerase Chain Reaction,\\
\item COVID:  Coronavirus disease .
\end{enumerate}
\section*{Acknowledgements}
The support of this research through the German Academic Exchange Service (DAAD) and the anonymous reviewers are hereby acknowledged.
\section*{Conflict of interest}
The authors declare that they have no conflict of interest.
\section*{Author's contributions}
All authors evenly contributed to the whole work. All authors read and approved the final manuscript.
\section*{Availability of data and materials}
The software used for the simulations in this paper is contained in the project ``COVID19 Cost Effectiveness Of Control Measures",  DOI:10.5281/zenodo.4474866 (\url{https://github.com/LuisEVT/COVID19_Cost_Effectiveness_Of_Control_Measures}).

\bibliographystyle{bmc-mathphys}
\bibliography{BiblioCOVID}
\addcontentsline{toc}{chapter}{References}
\end{document}